\documentclass[10pt,aps,prx,twocolumn,amsmath,amssymb]{revtex4-2}
\usepackage[table]{xcolor}
\usepackage[colorlinks=false]{hyperref}
\usepackage[utf8]{inputenc}
\usepackage{color}
\usepackage{bbm} 
\usepackage[english]{babel}
\usepackage{multirow}
\usepackage{amsfonts,amsmath,amssymb,stmaryrd}

\usepackage{graphicx}
\usepackage{filecontents}

\usepackage{epsfig}
\usepackage{mathrsfs}
\usepackage{verbatim}
\usepackage{centernot}
\usepackage{ulem}
\usepackage{nicefrac}
\usepackage[framemethod=tikz]{mdframed}
\usepackage{siunitx}
\usepackage[nolist]{acronym}

\usepackage{float} 
\usepackage{soul}
\usepackage{array}

\usepackage{cancel,ifthen}
\newcommand{\cmnt}[2][NoInPuT]{\ifthenelse{\equal{#1}{NoInPuT}}{}{{\color{red}\sout{#1}}} {\color{blue} #2}}

\usepackage{bm}	% bold in math mode

\newcommand{\q}{\mathbf{q}}
\newcommand{\dt}[1]{\frac{\partial #1}{\partial t}}

\newcommand{\dz}[1]{\frac{\partial #1}{\partial z}}

\begin{document}

\normalem	% changes \emph back to normal after introducing ulem package.

\title{Huge ultrafast spin Seebeck effect 
mediated by laser-excited superdiffusive
magnon currents 
%and ballistic-to-diffusive crossover of laser-excited nonequilibrium 
%{\red (?)}
%Boltzmann-theory simulation of laser-excited superdiffusive magnon current
%Ultrafast spin Seebeck effect carried by laser-excited superdiffusive magnon currents
}

\author{Luca Mikadze}
\email[]{luca.mikadze@physics.uu.se}
\affiliation{Department of Physics and Astronomy, Uppsala University, P.\ O.\ Box 516, S-751 20 Uppsala, Sweden}

\author{Peter M. Oppeneer}
\affiliation{Department of Physics and Astronomy, Uppsala University, P.\ O.\ Box 516, S-751 20 Uppsala, Sweden}

\author{Markus Wei{\ss}enhofer}
 \affiliation{Department of Physics and Astronomy, Uppsala University, P.\ O.\ Box 516, S-751 20 Uppsala, Sweden}


\date{\today}

\begin{abstract}
Subpicosecond laser excitation of ferromagnetic metals induces strongly nonequilibrium dynamics involving scattering and transport of electrons, phonons, and magnons. Widely used theoretical approaches, such as the three-temperature model and diffusion equations, are ill-suited to capture these processes on ultrafast timescales. Here, we present an \textit{ab initio}-parameterized microscopic framework that incorporates nonthermal magnon scattering and transport via the quantum Boltzmann equation. We apply this approach to simulate ultrafast laser-induced demagnetization in bcc Fe films. The model predicts an ultrafast spin Seebeck effect, characterized by a strong burst of fast-moving magnonic spin current reaching technologically relevant amplitudes.
Furthermore, we identify a superdiffusive transport regime: a crossover from initially ballistic magnon transport to a diffusive regime at later times. To connect our theoretical predictions to experimentally accessible observables, we calculate the magneto-optical Kerr angles resulting from the predicted depth-resolved magnetization profiles.
Our framework provides a route to describe ultrafast nonthermal magnon transport beyond diffusive models and will aid in the design and interpretation of time-resolved spin-transport experiments.

\end{abstract}

\maketitle

\begin{acronym}
\acro{AFM}[AFM]{antiferromagnetic}
\acro{FM}[FM]{ferromagnetic}
\acro{NM}[NM]{nonmagnetic}
\acro{AFMR}[AFMR]{antiferromagnetic resonance}
\acro{LLG}[LLG]{Landau-Lifshitz-Gilbert}
\acro{STT}[STT]{spin-transfer torque}
\acro{SSW}[SSW]{standing spin wave}
\acro{FWHM}[FWHM]{full width at half maximum}
\acro{LSWT}[LSWT]{linear spin-wave theory}
\acro{3TM}[3TM]{three-temperature model}
\end{acronym}

%%%%%%%%%%%%%%%%%%%%%%%%%%%%%%%%%%%%%%
\section{Introduction}

The field of {light-induced ultrafast magnetization dynamics} emerged in 1996 when Beaurepaire \textit{et al.}~\cite{Beaurepaire1996} first observed the sub-picosecond demagnetization of a ferromagnetic metal when radiated by a short and intense laser pulse. As the magnetization dynamics occurs much more rapidly than when using conventional methods of manipulating magnetism, this phenomenon holds great promise for advancing information storage technologies, enabling faster and more efficient data processing~\cite{Kirilyuk2010, Scheid2022, Chen2025}.

In spite of decades of intensive research, the microscopic origin of this phenomenon remains under debate~\cite{Kirilyuk2010,Carva2017,Scheid2022,Chen2025}. This is largely because many well-established physical equilibrium paradigms cannot be simply transferred to the ultrafast regime, due to the strongly nonequilibrium dynamics present on these timescales. 

Several studies~\cite{Carpene2008,Schmidt2010,Carpene2015,Turgut2016,Eich2017,Yamamoto2019,Frietsch2020} suggest that transverse spin excitations play a dominant role. These correspond to a reduction of the average magnetization due to mutual tilting of atomic moments. In contrast, other works~\cite{Rhie2003,Cinchetti2006,Koopmans2010,Schellekens2013b,Griepe2023} point toward longitudinal excitations, in which the magnitude of the atomic magnetic moments are reduced by spin-flip processes that decrease the exchange splitting. Beyond these local mechanisms, nonlocal contributions driven by superdiffusive spin currents~\cite{Battiato2010,Battiato2012} can also be significant, particularly in metallic heterostructures~\cite{Malinowski2008,Rudolf2012,Bergeard2016,Xu2017,Hofherr2017,Gupta2023}.

Theoretical descriptions of ultrafast dynamics in metals commonly separate electronic, phononic, and, if magnetization dynamics are relevant, spin degrees of freedom. A prominent example is the \ac{3TM}, introduced by Beaurepaire \textit{et al.}~\cite{Beaurepaire1996}. The model assumes that each subsystem is internally thermalized, with electrons described by a Fermi–Dirac distribution, and phonons and spin excitations by Bose–Einstein distributions. The dynamics are governed by three coupled differential equations for the corresponding temperatures. While successfully reproducing many experimental results, the \ac{3TM} lacks a description of the angular momentum transfer associated with demagnetization~\cite{Dornes2019,Tauchert2022}. Moreover, numerous studies have demonstrated that on ultrafast timescales, these distributions deviate significantly from thermal equilibrium~\cite{Sun1993,Sun1994,DelFatti1998,DelFatti2000,Guo2001,Carpene2006,Mueller2013,Tveten2015,Maldonado2017,Maldonado2020,Wilson2020,Ritzmann2020,Beens2020,Beens2022,Shokeen2024}.

In a recent work, Wei{\ss}enhofer and Oppeneer~\cite{Weissenhofer2024} demonstrated the importance of nonthermal spin dynamics during ultrafast demagnetization. They developed an \textit{ab initio}-parametrized model based on quantum kinetic theory for electron-magnon scattering, coined the \textit{(N+2)-temperature model} ((N+2)TM), where $N$ denotes the number of magnon modes. This model is an extension of the \ac{3TM}, in which transverse spin excitations (magnons) are no longer assumed to thermalize instantaneously on these short femtosecond timescales, and demagnetization is a result of \textit{local} changes in magnon populations. The model reproduces demagnetization curves in remarkable agreement with experimental measurements for thin (\SI{7}{\nano\metre}) Fe films for pump-probe delay times up to \SI{10}{\pico\second}~\cite{Carpene2008,Weissenhofer2024}. However, in thicker films, transport effects are expected to contribute to the demagnetization in a \textit{nonlocal} manner. To date, theoretical efforts have primarily focused on out-of-equilibrium electron transport~\cite{Battiato2010,Battiato2012,Balaz2018,Balaz2023}, while magnon transport remains largely unexplored, particularly beyond diffusive descriptions~\cite{Xiao2010,Cornelissen2016,Beens2022}.

In this work, we extend the (N+2)TM to include thermal and superdiffusive magnon transport, enabling \textit{ab initio} predictions of demagnetization dynamics in films of arbitrary thickness, as well as the magnitude and spatial extent of propagating magnon currents. The magnon dynamics is described by the quantum Boltzmann equation, going beyond diffusive transport treatments. Using this extended model, we investigate the impact of particle transport on the ensuing demagnetization dynamics, adopting Fe films as model example. For films thicker than the laser penetration depth, we obtain depth-resolved demagnetization profiles and a significant {ultrafast magnonic spin Seebeck effect}, driven by strong temperature gradients arising from the inhomogeneous absorption of the laser pulse.

The paper is structured as follows. In Sec.~\ref{sMethods}, we present the extended (N+2)TM, the numerical implementation, and the parameters used in our study. In Sec.~\ref{sResults}, we report the resulting ultrafast dynamics. Finally, Sec.~\ref{sConclusions} summarizes our findings and conclusions.

\section{Methodology}
\label{sMethods}

In this section, we present the theoretical framework used in this work. Building on the (N+2)TM introduced in Ref.~\cite{Weissenhofer2024}, we extend the model to include spatial dynamics and transport effects. To distinguish between the two approaches, we refer to the original formulation as the \textit{local} (N+2)TM and to the extended version as the \textit{nonlocal} (N+2)TM. We first briefly review the local model before introducing its nonlocal extension and the associated numerical implementation.

\subsection{Local (N+2)-temperature model}
\label{slN+2TM}

\begin{figure}[t]
    \centering
\includegraphics[width=0.99\linewidth]{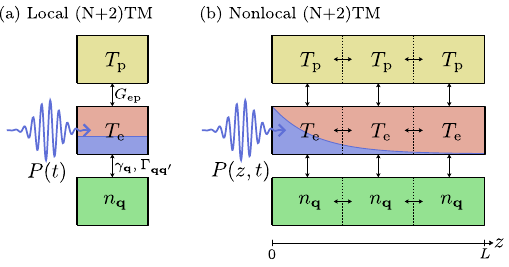}
    \caption{Illustration of the method. (a) In the local (N+2)TM, the laser pulse $P$ impinges on a metal film and heats the thermal electronic system $T_e$ homogeneously, which, in turn, is coupled to a thermal phonon bath $T_p$ and nonthermal magnon occupations $n_\q$. (b) The nonlocal (N+2)TM allows for a heterogeneous laser absorption profile resulting in temperature gradients. Transport occurs within each subsystem across the isolated film domain $z \in [0,L]$, and the subsystems exchange heat locally.}
    \label{fsetup}
\end{figure}

The local (N+2)TM  [Fig.~\ref{fsetup}(a)] describes energy transfer between electrons, phonons, and magnons based on the following set of coupled differential equations~\cite{Weissenhofer2024},
\begin{align}
    \dot{n}_\q &=  s_\q(\{n_{\q}\},T_\mathrm{e}) ,\label{eq:lN+2TMmag} \\
    C_\mathrm{e}\dot{T}_\mathrm{e} &=   P(t) - G_\mathrm{ep}(T_\mathrm{e}-T_\mathrm{p}) -\frac{1}{NV}\sum_\q \hbar\omega_\q s_\q ,\label{eq:lN+2TMe}\\
    C_\mathrm{p}\dot{T}_\mathrm{p} &= G_\mathrm{ep}(T_\mathrm{e}-T_\mathrm{p}) \label{eq:lN+2TMph}.
\end{align}
The dynamical variables are the electron/phonon temperatures $T_\mathrm{e/p}$, and the magnon occupation numbers $n_\q$ (here for a single magnon band, but generalization to multiple bands is possible), where $\q$ denotes a point on an equidistant grid in the first Brillouin zone. $N$ is the number of magnon modes, $V$ is the volume per spin site, $C_\mathrm{e/p}$ are the electron/phonon specific heat capacities, $G_\mathrm{ep}$ is the effective electron-phonon coupling constant, and $P(t)$ is the laser power absorbed by the electrons. Note that direct magnon-phonon coupling is not considered, which has been shown to be a reasonable approximation for $3d$ ferromagnets~\cite{Zahn2021,Zahn2022}.

The local (N+2)TM is an extension of the \ac{3TM} that also captures nonthermal magnon dynamics: the occupation numbers are not restricted to a specific distribution. This is in stark contrast to the \ac{3TM}, where the magnons are assumed to always follow a Bose-Einstein distribution for some temperature.

In the \textit{local} (N+2)TM, the dynamical variables depend only on time $t$ but not on position $\mathbf{r}$, i.e., they are assumed to be homogeneous throughout the system. As we will show below, this assumption is only valid in thin films.

The electron-magnon scattering terms 
\begin{eqnarray}
\label{eSourceMag}
%\begin{split}
    &&s_\q(\{n_{\q}\},T_\mathrm{e})
    = 
    \left[n^0(\omega_\q,T_\mathrm{e}) - n_\q\right]\gamma_\q  + \nonumber
    \\
   &&
    \sum_{\q'} \left[(n_\q+1)n_{\q'}n^0(\omega_\q-\omega_{\q'},T_\mathrm{e}) + (\q \leftrightarrow \q') \right]\Gamma_{\q\q'} ,~~~
%\end{split}
\end{eqnarray}
are derived from an \textit{sp-d} Hamiltonian-based description of the interactions between electrons and magnons, a common approach~\cite{Zener1951,Zener1951b,Mitchell1957,Heinrich1967,Tserkovnyak2004,Zhang2004,Manchon2012,Tveten2015,Brener2017,Barbeau2022,Remy2023}, together with Fermi's golden rule~\cite{Weissenhofer2024}.
Here, $n^0(\omega,T) = [\mathrm{e}^{\hbar\omega/k_\mathrm{B}T}-1]^{-1}$ is the Bose-Einstein distribution, and $\omega_\q$ is the angular frequency of the magnons in state $\q$. The first term in Eq.~\eqref{eSourceMag} corresponds to electron-magnon scattering events in which an electron flips its spin and absorbs or emits a magnon. The term drives the magnon state $\q$ toward thermal equilibrium with the electrons, at a first-order scattering rate $\gamma_\q$. The derivation of this term~\cite{Weissenhofer2024} is analogous to what was laid out by Allen in his seminal work~\cite{Allen1987} on electron-phonon scattering. The second term represents magnon-number-conserving scattering events in which electrons and magnons scatter from one state to another. It redistributes magnons among their states at a second-order scattering rate $\Gamma_{\q\q'}$. Because of this second-order process, the scattering term at point $\q$ depends on all magnon occupation numbers, as indicated by $\{n_{\q}\}$.

A potential source of confusion is the name (N+2)-\textit{temperature} model, as the formulation employs $N$ magnon occupation numbers. This terminology originates from the \ac{3TM}, since the occupations $n_\q$ can be mapped onto mode-dependent effective temperatures via inversion of the Bose-Einstein distribution: $T_\q=\hbar \omega_\q/(k_\mathrm{B} \ln (n_\q^{-1} + 1))$.

We want to point out that the local (N+2)TM is a parameter-free model---all coefficients are obtained from first-principles calculations---that provides excellent agreement with experimental demagnetization measurements for ultrathin Fe films~\cite{Carpene2008,Weissenhofer2024}.

We briefly discuss the range of applicability and limitations of the (N+2)TM. Within this framework, the distribution of magnons in their quantum states is not constrained and is free to become nonthermal. In contrast, and the same as in the \ac{3TM}, electrons and phonons are each individually assumed to be in thermal equilibrium and therefore follow Fermi-Dirac/Bose-Einstein statistics with temperatures $T_\mathrm{e/p}$, that may differ from one another, at each point in time. This is clearly an approximation to the actual dynamics after ultrafast laser-excitation, where all occupations are initially nonthermal and thermalize on different characteristic timescales. For electrons in metals, typical relaxation times are on the order of a few tens of femtoseconds~\cite{Aeschlimann1997}, so this approach is justified for timescales longer than this, while for phonons they are of the order of several picoseconds~\cite{Maldonado2017,Ritzmann2020,Maldonado2020,Shokeen2024}. However, given that (i) the direct coupling between magnons and phonons is very weak in $3d$ transition-metal ferromagnets~\cite{Zahn2021,Zahn2022}, to which we apply this model here, and (ii) the focus of this work is on magnon dynamics, we nevertheless believe that this assumption is a reasonable approximation here.

A change in magnetization is related to a transfer of angular momentum. The (N+2)TM implicitly assumes that angular momentum transferred to the electronic system via electron–magnon scattering is rapidly dissipated---e.g., through coupling to the phonons---so that no appreciable angular momentum accumulation occurs to bottleneck electron–magnon interactions.

Since higher order magnon terms, which give rise to magnon-magnon scattering and magnon softening, are neglected, the model is not applicable at temperatures close to the Curie temperature of the material and can only describe demagnetization for moderate laser fluences.

The (N+2)TM can be viewed as a quantum analog of the stochastic Landau-Lifshitz-Gilbert (LLG) equation~\cite{Nowak2007} coupled to a two-temperature model, a framework that has been widely used in describing ultrafast, out-of-equilibrium magnetization phenomena~\cite{Kazantseva_2008,Ostler2012,Frietsch2020,Zahn2021,Zahn2022}. We note that the $\q$-dependent damping processes of the (N+2)TM~\cite{Weissenhofer2024}, as described by Eq.~\eqref{eSourceMag}, go beyond the constant Gilbert-type damping of the LLG.

\subsection{Modeling (quasi)-particle transport}
\label{snlN+2TM}

In a realistic experimental scenario, interfaces and spatial variations in laser absorption lead to inhomogeneous dynamics~\cite{Malinowski2008,Rudolf2012,Schellekens2013,Lee2021,Qiu2025,Chardonnet2026}. In the framework of the (N+2)TM, the dynamical variables $T_\mathrm{e}$, $T_\mathrm{p}$, and $n_\q$, become position-dependent. In this work, we model these spatial dynamics using the quantum Boltzmann equation for magnons and the heat diffusion equation for the electrons and phonons.

In general, the quantum Boltzmann equation (QBE)~\cite{Uehling1933}
\begin{equation}
    \left(\dt{} + \mathbf{v} \cdot \nabla_\mathbf{r} + \frac{1}{\hbar} \mathbf{F} \cdot \nabla_\q\right)n = \left(\dt{n}\right)_{\mathrm{scatter}}, \label{eBoltzGeneral}
\end{equation}
describes the time evolution of an occupation probability density $n(\mathbf{r},\q,t)$, due to drift with velocity $\mathbf{v}$, acceleration induced by external forces $\mathbf{F}$, and scattering.
The QBE is widely used to describe ultrafast dynamics of electrons and phonons~\cite{Maldonado2017,Maldonado2020,Zhou2021,Caruso2022,Lee2023}.

Given that the laser spot size is typically much larger than the propagation lengths of the excitations, in most scenarios it is sufficient to consider only the coordinate perpendicular to the sample surface (denoted as $z$). Moreover, external forces $\mathbf{F}$ are absent for magnons. The QBE describing magnon dynamics can hence be formulated as
\begin{equation}\label{eMagnon}
    \dot{n}_\q = - v_\q \dz{n_\q} + s_\q(\{n_{\q}\},T_\mathrm{e}),
\end{equation}
where we have identified $(\dt{n})_\mathrm{scatter} = s_\q$. This is simply an extension of the equation for the magnon dynamics of the (N+2)TM, Eq.~\eqref{eq:lN+2TMmag}, by a drift term. Eq.~\eqref{eMagnon} is a set of coupled advection equations that model magnon propagation due to spatial gradients in the magnon occupations along $z$, with the group velocities $v_\q=\partial \omega_\q/\partial q_z$. Note that our description of magnon dynamics based on the QBE goes beyond the diffusive descriptions used in earlier studies~\cite{Xiao2010,Cornelissen2016,Beens2022}.

To obtain equations describing the transport of electrons and phonons, we start from continuity equations for their respective energy densities. The (electron or phonon) volumetric energy density $\varepsilon(z,t)$ changes due to heat transport via a current density $U(z,t)$ and local source/sink terms $S(z,t)$,
\begin{equation}\label{eHE2}
   \dt{} \varepsilon = - \dz{U} + S.
\end{equation}
We invoke Fourier's law~\cite{FouriersLaw}, which states that $U = -\kappa\partial T/\partial z$, where $\kappa$ is the thermal conductivity. While this approximation is known to break down in nanoscale and ultrafast transport regimes~\cite{Wilson2014}, it provides a tractable description of heat flow (see Appendix~\ref{aFourier} for a detailed discussion). A more complete treatment would require going beyond Fourier transport, which is outside the scope of the present work.

Both the thermal conductivity and the volumetric specific heat capacity $C$ are assumed to be constant. The energy density can therefore be rewritten as $\varepsilon = CT$, and Eq.~\eqref{eHE2} becomes
\begin{equation}\label{eHE}
    C \dot{T} = \kappa \frac{\partial^2T}{\partial z^2} + S,
\end{equation}
which is the heat diffusion equation with a source/sink term. Identifying the right-hand side of Eqs.~\eqref{eq:lN+2TMe} and \eqref{eq:lN+2TMph} with heat source/sink terms, we get  
\begin{align}
    \begin{split}
    C_\mathrm{e} \dot{T}_\mathrm{e} &= \kappa_\mathrm{e} \frac{\partial^2T_\mathrm{e}}{\partial z^2} +
    P(z,t) - G_\mathrm{ep}(T_\mathrm{e}-T_\mathrm{p}) \\
    &-\frac{1}{NV}\sum_\q \hbar\omega_\q s_\q  , 
    \end{split}
    \label{eEle}
    \\
    C_\mathrm{p} \dot{T}_\mathrm{p} 
    &= 
    \kappa_\mathrm{p} \frac{\partial^2T_\mathrm{p}}{\partial z^2} + G_\mathrm{ep}(T_\mathrm{e}-T_\mathrm{p}).
    \label{ePho}
\end{align}
These two equations, along with Eq.~\eqref{eMagnon}, form the foundation of the \textit{nonlocal} (N+2)TM [Fig.~\ref{fsetup}(b)].

The presence of transport requires boundary conditions for the differential equations. In this work we assume that the laser-excited sample is thermally isolated, i.e., there is no energy transport across the boundaries. Since Fourier's law is used for the electrons and phonons, this condition is equivalent to vanishing temperature gradients at the boundaries $z = 0$ and $z = L$:
\begin{equation}\label{eBoundElePho}
    \dz{T_\mathrm{e/p}(0)} = \dz{T_\mathrm{e/p}(L)} = 0 \, .
\end{equation}

For the magnons, reflective boundary conditions are imposed. When a magnon reaches a boundary, it is reflected into a mode with the same momentum except for the sign of its $z$-component. By symmetry, the incident and reflected modes differ only in the sign of their velocity along $z$. Mathematically, the boundary condition is expressed as:
\begin{equation}\label{eBoundMag}
\begin{array}{l}
    n_{(q_x,q_y,q_z)}(0) = n_{{(q_x,q_y,-q_z)}}(0) \, ,\\
    n_{(q_x,q_y,q_z)}(L) = n_{(q_x,q_y,-q_z)}(L) \, .
\end{array}
\end{equation} 

Real samples are never entirely isolated. A close physical realization of these boundary conditions is a metallic ferromagnet on an insulating, nonmagnetic substrate, such as the commonly used sapphire~\cite{Schellekens2013}. In this system, the primary sources of loss---aside from minor, unavoidable radiative losses---occur at the substrate interface due to phonon transmission. This process leads to cooling of the ferromagnet on timescales of tens of picoseconds~\cite{Awsaf2025}. Additionally, a minor dissipation channel arises from the {phonon pumping effect}, which enables energy transfer from magnons to the substrate~\cite{Streib2018}.

\subsection{Inhomogeneous laser absorption}\label{sLaserProfile}

In the nonlocal (N+2)TM, translational symmetry is broken by the inhomogeneous laser absorption; assuming a uniform absorption recovers the local (N+2)TM.

We model the laser absorption profile using pyGTM~\cite{pyGTM,Passler2017,Passler2019,Passler2020}, a Python implementation of the generalized transfer matrix method for multilayer films. This method solves Maxwell’s equations in each homogeneous layer by enforcing the continuity of the tangential electric and magnetic fields at every interface. In addition to the frequency and angle of incidence of the light, it requires the thicknesses and complex permittivity tensors of each layer, which are assumed to fully characterize the optical response.

From pyGTM, we obtain the absorption density profile $\rho(z)$, which gives the fraction of incident energy absorbed per unit depth and satisfies $\int_0^L \rho(z)\,\mathrm{d}z < 1$ due to reflection and transmission losses. Assuming a temporally Gaussian laser pulse with standard deviation $\tau_l$ and total incident fluence $F$, the absorbed power density is given by
\begin{equation}\label{ePulse}
    P(z,t) = F  \frac{\mathrm{e}^{-(t/\tau_l)^2/2}}{\tau_l \sqrt{2\pi}} \rho(z).
\end{equation}

\subsection{Numerical methods}
\label{snumerics}

The film is spatially discretized into $N_z$ equidistant points over the domain $z \in [0,L]$, resulting in a spatial step size $\Delta z = L/(N_z-1)$. Time is advanced in steps of size $\Delta t$.

The electron and phonon heat diffusion equations are solved using the unconditionally stable implicit Crank-Nicolson scheme~\cite{Crank-Nicholson}. Achieving second-order accuracy in time requires evaluation of the source term at both the current and subsequent time step. Since the source term depends on the electron and phonon temperatures as well as the magnon scattering contribution, it cannot be evaluated explicitly at the next time step because these quantities are not yet known. To address this, the magnon scattering terms at $t+\Delta t$ are approximated using a linear extrapolation $s_\q(t+\Delta t) \approx 2 s_\q(t) - s_\q(t-\Delta t)$, and the electron and phonon temperatures are obtained by performing a fixed-point iteration within each time step of the Crank-Nicolson scheme to achieve self-consistency of the temperature-dependent source term. The magnon scattering terms are not fixed-point iterated, as their evaluation constitutes the computational bottleneck of the simulation.

The advection equations governing the magnon occupation numbers, Eq.~\eqref{eMagnon}, are solved using an explicit total variation diminishing (TVD) scheme with the monotonized central (MC) slope limiter~\cite{TVD}. This method is subject to the Courant-Friedrichs-Lewy (CFL) stability condition $\Delta t < \Delta z / v_{\mathrm{max}}$, where $v_{\mathrm{max}}$ is the maximum magnon group velocity among all $v_\q$.

\subsection{Parameter values}
\label{sparameters}

For this study, we consider body-centered cubic ferromagnetic Fe(001) films with thicknesses $L = \SI{100}{\nano\meter}$ and $\SI{1}{\micro\meter}$, initially in thermal equilibrium at \SI{300}{K}. We use the room-temperature values $\kappa_\mathrm{e}=\SI{70}{\watt\metre^{-1}\kelvin^{-1}}$ and $\kappa_\mathrm{p}=\SI{5}{\watt\metre^{-1}\kelvin^{-1}}$, obtained from Ref.~\cite{ThermalCond}, as well as $C_\mathrm{e}=\SI{1.013e5}{\joule\metre^{-3}\kelvin^{-1}}$, $C_\mathrm{p}=\SI{3.177e6}{\joule\metre^{-3}\kelvin^{-1}}$, and $G_\mathrm{ep}=\SI{1.051e18}{\watt\metre^{-3}\kelvin^{-1}}$ obtained in Refs.~\cite{Maldonado2017,Ritzmann2020} from \textit{ab initio} calculations. The volume per spin site $V$ is given by $\frac{1}{2}a^3$, where $a = \SI{0.286}{nm}$ is the bcc Fe lattice parameter.

The magnon frequencies $\omega_\q$ and first- and second-order scattering rates $\gamma_\q$ and $\Gamma_{\q\q'}$ are taken from Ref.~\cite{Weissenhofer2024}, where they were computed from first principles on an equidistant $20 \times 20 \times 20$ grid in the first Brillouin zone. These quantities are assumed to be independent of temperature and interface proximity.

To compute the laser absorption density profile $\rho(z)$ using pyGTM, we assume a laser wavelength of \SI{800}{nm} (a value that is typically used in experiments~\cite{800nm_laser}) incident normally on the Fe film. The film is assumed to be grown on a sapphire substrate, and we use the complex refractive indices $n_\mathrm{air}=1$, $n_\mathrm{Fe}=4.11 + 3.44\mathrm{i}$, and $n_\mathrm{sapphire}=1.76$~\cite{Oppeneer1992,Refractiveindex.info} to construct diagonal permittivity tensors for the respective layers. The temporal standard deviation of the laser pulse is set to $\tau_l = \SI{60}{fs}$, and simulations are performed with an incident fluence of $F=\SI{1}{mJ/cm^2}$ unless otherwise specified.

\section{Results}
\label{sResults}
Using the framework described above, we study ultrafast laser-excited dynamics in bcc Fe. Section~\ref{sTemp} examines the evolution of the temperature distributions and the heat propagation through the system. We identify two stages of the dynamics: an initial transient stage governed primarily by local processes, followed by a quasi-steady stage governed mainly by nonlocal dynamics. In Sec.~\ref{sCurrent}, we present the magnonic angular-momentum currents generated by the laser pulse, which constitute an \textit{ultrafast spin Seebeck effect}. Our transport analysis, detailed in Sec.~\ref{sCrossover}, reveals an initial ballistic propagation of magnons that subsequently yields to diffusive behavior—a signature of the superdiffusive transport regime. Section~\ref{sDemag} discusses the depth-resolved demagnetization curves, attributing their characteristics to distinct physical mechanisms and comparing them with the demagnetization observed in the local (N+2)TM. Finally, Sec.~\ref{sMOKE} presents simulations of time-resolved magneto-optical Kerr signals based on our depth-dependent demagnetization profiles, aiming to bridge the gap between theory and experiment.

\subsection{Temperature dynamics}\label{sTemp}

\begin{figure*}[t!]
    \centering
    \includegraphics[width=\textwidth]{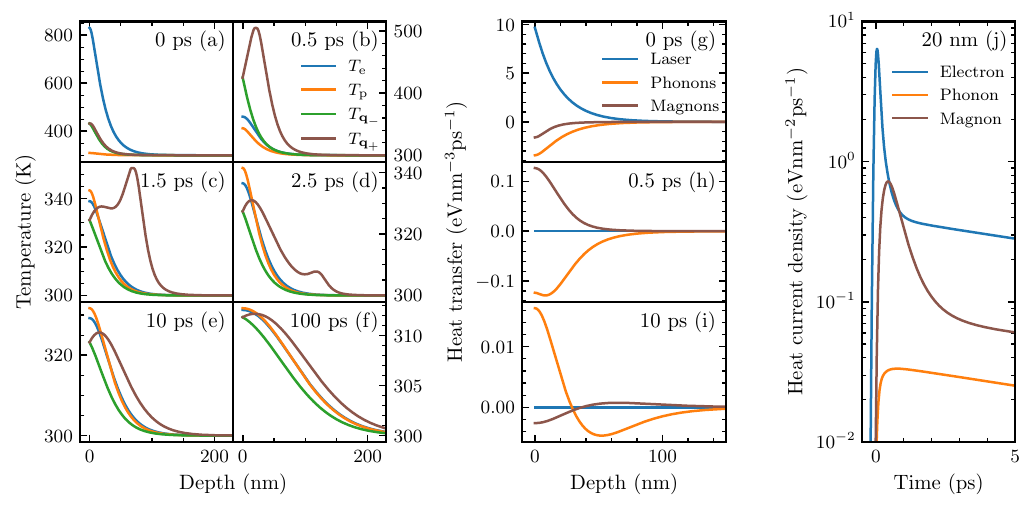}
    \vspace*{-0.9cm}
    \caption{Nonequilibrium electron, phonon, and magnon dynamics of a laser-excited bcc Fe(001) film of thickness $L=\SI{1}{\micro\meter}$. (a)-(f) Temperature distributions at various times. $T_{\q_\pm}$ denote the temperatures of the magnon modes at $\q_\pm = \frac{2\pi}{a}\frac{1}{20}(3,3,\pm 8)$. (g)-(i) Heat transfer to the electrons from the laser pulse, phonons, and magnons. (j) Heat current densities at a depth of $\SI{20}{nm}$. %Heat propagation is predominantly mediated by the electrons.
    }
    \label{fThermal}
\end{figure*}

To illustrate the nonequilibrium dynamics captured by our model, we simulate the excitation of a film with thickness $L=\SI{1}{\micro\meter}$, and study its thermal evolution. The results effectively represent the bulk response, since the temperatures are negligibly perturbed at the $z=L$ boundary for at least $\SI{100}{ps}$ following excitation.

Figures~\ref{fThermal}(a)-(f) show the temperature profiles of the electrons, phonons, and two representative magnon modes, at various times. We consider magnon modes at $\q_\pm = \frac{2\pi}{a}\frac{1}{20}(3, 3, \pm8)$, which have velocities $v_{\q_\pm} = \SI{\pm49.6}{nm/ps}$. Their effective temperatures $T_{\q_\pm}$ are obtained by inverting the Bose-Einstein distribution (see Sec.~\ref{slN+2TM}).

At $t = 0$, corresponding to the peak of the laser pulse, the laser heats the electrons, whose temperature distribution---due to transport---already deviates from the purely exponential absorption profile. Through coupling to the electrons, the temperatures of the magnons and phonons also begin to rise. By symmetry, the magnon modes at $\q_\pm$ share the same $\omega_\q, \gamma_\q$, and $\Gamma_{\q\q'}$, and are therefore initially excited identically.

The initial burst of magnons, which is excited close to the surface during the transient stage of the dynamics, propagates into the depth of the film, as shown in Figs.~\ref{fThermal}(b)-(d). The magnons are progressively damped as they move deeper into the sample, where they transfer their energy to cooler electrons and magnon modes. 

After several \si{\pico\second}, the system reaches a quasi-steady state [Figs.~\ref{fThermal}(e)-(f)], where the dynamics evolve on longer timescales. This happens because the phonons---with their large heat capacity---absorb most of the deposited energy, leading to reduced temperature gradients and weaker heat currents. In this regime, the phonons act as a heat source for the electrons and magnons near the irradiated surface, and as a sink deeper inside. As will be seen in Sec.~\ref{sCurrent}, this process fuels a continued generation of spin currents, with a weaker amplitude compared to the currents driven by the initial burst of magnons excited in the transient stage. 

In our model, the electrons are the only particles directly coupled to all others (photons, phonons, and magnons). Figs.~\ref{fThermal}(g)-(i) provide a depth-resolved view of the local heat flow, depicting the energy transfer between the electrons and the other subsystems. Initially, as the laser deposits energy into the electrons, the phonons and magnons conversely draw heat away from the electronic system. 
Following the rapid cooling of the electrons via energy transfer to the phonons, the magnons subsequently return part of their energy to the electrons. This local process during the transient stage of the dynamics is responsible for the prompt remagnetization.

During the quasi-steady stage [Fig.~\ref{fThermal}(i)], the phonons heat the electrons near the surface and cool them deeper inside. The electrons, in turn, heat the magnons at the surface and cool them at depth. As will be shown in Sec.~\ref{sDemag}, this heat transport into the film depth gradually remagnetizes the surface and further demagnetizes the deeper regions.

The rate of heat transport influences the magnitude of the initial demagnetization and governs the evolution of the system in the quasi-steady stage. Figure~\ref{fThermal}(j) shows the heat current densities carried by electrons, phonons, and magnons at a depth of \SI{20}{nm}, which corresponds approximately to the depth at which the laser amplitude has decayed to $1/e$. The electronic and phononic heat current densities are evaluated as $-\kappa_{\mathrm{e/p}}\partial T_{\mathrm{e/p}}/\partial z$, while the magnonic contribution is given by $\frac{1}{NV}\sum_\q \hbar \omega_\q v_\q n_\q$. During and immediately after the laser pulse, before substantial energy transfer to the phonons occurs, the steep electron temperature gradient drives a pronounced electronic heat current. As will be shown below, this has a significant impact on the initial demagnetization. Magnons also contribute considerably to the heat current during their initial excitation, but at later times---during the second stage of the dynamics---heat transport is predominantly mediated by electrons.

\subsection{Laser-excited magnonic spin currents}\label{sCurrent}

\begin{figure*}
\centering
\includegraphics[width=\textwidth]{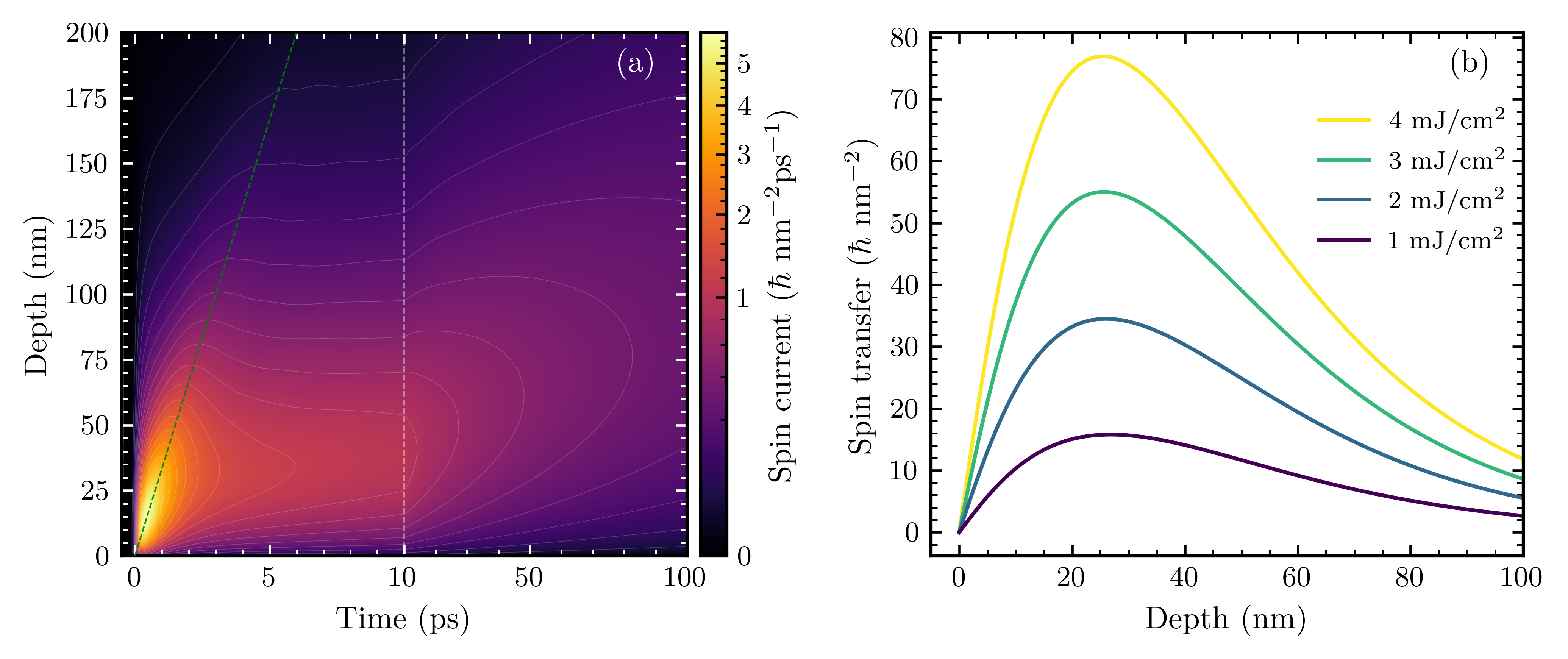}
\caption{Laser-excited magnonic spin currents in a \SI{1}{\micro\meter} film. (a) Magnonic spin current density following irradiation with a \SI{1}{mJ/cm^2} laser pulse. Note that the time axis is piecewise linear with a breakpoint at \SI{10}{ps} and only the relevant first \SI{200}{\nano\meter} are shown. The dashed green line from the origin to (\SI{6}{ps}, \SI{200}{nm}) is a guide for the eye to identify the ballistic motion of the initial burst of magnons. (b) Magnonic spin transfer across every depth during the first \SI{10}{ps} for fluences from \SI{1}{mJ/cm^2} to \SI{4}{mJ/cm^2}.}

\label{fSpinCurrent}
\end{figure*}

The transient dynamics and thermal gradients discussed in the preceding section generate magnon currents. Since each magnon carries a spin moment of $\hbar$, these magnons contribute to a \textit{spin Seebeck effect} on ultrafast timescales~\cite{Uchida2010, Kimling2017}. We quantify the resulting spin transport through the magnonic spin current density $\frac{1}{NV}\sum_\q \hbar v_\q n_\q$.

Figure~\ref{fSpinCurrent}(a) presents a color map of the magnonic spin current density for the \(L=\SI{1}{\micro\meter}\) system. The spin current reaches a maximum value of roughly \(\SI{5}{\hbar\,\mathrm{nm}^{-2}\mathrm{ps}^{-1}}\) at a depth of about \(\SI{20}{nm}\) within the first picosecond. This initial burst of surface-generated magnons produces a pronounced spin-current pulse that propagates ballistically into the bulk of the film while undergoing progressive attenuation. It can be clearly identified by the ridge extending from the origin to $(\SI{6}{ps}, \SI{200}{nm})$, corresponding to an average velocity of about \SI{35}{nm/ps}, which is comparable to experimentally measured ballistic magnon velocities in antiferromagnetic NiO~\cite{Lee2021,Qiu2025}. 
%{\red PMO: compare to Fe magnon group velocity?}

As the system transitions into the quasi-steady regime, the current amplitude reduces while extending over a broader spatial region. This evolution reflects the widening electron-temperature gradient, which continues to generate substantial magnonic spin currents even tens of picoseconds after excitation.

To put the magnitude of the spin current in perspective, we compare it to the magnonic spin Seebeck effect in the linear-response regime \cite{Rezende2016,Schmidt2021}. The spin Seebeck coefficient $S_\mathrm{m}$ is derived using linear-response theory as outlined in Appendix~\ref{aLinearReponse}. Evaluating  Eq.~\eqref{eSeebeckCoefficient} at $T=\SI{300}{K}$, we obtain $S_\mathrm{m} = \SI{0.005}{\hbar\,nm^{-2}ps^{-1}/K\micro\meter^{-1}}$. Even an extreme temperature gradient of \SI{1000}{K/mm} (experiments typically use gradients on the order of \SI{10}{K/mm} \cite{Uchida2010,Rezende2016}) only generates a magnonic spin current of approximately \SI{0.005}{\hbar\,nm^{-2}ps^{-1}}. This is three orders of magnitude 
smaller than the currents achieved in the simulation.

The average velocity of the magnons participating in transport is given by Eq.~\eqref{eTransportVelocityLinear} in the linear-response regime, which evaluates to \SI{15}{nm/ps} at \SI{300}{K}. In contrast, the laser-excited magnons achieve an average velocity of up to \SI{30}{nm/ps} (calculated using Eq.~\eqref{eTransportVelocityGeneral}). This is because magnon transport is primarily mediated by low-energy magnons in the linear-response regime, which generally have lower velocities. The laser-excited magnons are of higher energy and move faster on average. 

Controlled manipulation or reversal of magnetization on ultrafast timescales is of technological interest~\cite{Kirilyuk2010, Scheid2022, Chen2025}. Such control can, in principle, be achieved by delivering a large spin current over a short time interval. Figure~\ref{fSpinCurrent}(c) shows the time-integrated spin current density (i.e., the \textit{spin transfer}) over the first \SI{10}{ps} for different laser fluences. For all fluences, the maximum spin transfer occurs at a depth of about \SI{25}{nm}. At \SI{1}{mJ/cm^2} and \SI{4}{mJ/cm^2}, this maximum reaches approximately \SI{16}{\hbar \,nm^{-2}} and \SI{77}{\hbar \, nm^{-2}}, respectively, revealing a slight nonlinear dependence on fluence. 

A \SI{1}{nm}-thick Fe layer contains roughly \SI{85}{atoms/nm^{2}}. The required spin transfer to reverse the magnetization of such a layer can be estimated as maximally  \SI{187}{\hbar \,nm^{-2}} under ideal absorption. This follows from the saturation moment of Fe being about $\mu_s =2.2\mu_B$ per atom along with the fact that the magnetic moment is predominantly of spin origin due to the quenching of the orbital momentum~\cite{Danan1968,Magnetism}. Using $\mu_s = g_s\mu_B S/\hbar$ with $g_s\approx 2$, this corresponds to a spin angular momentum of approximately $S = \SI{1.1}{\hbar/atom}$. Reversing the magnetization requires twice this amount, i.e., $\SI{2.2}{\hbar/atom}$. 

In addition to the magnitude requirement, the spin transfer must occur on a timescale short compared to the magnetic damping time. We can estimate the characteristic damping time of coherent excitations of the magnetic sites from the magnon lifetime at the $\Gamma$-point, which for bcc Fe is $\gamma_{\q=\Gamma}^{-1} = \SI{2.4}{ns}$~\cite{Weissenhofer2024}. These order-of-magnitude estimates indicate that ultrafast laser-induced magnon currents of the calculated amplitude, delivered within \SI{10}{ps}, can potentially induce substantial reorientation---or even reversal---of the magnetization in thin films. 

A more comprehensive analysis would model the magnetic evolution of an adjacent layer by treating our magnonic spin transfer as a spin-transfer torque (STT) on the local magnetization. This type of magnonic spin current-induced switching has been previously proposed in a theoretical study~\cite{Cheng2018}. 

\subsection{Ballistic-to-diffusive crossover}\label{sCrossover}

A central feature of our approach is its ability to capture the ballistic to diffusive crossover of magnon transport. Unlike the heat diffusion equation---which is diffusive by construction at all times---the Boltzmann equation considers scattering events explicitly. As a result, transport is ballistic at short times (prior to significant collisions), and gradually transitions toward diffusive behavior. This provides a consistent description across the full temporal range.

\begin{figure}
\centering
\includegraphics[width=1\linewidth]{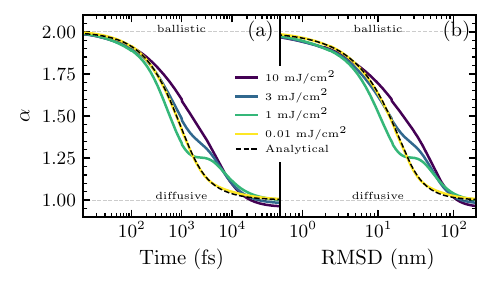}
\caption{Transition from ballistic to diffusive transport of magnons generated at the irradiated surface during the peak of the laser pulse. (a) The instantaneous transport exponent $\alpha$ as a function of elapsed time. (b) The instantaneous transport exponent as a function of the average propagated distance (root mean squared displacement).}
\label{fCrossover}
\end{figure}

This crossover is quantified in Fig.~\ref{fCrossover}, which shows the evolution of the instantaneous transport exponent $\alpha$ for magnons generated at the irradiated surface during the peak of the laser pulse. The exponent $\alpha$, defined and computed as detailed in Appendix~\ref{aCrossover}, characterizes the scaling of the mean squared displacement with time ($\mathrm{MSD} \sim t^\alpha$): $\alpha=2$ corresponds to ballistic transport, $\alpha=1$ to diffusive transport, and $1<\alpha<2$ to superdiffusive dynamics. In the context of active matter, superdiffusion refers to a steady-state transport regime where $\alpha$ is constant in time~\cite{Metzler2004}. Conversely, in condensed matter physics, the term is frequently used to describe a dynamical regime which transitions from ballistic to diffusive transport. This is the case for the superdiffusive electron spin transport model~\cite{Battiato2010,Battiato2012}, to which our model can be viewed as the magnonic analog. 

The analytical reference curve in Fig.~\ref{fCrossover} assumes particles with constant speed $v$ that undergo direction-randomizing scattering events separated by an average time $\tau$  (see App.~\ref{aCrossover}). Fitting this analytical model to the $F=\SI{0.01}{mJ/cm^2}$ simulation yields $\tau \approx \SI{370}{fs}$ and $v \approx \SI{42}{nm/ps}$. The extracted scattering time agrees well with the energy-weighted average magnon lifetime $\sum_\q \omega_\q \gamma_\q^{-1} / \sum_\q \omega_\q = \SI{354}{fs}$, supporting the internal consistency of our model.

At higher fluences, the numerical model predicts a delayed crossover to diffusive behavior, suggesting that magnons may persist in the superdiffusive regime for longer times as the fluence increases.

Since magnon transport remains nondiffusive for up to $\SI{10}{ps}$ and over distances up to $\SI{100}{nm}$, purely diffusive models~\cite{Xiao2010,Cornelissen2016,Beens2022} fail to capture the transient dynamics revealed here.

\subsection{Depth-resolved demagnetization dynamics}\label{sDemag}

\begin{figure*}
    \centering
    \includegraphics[width=\textwidth]{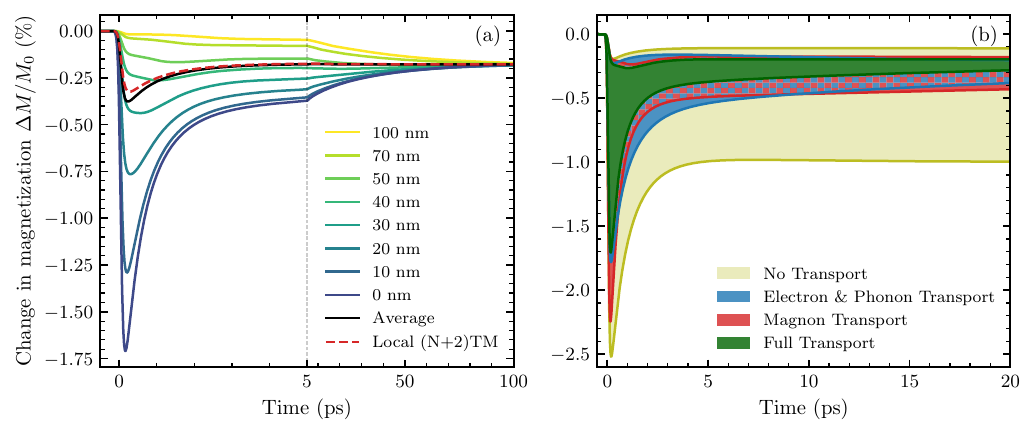}
    \caption{Calculated demagnetization dynamics. (a) Depth-resolved relative change in magnetization for a  \SI{100}{nm} film. Note that the time axis is piecewise linear with a breakpoint at 5 ps, allowing both the short- and long-term dynamics to be seen clearly. (b) Comparison of demagnetization curves with various transport contributions included ($L=\SI{1}{\micro\meter}$). Each color region depicts the range of demagnetization in $z \in [0,40]$ nm.}
    \label{fDemagnetization}
\end{figure*}

Having established an understanding of the superdiffusive magnon currents and temperature dynamics of the nonlocal (N+2)TM, we now move on to investigate the demagnetization dynamics. The relative change in magnetization $\Delta M/M_0$ is computed using the magnon occupations according to Eq.~\eqref{eDemag} in Appendix~\ref{aDemag}.

Figure~\ref{fDemagnetization}(a) shows the depth-resolved change in magnetization in a \SI{100}{nm}-thick Fe film. In the transient stage ($t \lesssim \SI{3}{ps}$), the near-surface region exhibits a rapid demagnetization followed by a prompt partial recovery. This resembles ultrafast demagnetization observed in ultrathin films~\cite{Beaurepaire1996,Carpene2008}.
Deeper within the film, local energy transfer processes lead to an initial drop in magnetization, followed by additional demagnetization as heat and magnons generated near the surface propagate into the material.

During the quasi-steady stage, remagnetization occurs near the surface, while additional demagnetization develops deeper within the film, gradually approaching spatial uniformity. 
Thinner films reach this uniform state more rapidly; for example, a \SI{40}{nm} film becomes nearly homogeneous after approximately \SI{5}{ps}, while films thinner than \SI{20}{nm} achieve uniformity almost instantaneously, demonstrating the validity of the local (N+2)TM~\cite{Weissenhofer2024} for ultrathin films. 
%The resulting homogeneous demagnetization matches that obtained from the local (N+2)TM, as it arises merely from the increased temperature of the system which has reached thermal equilibrium. 

The spatially averaged demagnetization obtained from the nonlocal simulation is initially slightly larger than that predicted by the local model [Fig.~\ref{fDemagnetization}(a)]. This discrepancy is likely due to the nonlinearity of demagnetization with respect to absorbed fluence, as noted in the original work~\cite{Weissenhofer2024} on the local (N+2)TM: more magnons are generated when the deposited energy is spatially concentrated rather than uniformly distributed.

Transport draws magnetic angular momentum and heat away from the surface, thereby reducing the surface demagnetization while enhancing it at depth. To disentangle the roles of the different transport channels, we perform simulations for a system with $L=\SI{1}{\micro\meter}$ with selectively enabled magnon and electron transport. The contribution of phonon transport is negligible and is included when electron transport is active. The corresponding demagnetization profiles are displayed in Fig.~\ref{fDemagnetization}(b), showing the full range of demagnetization values from the surface to depths of \SI{40}{nm}.

Magnon transport continuously transfers angular momentum from the surface region into the bulk, quickly flattening the demagnetization profile. However, due to the relatively small magnonic heat current, the temperature and demagnetization profiles quickly stabilize, and the subsequent evolution remains slow.

When magnon transport is disabled, the demagnetization evolves primarily in response to changes in the local electron temperature. Electron transport therefore drives the demagnetization towards homogeneity by flattening the temperature profile.

Although not accounted for in the current model, superdiffusive electron spin transport~\cite{Battiato2010,Battiato2012} represents an additional channel that can influence the demagnetization profile, particularly in thicker films.

\subsection{Simulating magneto-optical Kerr effect signals from depth-resolved magnetization dynamics}\label{sMOKE}

Ultrafast laser-induced demagnetization is often studied using the magneto-optical Kerr effect (MOKE), in which changes in the polarization or intensity of reflected light are measured. In the semi-infinite limit~\cite{OppeneerMOKE}, the MOKE signal is proportional to the sample’s magnetization~\cite{Carpene2008,Beaurepaire1996}, although the validity of this relation in the strongly nonequilibrium regime has been debated~\cite{Koopmans2000,Oppeneer2004,Chen2025,Richter2024}. 
In thin films and multilayer structures, internal reflections and a depth-dependent magnetization further complicate this relation~\cite{Wieczorek2015,Razdolski2017-MOKE,You2018}.

Depth-sensitive magneto-optical measurement methods typically reconstruct magnetization profiles by measuring the Kerr response across multiple reflection angles~\cite{Hamrle2002,Wieczorek2015,Hofherr2017}. To circumvent the need for such complex experimental reconstructions, we instead directly relate our calculated inhomogeneous magnetization profiles (Sec.~\ref{sDemag}) to simulated MOKE signals. This is achieved using the generalized transfer matrix method implemented in pyGTM (see Sec.~\ref{sLaserProfile}).

The MOKE response depends on the orientation of the magnetization relative to the sample surface and the plane of incidence. In our simulations, we consider a longitudinal MOKE (L-MOKE) geometry, where the magnetization lies in the film plane and within the plane of incidence (the $xz$-plane). The response is characterized by the complex Kerr angle~\cite{OppeneerMOKE},
\begin{equation}
\begin{aligned}
    \theta_K^s + i\varepsilon_K^s &= \frac{r_{ps}}{r_{ss}}, \\
    \theta_K^p + i\varepsilon_K^p &= \frac{r_{sp}}{r_{pp}},
\end{aligned}
\end{equation}
for $s$ and $p$ polarized light, respectively, which can be decomposed into the Kerr rotation $\theta_K$ and ellipticity $\varepsilon_K$. The coefficients $r_{pp}$ and $r_{ss}$ describe reflection of the same polarization, while $r_{ps}$ and $r_{sp}$ quantify the magneto-optically induced cross-polarization. These ratios therefore directly encode the magneto-optical response.

\begin{figure*}
    \centering
    \includegraphics[width=\textwidth]{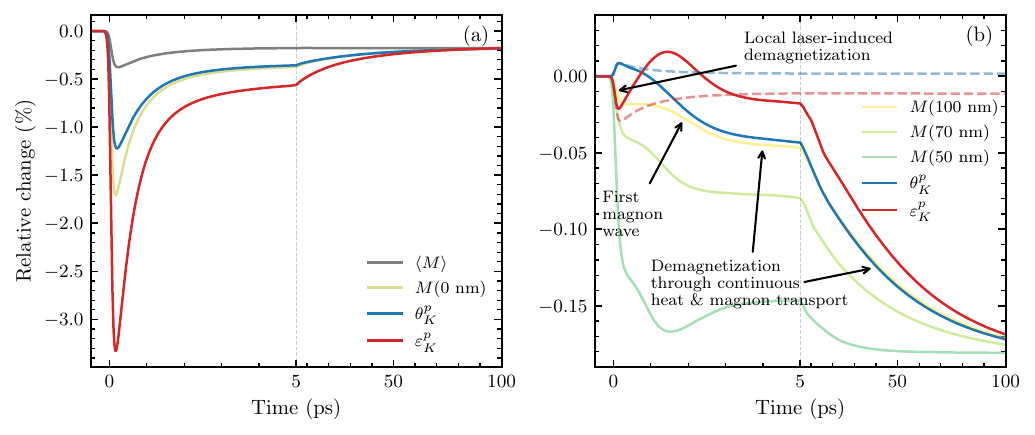}
    \caption{Calculated L-MOKE signals of the demagnetization in an $L =$\SI{100}{nm} film using a $p$-polarized probe of wavelength \SI{800}{nm} with a \SI{45}{\degree} angle of incidence. Kerr rotation $\theta_K^p$, ellipticity $\varepsilon_K^p$, and average demagnetization $\langle M\rangle = \frac{1}{L}\int_0^LM \mathrm{d} z$. (a) Probe on the pumped surface. (b) Probe on the opposite side of the film. Dashed lines are Kerr rotation and ellipticity from a simulation without transport.}
    \label{fMOKE}
\end{figure*}

The (magneto-)optical properties of the material are captured by the permittivity tensor. To linear order in the magnetization, the permittivity tensor for bcc Fe can be written as
\begin{equation}
    \boldsymbol{\epsilon}(z,t) =
    \begin{pmatrix}
        \epsilon & 0        & 0        \\
        0        & \epsilon & \epsilon^{\prime}(z,t) \\
        0        & -\epsilon^{\prime}(z,t) & \epsilon
    \end{pmatrix},
\end{equation}
which contains small off-diagonal elements $\epsilon^{\prime}(z,t)$ representing the magneto-optical coupling. To lowest order, they scale linearly with magnetization according to $\epsilon^{\prime}(z,t) = \epsilon_{xy}\frac{M(z,t)}{M_0}$, where $\epsilon_{xy}$ is the value of the off-diagonal permittivity when saturated at room temperature~\cite{OppeneerMOKE}. The diagonal elements $\epsilon = n_\mathrm{Fe}^2$ are given by the complex refractive index. 

We use $\lambda = \SI{800}{nm}$ probe light incident at a \SI{45}{\degree} angle, and $\omega\sigma_{xy} = (2.8 + 6.3\,i) \times 10^{29} \;\mathrm{s}^{-2}$ as obtained from~\cite{Oppeneer1992}, from which $\epsilon_{xy}$ is obtained via $\epsilon_{xy} = \frac{4\pi i}{\omega}\sigma_{xy}$~\cite{OppeneerMOKE}. By discretizing the film into many thin slices, each with a dielectric tensor determined by the local magnetization $M(z,t)$, we can compute the time evolution of the Kerr angle.

Figure~\ref{fMOKE}(a) displays the change in Kerr rotation and ellipticity for L-MOKE using $p$-polarized light on the laser-irradiated side of a \SI{100}{nm} bcc Fe film. For the initial homogeneous magnetization, we obtain $\theta_K^p = \SI{-1.48}{\milli\radian}$, $\varepsilon_K^p=\SI{0.31}{\milli\radian}$. For comparison, we also show the changes in surface magnetization $M(\SI{0}{nm})$, and spatially averaged magnetization $\langle M\rangle = \frac{1}{L}\int_0^LM \mathrm{d} z$.

In line with earlier studies~\cite{Wieczorek2015,Richter2024,Ashok2025}, we observe that relative changes in the Kerr angle do not necessarily reflect the relative demagnetization, and that its individual components exhibit distinct behaviors on ultrafast timescales. In particular, the change in Kerr ellipticity exceeds even the surface magnetization, which is the maximum within the film. Only on timescales of tens of picoseconds, when the magnetization becomes homogeneous, do both Kerr rotation and ellipticity track the transient average magnetization.

The results shown in Fig.~\ref{fMOKE}(a) can be directly compared to experimental data, allowing one to assess whether the nonlocal (N+2)TM accurately captures the demagnetization and transport dynamics. A particularly sensitive test of nonlocal transport would be MOKE probing from the side opposite to the pumped surface, as has been done previously~\cite{Razdolski2017}. 

In Fig.~\ref{fMOKE}(b), we show a simulation of L-MOKE probing performed on the opposite side of the bcc Fe film. Three distinct stages can be identified in the demagnetization and corresponding Kerr response. The first is the local, laser-induced demagnetization resulting from the small fraction of pump light absorbed at this depth. The second stage is a rapid additional demagnetization caused by the arrival---after a ballistic delay time---of the burst of magnons excited near the pumped surface during the laser excitation. In the third stage, the continued demagnetization is driven by the ongoing transport of heat and magnons deeper into the film.

Interestingly, the Kerr rotation and ellipticity exceed their original value despite the overall reduction in magnetization. This behavior can be understood from the analytical expression for the Kerr angle from a depth-dependent magnetization,
\begin{equation}
    \theta_K + i\varepsilon_K = \int_0^L W(z)M(z)\mathrm{d}z,
\end{equation}
where $W(z) \approx W(0)\mathrm{e}^{-4 \pi i n z/\lambda}$ is the complex depth sensitivity function~\cite{You2018,MOKEAnalytical}. For the parameters used here, $W(z)$ reverses sign approximately every \SI{50}{nm}. Consequently, demagnetization occurring at certain depths can actually lead to an increase in the magnitude of the Kerr angle components. Nonlocal demagnetization dynamics arising from magnon transport even leads to a sign change in the relative change of the Kerr ellipticity, a feature that is absent without transport (dashed lines in Fig.~\ref{fMOKE}(b)). This highlights that MOKE signals in ultrafast experiments can exhibit counterintuitive behavior that directly reflects the underlying inhomogeneous magnetization and nonlocal transport dynamics.

If a comparably delayed second demagnetization stage were observed experimentally, it would provide a clear signature of magnon-mediated transport---specifically, the arrival of a ballistic magnon population generated near the pumped surface. Such an observation would therefore lend strong support to attributing ultrafast laser-induced demagnetization to transverse spin excitations.

%{\red \tiny LM: This is in regards to this paper~\cite{Ashok2025}. If the electron transport is fully ballistic (next to no scattering), then yes, the temperature and demagnetization would be homogeneous (within a few tens of femtoseconds). However, this does not imply ballistic \textit{magnon} motion is out of the question. Furthermore, the mean squared displacement scales as $t^2$ for ballistic motion and $t$ for diffusive motion. This means that for a short time, diffusive propagation is actually faster. Overall, I feel like their assumption that ballistic motion instantly homogenizes the film is extremely crude. }

\section{Conclusion}
\label{sConclusions}

We have developed a theoretical model to study nonlocal magnon dynamics on ultrafast timescales within a quantum Boltzmann framework. This approach is an extension of the widely used three-temperature model~\cite{Beaurepaire1996} that also captures \textit{nonthermal} magnon distributions as well as transport effects. We applied this method to investigate magnon generation and propagation following femtosecond laser excitation of Fe films. 

We observe an {ultrafast magnonic spin Seebeck effect} with a two-stage behavior. Immediately after excitation, a transient burst of nonthermal magnons is generated near the irradiated surface and propagates superdiffusively into the sample. This is followed by a weaker but continuous magnon flow sustained by strong temperature gradients during a quasi-steady stage lasting several tens of picoseconds. The laser-induced magnon currents are faster and orders of magnitude larger than those obtainable by steady-state temperature gradients, owing to the extreme interfacial gradients and nonequilibrium populations generated during ultrafast excitation. The calculated currents suggest that laser-induced magnonic spin transport can reach technologically relevant magnitudes on picosecond timescales.

Going beyond earlier approaches~\cite{Beens2022}, our model captures the temporal and spatial crossover from ballistic to diffusive magnon transport. This superdiffusive regime is analogous to that observed for electrons~\cite{Battiato2010,Battiato2012}.

The simulations further provide depth-resolved magnetization profiles. While the surface exhibits the typical demagnetization within a few hundred femtoseconds followed by partial remagnetization, deeper regions initially demagnetize less strongly but continue to demagnetize due to the influx of heat and magnons from the irradiated surface. This behavior is particularly relevant in thicker films, where a homogeneous magnetization profile is reached only after tens of picoseconds.

Finally, we use the resulting magnetization profiles to compute time-resolved Kerr angles. This provides a path for direct comparison with experiment, without requiring depth-sensitive magneto-optical reconstruction of the magnetization~\cite{Hamrle2002,Wieczorek2015,Hofherr2017}.

The framework developed here can be applied to a wide range of ferromagnetic materials and extended to layered heterostructures. It thus provides a route to study ultrafast magnon transport beyond diffusive models and supports the interpretation of time-resolved spin-transport experiments.

%%%%%%%%%%%%%%%%%%%%%%%%%%%%%%%%%%%%%%%
%%%%%%%%%%%%%%%%%%%%%%%%%%%%%%%%%%%%%%%

%\acknowledgements
\begin{acknowledgments}
 This work has been supported by the Swedish Research Council (VR), the German Research Foundation (Deutsche Forschungsgemeinschaft) through CRC/TRR 227 ``Ultrafast Spin Dynamics'' (project MF, project-ID: 328545488)), and the K.\ and A.\ Wallenberg Foundation (Grants No.\ 2022.0079 and 2023.0336).  The computational resources were provided by the National Academic Infrastructure for Supercomputing in Sweden (NAISS) at NSC Link\"oping, partially funded by VR through Grant Agreement No.\ 2022-06725.
\end{acknowledgments}

%%%%%%%%%%%%%%%%%%%%%%%%%%%%%%%%%%%%%%%%

\section*{Appendix}

\subsection{Linear-response transport coefficients}\label{aLinearReponse}

In order to contrast our results to steady-state transport coefficients, we show the linearized version of Eq.~\eqref{eMagnon}. For this, we set $n_\q = n^0(\omega_\q,T) + \delta n_\q$ and assume $\delta n_\q \ll 1$ (meaning the magnons are nearly Bose-Einstein distributed). In steady-state, to lowest order, and far away from interfaces, we obtain $0 = - v_\q \dz{n^0(\omega_\q,T)} - \delta n_\q \gamma_\q$. Rearranging, applying the chain rule, and defining the mode-dependent magnon lifetimes $\tau_\q = \gamma_\q^{-1}$, we obtain
\begin{equation}\label{eMagnonLinear}
    \delta n_\q = - \tau_\q v_\q\frac{\partial n^0(\omega_\q,T)}{\partial T}\frac{\partial T}{\partial z}.
\end{equation}

Transport quantities such as, for example, the magnonic spin current density, can be computed as $J = \frac{1}{NV}\sum_\q \hbar v_\q n_\q = \frac{1}{NV}\sum_\q \hbar v_\q \delta n_\q$, where we have used that the total contribution of $n^0$ cancels out (no transport under equilibrium).

The magnonic spin Seebeck effect is defined as a spin current arising from a temperature gradient $J = -S_\mathrm{m}\dz{T}$, from which the magnonic spin Seebeck coefficient becomes~\cite{Rezende2016}
\begin{equation}\label{eSeebeckCoefficient}
    S_\mathrm{m} = \frac{1}{NV}\sum_\q \hbar v_\q^2 \tau_\q\frac{\partial n^0(\omega_\q,T)}{\partial T}.
\end{equation}

Similarly, the magnonic thermal conductivity is
\begin{equation}\label{eThermalConductivity}
    \kappa_\mathrm{m} = \frac{1}{NV}\sum_\q \hbar\omega_\q v_\q^2 \tau_\q\frac{\partial n^0(\omega_\q,T)}{\partial T}.
\end{equation}

Magnons at $\q$ and $-\q$ contribute to transport in an equal and opposite fashion. Consequently, the net transport is determined by the asymmetry in the distribution. The average velocity (along $z$) of magnons whose contribution is not canceled by an oppositely moving counterpart is given by
\begin{equation}\label{eTransportVelocityGeneral}
    \langle v\rangle_\mathrm{tr} = \frac{\sum_\q v_\q (n_{(q_x,q_y,q_z)}-n_{(q_x,q_y,-q_z)})}{\sum_\q |n_{(q_x,q_y,q_z)}-n_{(q_x,q_y,-q_z)}|},
\end{equation}
which, in the linear-response regime, simplifies to
\begin{equation}\label{eTransportVelocityLinear}
    \langle v\rangle_\mathrm{tr} = \frac{\sum_\q v_\q \delta n_\q}{\sum_\q |\delta n_\q|}.
\end{equation}

For bcc Fe at $T=\SI{300}{K}$, we obtain in the linear-response regime $S_{\mathrm{m}} = \SI{5e-3}{\hbar\, nm^{-2}ps^{-1}}, \kappa_{\mathrm{m}} = \SI{39}{Wm^{-1}K^{-1}}$, and $\langle v \rangle_\mathrm{tr} = \SI{15}{nm/ps}$. Similar to what was found in Ref.~\cite{Xufei2018}, the magnonic thermal conductivity we obtain is relatively high given the experimentally measured total thermal conductivity of \SI{77.3}{Wm^{-1}K^{-1}}, of which roughly \SI{70}{Wm^{-1}K^{-1}} is attributed to electrons. This could potentially be a consequence of the absence of magnon-phonon~\cite{Schmidt2021}, magnon-magnon, and magnon-defect scattering in our model.

\subsection{Calculating the demagnetization from magnon occupations}\label{aDemag}

The magnetization in a ferromagnet is proportional to the spin per site $S$ minus the number of magnons per site $\frac{1}{\Omega}\int_\Omega n(\q)\mathrm{d} \q \approx \frac{1}{N}\sum_\q n_\q$, where the integral and sum are over a reciprocal unit cell $\Omega$, and $N$ is the number of equidistantly spaced magnon modes sampled over a reciprocal unit cell. From this, a relative change in magnetization can be written as
\begin{equation}\label{eDemagNaive}
    \frac{M-M_0}{M_0} = \frac{\sum_\q (n^{\mathrm{init}}_\q-n_\q)}{(NS-\sum_\q n^{\mathrm{init}}_\q)},
\end{equation}
where $n_\q^{\mathrm{init}} = n^0(\omega_\q,\SI{300}{K})$ is the equilibrium occupancy prior to laser excitation.

This expression assumes that each sampled point $\q'$ represents an average over its associated reciprocal-space volume $\Omega_{\q'}$, i.e. $n_{\q'} \approx \frac{N}{\Omega}\int_{\Omega_{\q'}} n(\q)\mathrm{d} \q$, which holds if $n(\q)$ varies slowly or monotonously within $\Omega_{\q'}$. While generally reasonable, this assumption fails near the $\Gamma$-point.

The magnetocrystalline anisotropy energy of Fe films is small (\SI{6.97}{\micro\electronvolt} per atom~\cite{Razdolski2017}), due to which $\Gamma$-point magnons have energies close to zero. Consequently, their Bose-Einstein occupation becomes very large and decreases rapidly within the immediate surrounding region. Approximating the occupancy in this volume by its center value therefore overestimates its contribution.

Numerically, this manifests as slow convergence of Eq.~\eqref{eDemagNaive} with respect to reciprocal-space sampling density. To correct for this, the $\Gamma$ point was treated separately:
\begin{equation}
    \tilde n_{\Gamma} = \frac{N}{\Omega}\int_{\Omega_{\Gamma}} n^0(\omega(\q),T_\Gamma) \mathrm{d}\q,
\end{equation}
where $T_\Gamma$ is the magnon temperature at the $\Gamma$ point. This integral was evaluated numerically. The demagnetization expression then becomes,
\begin{equation}\label{eDemag}
    \frac{\Delta M}{M_0} = \frac{\tilde n_\Gamma^\mathrm{init} - \tilde n_\Gamma + \sum_{\q\neq\Gamma} (n^{\mathrm{init}}_\q-n_\q)}{(NS- \tilde n_\Gamma^\mathrm{init}-\sum_{\q\neq\Gamma} n^{\mathrm{init}}_\q)},
\end{equation} 
which converges at a considerably coarser sampling than Eq.~\eqref{eDemagNaive}. 

In principle, these arguments apply to other magnon-related integrated quantities, such as the spin current. However, transport quantities requires a symmetry-breaking temperature interpolation across the $\q$-grid---a task that lies beyond the scope of this work.

%{\red\tiny LM: This treatment probably needs to be done for all magnon-related quantities, such as for example the spin current! This is also important if we want to compare with linear-response stuff, which we should also compute more exactly by means of interpolation. However, for this method to have an effect transport quantities, we need to interpolate the temperature to break symmetry. Will take quite some time to implement.}

\subsection{Quantifying the magnon ballistic-to-diffusive crossover}\label{aCrossover}

To characterize the transport regime of a particle ensemble, we can monitor the evolution of their mean squared displacement (MSD), $\langle \Delta z^2 \rangle$. We define the instantaneous transport exponent $\alpha(t)$ through the local logarithmic slope
\begin{equation}\label{ealphaslope}
\alpha(t) = \frac{\mathrm{d} \ln \langle \Delta z^2 \rangle}{\mathrm{d} \ln t}.
\end{equation}
If the MSD obeys a pure power law,
\begin{equation}
\langle \Delta z^2 \rangle = K t^{\alpha},
\end{equation}
with constants $K$ and $\alpha$, then $\alpha(t)=\alpha$ is constant. Deviations from a constant exponent signals a dynamic transport regime. $\alpha(t)$ provides a time-resolved measure of the transport character directly from the MSD.

An analytical expression for the MSD can be obtained under the assumptions of constant particle speed $v$ and exponentially distributed scattering times with mean $\tau$, where each scattering event randomizes the propagation direction. These assumptions yield the velocity autocorrelation function $\langle \mathbf{v}(t_1)\cdot \mathbf{v}(t_2)\rangle = v^2\mathrm{e}^{-|t_1-t_2|/\tau}$, from which one finds~\cite{MSDAnalytical}
\begin{equation}\label{eMSDAnalytical}
\langle \Delta z^2 \rangle
= \frac{2}{3}v^2\tau^2\left(\frac{t}{\tau}-1+\mathrm{e}^{-t/\tau}\right).
\end{equation}

From this expression, one finds that $\alpha(t)=2$ in the ballistic limit ($t\ll \tau$), and $\alpha(t)=1$ in the diffusive limit ($t\gg \tau$). In general, the analytical transport exponent becomes
\begin{equation}\label{ealpha_ana}
\alpha_{\mathrm{analytic}}(t)
= \frac{t\left(1-\mathrm{e}^{-t/\tau}\right)}
{t - \tau\left(1-\mathrm{e}^{-t/\tau}\right)},
\end{equation}
which is notably independent of the particle speed $v$.

To express the transport exponent as a function of displacement rather than time, we invert the root mean squared displacement $\ell(t)=\sqrt{\langle \Delta  z^2 \rangle}$. From the analytical MSD, this becomes
\begin{equation}
t(\ell)
= \tau\left(1+\frac{3\ell^2}{2v^2\tau^2}+W\left(-\mathrm{e}^{-1-\frac{3\ell^2}{2v^2\tau^2}}\right)\right),
\end{equation}
where $W$ denotes the Lambert $W$ function. Substitution into Eq.~\eqref{ealpha_ana} yields $\alpha_{\mathrm{analytic}}(\ell)$.

We now turn to the numerical computation of the MSD for magnons in our simulations, allowing us to extract both $\alpha(t)$ and $\alpha(\ell)$, and thereby directly characterize the crossover from ballistic to diffusive transport.

If all particles originate at $(t_0,z_0)=(0,0)$, the MSD is formally given by
\begin{equation}
\langle \Delta  z^2 \rangle
= \frac{\int n(z,t) z^2 \mathrm{d} z}{\int n(z,t) \mathrm{d} z},
\end{equation}
where $n$ is the particle number density. However, this expression cannot be applied directly to $\sum_\mathbf{q} n_\mathbf{q}$, since $n_\mathbf{q}$ contains magnons originating from all positions and times. The simulation does not track the individual origin of each magnon; all are aggregated in $n_\mathbf{q}$.

To overcome this limitation, we inject a small population of virtual magnons at $(t_0,z_0)=(0,0)$ that evolve within the background solution $(T_\mathrm{e}, n_\mathbf{q})$ without modifying it. These tracked magnons all have a common origin and can be tracked explicitly. Because magnons in the model scatter exclusively with electrons, the corresponding electron perturbation must also be tracked. This allows us to ensure energy conservation.

The tracked magnons $\delta n_\mathbf{q}$ and the perturbation to the electron temperature $\delta T_\mathrm{e}$ obey
\begin{align}
    \delta \dot n_\q &= -v_\q\frac{\partial \delta n_\q}{\partial z} + \delta s_\q +  \epsilon s_\q \delta(z,t) ,\\
    C_\mathrm{e}\delta \dot T_\mathrm{e} &= - \frac{1}{NV} \sum_\q \hbar \omega_\q\delta s_\q ,\\
    \delta s_\q &= s_\q(\{n_\q + \delta n_\q\}, T_\mathrm{e}+\delta T_\mathrm{e})-s_\q(\{n_\q\}, T_\mathrm{e}).
    \label{eSourceMagTracked}
\end{align}

Initially, $\delta n_\mathbf{q}=0$ and $\delta T_\mathrm{e}=0$, implying $\delta s_\mathbf{q}=0$. At $(z,t)=(0,0)$ -- corresponding to the surface during the peak of the laser pulse -- a small number of tracked magnons is injected, proportional to the magnon source term and scaled by a small number, $\epsilon \ll 1$. These magnons subsequently propagate and scatter with the tracked electrons.

The impact of $\delta T_\mathrm{e}$ on electron transport is intentionally disabled in the tracked subsystem, as it would obscure the magnon transport signal of interest. Instead, $\delta T_\mathrm{e}$ serves only as a temporary energy reservoir: it can absorb magnons and later re-emit new magnons in other directions.

Conceptually, this procedure resembles subtracting two simulations (with and without the impulsive magnon source $\epsilon s_\mathbf{q}\delta(z,t)$). However, in such a subtraction approach, electron diffusion and phonon absorption could not be selectively disabled, leading to different dynamics.

Because magnon numbers are not conserved and part of the transported energy resides in the electrons, it is more appropriate to use the mean squared energy displacement to compute $\alpha(t)$ and $\alpha(\ell)$,
\begin{equation}
\langle \Delta z^2 \rangle_E
= \frac{\int \varepsilon(z,t) z^2 \mathrm{d} z}
{\int \varepsilon(z,t) \mathrm{d} z},
\end{equation}
where $\varepsilon = C_\mathrm{e}\delta T_\mathrm{e} + \frac{1}{NV}\sum_\q \hbar\omega_\q \delta n_\q$ is the energy density. 

For computational efficiency, we neglect the second-order electron-magnon scattering term in Eq.~\eqref{eSourceMag} when simulating the ballistic-to-diffusive crossover. Its inclusion shifts the crossover to slightly earlier times.

%The curves in Fig.~\ref{fCrossover} are obtained by stitching together the results of several simulations. This allows us to use fine discretization ($\Delta x,\Delta t$) to resolve the ballistic regime, and large system sizes $L$ to permit unimpeded magnon propagation at later times. 

The agreement seen in Fig.~\ref{fCrossover} between the analytical curve and the numerical results at low fluence confirms the validity of the approach outlined here.

\subsection{Estimating the validity of Fourier's law for electrons}\label{aFourier}

Fourier's law, which we used in the derivation of the heat diffusion equations for electrons and phonons, is not generally valid in nanoscale transport and in the strongly nonequilibrium regime~\cite{Wilson2014}. To illustrate this, we start from the Boltzmann transport equation (Eq.~\eqref{eBoltzGeneral}) in the absence of external forces.

Substituting $n = f^0 + g$, where $f^0$ is the Fermi-Dirac distribution and $g$ is a perturbation, and applying the relaxation time approximation $(\dt{n})_\mathrm{scatter} = -g/\tau$, with $\tau$ being the relaxation time, we obtain
\begin{equation}\label{eg1D}
    \left(\dt{} + v \dz{} + \frac{1}{\tau}\right) g = \mathcal{F}[f^0],
\end{equation}
where $\mathcal{F}[f^0]$ is a functional whose explicit form is not relevant here. Fourier's law can be derived~\cite{FourierDerivation} if the scattering term $g/\tau$ dominates over the drift and temporal terms on the left-hand side. This corresponds to a regime in which the distribution varies slowly in space and time compared to the scattering rate.

We define a characteristic time, taken as the width of the laser pulse, $\tau_c = 2.355\tau_l \approx \SI{140}{fs}$, and a characteristic length, taken as the laser penetration depth, $\Lambda \approx \SI{20}{nm}$. Using typical values for the velocity $v = \SI{1.5}{nm/fs}$ and the relaxation time $\tau = \SI{10}{fs}$, we can compare the magnitudes of the terms in Eq.~\eqref{eg1D}.

Assuming $g = 0$ at some initial time and $g = g$ after the characteristic time $\tau_c$, we estimate $\partial g / \partial t \approx g / \tau_c$. Similarly, assuming $g(z=0) = g$ and $g(z=\Lambda)=0$, we estimate $\partial g / \partial z \approx -g / \Lambda$. This yields
\begin{equation}
    \left( \dt{} + v\dz{} + \frac{1}{\tau}\right) g \approx \frac{1}{\SI{10}{fs}}(0.07 - 0.75+1)g.
\end{equation}

Since the spatial derivative term in particular is comparable in magnitude to the scattering term -- especially during the transient dynamics -- Fourier's law provides only a rough approximation of the electronic heat flow. Equivalently, the mean free path $\lambda = v\tau \approx \SI{15}{nm}$ is comparable to the characteristic length $\Lambda = \SI{20}{nm}$. Electrons can traverse such distances without scattering. Under these conditions, electrons propagate quasi-ballistically rather than diffusively.

While Fourier's law likely underestimates the rate and extent of heat transport into the film depth, it still provides a useful first approximation. A more accurate treatment is outside of the scope of the present work.


%%%%%%%%%%%%%%%%%%%%%%%%%%%%%%%%%%%%%%%%
% \bibliography{bibfile.bib}

\begin{thebibliography}{108}%
\makeatletter
\providecommand \@ifxundefined [1]{%
 \@ifx{#1\undefined}
}%
\providecommand \@ifnum [1]{%
 \ifnum #1\expandafter \@firstoftwo
 \else \expandafter \@secondoftwo
 \fi
}%
\providecommand \@ifx [1]{%
 \ifx #1\expandafter \@firstoftwo
 \else \expandafter \@secondoftwo
 \fi
}%
\providecommand \natexlab [1]{#1}%
\providecommand \enquote  [1]{``#1''}%
\providecommand \bibnamefont  [1]{#1}%
\providecommand \bibfnamefont [1]{#1}%
\providecommand \citenamefont [1]{#1}%
\providecommand \href@noop [0]{\@secondoftwo}%
\providecommand \href [0]{\begingroup \@sanitize@url \@href}%
\providecommand \@href[1]{\@@startlink{#1}\@@href}%
\providecommand \@@href[1]{\endgroup#1\@@endlink}%
\providecommand \@sanitize@url [0]{\catcode `\\12\catcode `\$12\catcode
  `\&12\catcode `\#12\catcode `\^12\catcode `\_12\catcode `\%12\relax}%
\providecommand \@@startlink[1]{}%
\providecommand \@@endlink[0]{}%
\providecommand \url  [0]{\begingroup\@sanitize@url \@url }%
\providecommand \@url [1]{\endgroup\@href {#1}{\urlprefix }}%
\providecommand \urlprefix  [0]{URL }%
\providecommand \Eprint [0]{\href }%
\providecommand \doibase [0]{https://doi.org/}%
\providecommand \selectlanguage [0]{\@gobble}%
\providecommand \bibinfo  [0]{\@secondoftwo}%
\providecommand \bibfield  [0]{\@secondoftwo}%
\providecommand \translation [1]{[#1]}%
\providecommand \BibitemOpen [0]{}%
\providecommand \bibitemStop [0]{}%
\providecommand \bibitemNoStop [0]{.\EOS\space}%
\providecommand \EOS [0]{\spacefactor3000\relax}%
\providecommand \BibitemShut  [1]{\csname bibitem#1\endcsname}%
\let\auto@bib@innerbib\@empty
%</preamble>
\bibitem [{\citenamefont {Beaurepaire}\ \emph {et~al.}(1996)\citenamefont
  {Beaurepaire}, \citenamefont {Merle}, \citenamefont {Daunois},\ and\
  \citenamefont {Bigot}}]{Beaurepaire1996}%
  \BibitemOpen
  \bibfield  {author} {\bibinfo {author} {\bibfnamefont {E.}~\bibnamefont
  {Beaurepaire}}, \bibinfo {author} {\bibfnamefont {J.-C.}\ \bibnamefont
  {Merle}}, \bibinfo {author} {\bibfnamefont {A.}~\bibnamefont {Daunois}},\
  and\ \bibinfo {author} {\bibfnamefont {J.-Y.}\ \bibnamefont {Bigot}},\
  }\bibfield  {title} {\bibinfo {title} {Ultrafast spin dynamics in
  ferromagnetic nickel},\ }\href {https://doi.org/10.1103/PhysRevLett.76.4250}
  {\bibfield  {journal} {\bibinfo  {journal} {Phys. Rev. Lett.}\ }\textbf
  {\bibinfo {volume} {76}},\ \bibinfo {pages} {4250} (\bibinfo {year}
  {1996})}\BibitemShut {NoStop}%
\bibitem [{\citenamefont {Kirilyuk}\ \emph {et~al.}(2010)\citenamefont
  {Kirilyuk}, \citenamefont {Kimel},\ and\ \citenamefont
  {Rasing}}]{Kirilyuk2010}%
  \BibitemOpen
  \bibfield  {author} {\bibinfo {author} {\bibfnamefont {A.}~\bibnamefont
  {Kirilyuk}}, \bibinfo {author} {\bibfnamefont {A.~V.}\ \bibnamefont
  {Kimel}},\ and\ \bibinfo {author} {\bibfnamefont {T.}~\bibnamefont
  {Rasing}},\ }\bibfield  {title} {\bibinfo {title} {Ultrafast optical
  manipulation of magnetic order},\ }\href
  {https://doi.org/10.1103/RevModPhys.82.2731} {\bibfield  {journal} {\bibinfo
  {journal} {Rev. Mod. Phys.}\ }\textbf {\bibinfo {volume} {82}},\ \bibinfo
  {pages} {2731} (\bibinfo {year} {2010})}\BibitemShut {NoStop}%
\bibitem [{\citenamefont {Scheid}\ \emph {et~al.}(2022)\citenamefont {Scheid},
  \citenamefont {Remy}, \citenamefont {Lebègue}, \citenamefont {Malinowski},\
  and\ \citenamefont {Mangin}}]{Scheid2022}%
  \BibitemOpen
  \bibfield  {author} {\bibinfo {author} {\bibfnamefont {P.}~\bibnamefont
  {Scheid}}, \bibinfo {author} {\bibfnamefont {Q.}~\bibnamefont {Remy}},
  \bibinfo {author} {\bibfnamefont {S.}~\bibnamefont {Lebègue}}, \bibinfo
  {author} {\bibfnamefont {G.}~\bibnamefont {Malinowski}},\ and\ \bibinfo
  {author} {\bibfnamefont {S.}~\bibnamefont {Mangin}},\ }\bibfield  {title}
  {\bibinfo {title} {Light induced ultrafast magnetization dynamics in metallic
  compounds},\ }\href
  {https://doi.org/https://doi.org/10.1016/j.jmmm.2022.169596} {\bibfield
  {journal} {\bibinfo  {journal} {J. Magn. Magn. Mater.}\ }\textbf {\bibinfo
  {volume} {560}},\ \bibinfo {pages} {169596} (\bibinfo {year}
  {2022})}\BibitemShut {NoStop}%
\bibitem [{\citenamefont {Chen}\ \emph {et~al.}(2025)\citenamefont {Chen},
  \citenamefont {Adam}, \citenamefont {B{\"u}rgler}, \citenamefont {Wang},
  \citenamefont {Lu}, \citenamefont {Pan}, \citenamefont {Heidtfeld},
  \citenamefont {Greb}, \citenamefont {Liu}, \citenamefont {Liu}, \citenamefont
  {Wang}, \citenamefont {Schneider},\ and\ \citenamefont {Cao}}]{Chen2025}%
  \BibitemOpen
  \bibfield  {author} {\bibinfo {author} {\bibfnamefont {X.}~\bibnamefont
  {Chen}}, \bibinfo {author} {\bibfnamefont {R.}~\bibnamefont {Adam}}, \bibinfo
  {author} {\bibfnamefont {D.~E.}\ \bibnamefont {B{\"u}rgler}}, \bibinfo
  {author} {\bibfnamefont {F.}~\bibnamefont {Wang}}, \bibinfo {author}
  {\bibfnamefont {Z.}~\bibnamefont {Lu}}, \bibinfo {author} {\bibfnamefont
  {L.}~\bibnamefont {Pan}}, \bibinfo {author} {\bibfnamefont {S.}~\bibnamefont
  {Heidtfeld}}, \bibinfo {author} {\bibfnamefont {C.}~\bibnamefont {Greb}},
  \bibinfo {author} {\bibfnamefont {M.}~\bibnamefont {Liu}}, \bibinfo {author}
  {\bibfnamefont {Q.}~\bibnamefont {Liu}}, \bibinfo {author} {\bibfnamefont
  {J.}~\bibnamefont {Wang}}, \bibinfo {author} {\bibfnamefont {C.~M.}\
  \bibnamefont {Schneider}},\ and\ \bibinfo {author} {\bibfnamefont
  {D.}~\bibnamefont {Cao}},\ }\bibfield  {title} {\bibinfo {title} {Ultrafast
  demagnetization in ferromagnetic materials: Origins and progress},\ }\href
  {https://doi.org/https://doi.org/10.1016/j.physrep.2024.10.008} {\bibfield
  {journal} {\bibinfo  {journal} {Phys. Rep.}\ }\textbf {\bibinfo {volume}
  {1102}},\ \bibinfo {pages} {1} (\bibinfo {year} {2025})}\BibitemShut
  {NoStop}%
\bibitem [{\citenamefont {Carva}\ \emph {et~al.}(2017)\citenamefont {Carva},
  \citenamefont {Bal{\'a}{\v{z}}},\ and\ \citenamefont {Radu}}]{Carva2017}%
  \BibitemOpen
  \bibfield  {author} {\bibinfo {author} {\bibfnamefont {K.}~\bibnamefont
  {Carva}}, \bibinfo {author} {\bibfnamefont {P.}~\bibnamefont
  {Bal{\'a}{\v{z}}}},\ and\ \bibinfo {author} {\bibfnamefont {I.}~\bibnamefont
  {Radu}},\ }\bibfield  {title} {\bibinfo {title} {{Laser-Induced Ultrafast
  Magnetic Phenomena}},\ }in\ \href@noop {} {\emph {\bibinfo {booktitle}
  {Handbook of Magnetic Materials}}},\ Vol.~\bibinfo {volume} {26}\ (\bibinfo
  {publisher} {Elsevier, Amsterdam},\ \bibinfo {year} {2017})\ pp.\ \bibinfo
  {pages} {291--463}\BibitemShut {NoStop}%
\bibitem [{\citenamefont {Carpene}\ \emph
  {et~al.}(2008{\natexlab{a}})\citenamefont {Carpene}, \citenamefont {Mancini},
  \citenamefont {Dallera}, \citenamefont {Brenna}, \citenamefont {Puppin},\
  and\ \citenamefont {De~Silvestri}}]{Carpene2008}%
  \BibitemOpen
  \bibfield  {author} {\bibinfo {author} {\bibfnamefont {E.}~\bibnamefont
  {Carpene}}, \bibinfo {author} {\bibfnamefont {E.}~\bibnamefont {Mancini}},
  \bibinfo {author} {\bibfnamefont {C.}~\bibnamefont {Dallera}}, \bibinfo
  {author} {\bibfnamefont {M.}~\bibnamefont {Brenna}}, \bibinfo {author}
  {\bibfnamefont {E.}~\bibnamefont {Puppin}},\ and\ \bibinfo {author}
  {\bibfnamefont {S.}~\bibnamefont {De~Silvestri}},\ }\bibfield  {title}
  {\bibinfo {title} {Dynamics of electron-magnon interaction and ultrafast
  demagnetization in thin iron films},\ }\href
  {https://doi.org/10.1103/PhysRevB.78.174422} {\bibfield  {journal} {\bibinfo
  {journal} {Phys. Rev. B}\ }\textbf {\bibinfo {volume} {78}},\ \bibinfo
  {pages} {174422} (\bibinfo {year} {2008}{\natexlab{a}})}\BibitemShut
  {NoStop}%
\bibitem [{\citenamefont {Schmidt}\ \emph {et~al.}(2010)\citenamefont
  {Schmidt}, \citenamefont {Pickel}, \citenamefont {Donath}, \citenamefont
  {Buczek}, \citenamefont {Ernst}, \citenamefont {Zhukov}, \citenamefont
  {Echenique}, \citenamefont {Sandratskii}, \citenamefont {Chulkov},\ and\
  \citenamefont {Weinelt}}]{Schmidt2010}%
  \BibitemOpen
  \bibfield  {author} {\bibinfo {author} {\bibfnamefont {A.~B.}\ \bibnamefont
  {Schmidt}}, \bibinfo {author} {\bibfnamefont {M.}~\bibnamefont {Pickel}},
  \bibinfo {author} {\bibfnamefont {M.}~\bibnamefont {Donath}}, \bibinfo
  {author} {\bibfnamefont {P.}~\bibnamefont {Buczek}}, \bibinfo {author}
  {\bibfnamefont {A.}~\bibnamefont {Ernst}}, \bibinfo {author} {\bibfnamefont
  {V.~P.}\ \bibnamefont {Zhukov}}, \bibinfo {author} {\bibfnamefont {P.~M.}\
  \bibnamefont {Echenique}}, \bibinfo {author} {\bibfnamefont {L.~M.}\
  \bibnamefont {Sandratskii}}, \bibinfo {author} {\bibfnamefont {E.~V.}\
  \bibnamefont {Chulkov}},\ and\ \bibinfo {author} {\bibfnamefont
  {M.}~\bibnamefont {Weinelt}},\ }\bibfield  {title} {\bibinfo {title}
  {{Ultrafast Magnon Generation in an Fe Film on Cu(100)}},\ }\href
  {https://doi.org/10.1103/PhysRevLett.105.197401} {\bibfield  {journal}
  {\bibinfo  {journal} {Phys. Rev. Lett.}\ }\textbf {\bibinfo {volume} {105}},\
  \bibinfo {pages} {197401} (\bibinfo {year} {2010})}\BibitemShut {NoStop}%
\bibitem [{\citenamefont {Carpene}\ \emph {et~al.}(2015)\citenamefont
  {Carpene}, \citenamefont {Hedayat}, \citenamefont {Boschini},\ and\
  \citenamefont {Dallera}}]{Carpene2015}%
  \BibitemOpen
  \bibfield  {author} {\bibinfo {author} {\bibfnamefont {E.}~\bibnamefont
  {Carpene}}, \bibinfo {author} {\bibfnamefont {H.}~\bibnamefont {Hedayat}},
  \bibinfo {author} {\bibfnamefont {F.}~\bibnamefont {Boschini}},\ and\
  \bibinfo {author} {\bibfnamefont {C.}~\bibnamefont {Dallera}},\ }\bibfield
  {title} {\bibinfo {title} {Ultrafast demagnetization of metals: Collapsed
  exchange versus collective excitations},\ }\href
  {https://doi.org/10.1103/PhysRevB.91.174414} {\bibfield  {journal} {\bibinfo
  {journal} {Phys. Rev. B}\ }\textbf {\bibinfo {volume} {91}},\ \bibinfo
  {pages} {174414} (\bibinfo {year} {2015})}\BibitemShut {NoStop}%
\bibitem [{\citenamefont {Turgut}\ \emph {et~al.}(2016)\citenamefont {Turgut},
  \citenamefont {Zusin}, \citenamefont {Legut}, \citenamefont {Carva},
  \citenamefont {Knut}, \citenamefont {Shaw}, \citenamefont {Chen},
  \citenamefont {Tao}, \citenamefont {Nembach}, \citenamefont {Silva},
  \citenamefont {Mathias}, \citenamefont {Aeschlimann}, \citenamefont
  {Oppeneer}, \citenamefont {Kapteyn}, \citenamefont {Murnane},\ and\
  \citenamefont {Grychtol}}]{Turgut2016}%
  \BibitemOpen
  \bibfield  {author} {\bibinfo {author} {\bibfnamefont {E.}~\bibnamefont
  {Turgut}}, \bibinfo {author} {\bibfnamefont {D.}~\bibnamefont {Zusin}},
  \bibinfo {author} {\bibfnamefont {D.}~\bibnamefont {Legut}}, \bibinfo
  {author} {\bibfnamefont {K.}~\bibnamefont {Carva}}, \bibinfo {author}
  {\bibfnamefont {R.}~\bibnamefont {Knut}}, \bibinfo {author} {\bibfnamefont
  {J.~M.}\ \bibnamefont {Shaw}}, \bibinfo {author} {\bibfnamefont
  {C.}~\bibnamefont {Chen}}, \bibinfo {author} {\bibfnamefont {Z.}~\bibnamefont
  {Tao}}, \bibinfo {author} {\bibfnamefont {H.~T.}\ \bibnamefont {Nembach}},
  \bibinfo {author} {\bibfnamefont {T.~J.}\ \bibnamefont {Silva}}, \bibinfo
  {author} {\bibfnamefont {S.}~\bibnamefont {Mathias}}, \bibinfo {author}
  {\bibfnamefont {M.}~\bibnamefont {Aeschlimann}}, \bibinfo {author}
  {\bibfnamefont {P.~M.}\ \bibnamefont {Oppeneer}}, \bibinfo {author}
  {\bibfnamefont {H.~C.}\ \bibnamefont {Kapteyn}}, \bibinfo {author}
  {\bibfnamefont {M.~M.}\ \bibnamefont {Murnane}},\ and\ \bibinfo {author}
  {\bibfnamefont {P.}~\bibnamefont {Grychtol}},\ }\bibfield  {title} {\bibinfo
  {title} {{Stoner versus Heisenberg: Ultrafast exchange reduction and magnon
  generation during laser-induced demagnetization}},\ }\href
  {https://doi.org/10.1103/PhysRevB.94.220408} {\bibfield  {journal} {\bibinfo
  {journal} {Phys. Rev. B}\ }\textbf {\bibinfo {volume} {94}},\ \bibinfo
  {pages} {220408} (\bibinfo {year} {2016})}\BibitemShut {NoStop}%
\bibitem [{\citenamefont {Eich}\ \emph {et~al.}(2017)\citenamefont {Eich},
  \citenamefont {Pl{\"o}tzing}, \citenamefont {Rollinger}, \citenamefont
  {Emmerich}, \citenamefont {Adam}, \citenamefont {Chen}, \citenamefont
  {Kapteyn}, \citenamefont {Murnane}, \citenamefont {Plucinski}, \citenamefont
  {Steil}, \citenamefont {Stadtm{\"u}ller}, \citenamefont {Cinchetti},
  \citenamefont {Aeschlimann}, \citenamefont {Schneider},\ and\ \citenamefont
  {Mathias}}]{Eich2017}%
  \BibitemOpen
  \bibfield  {author} {\bibinfo {author} {\bibfnamefont {S.}~\bibnamefont
  {Eich}}, \bibinfo {author} {\bibfnamefont {M.}~\bibnamefont {Pl{\"o}tzing}},
  \bibinfo {author} {\bibfnamefont {M.}~\bibnamefont {Rollinger}}, \bibinfo
  {author} {\bibfnamefont {S.}~\bibnamefont {Emmerich}}, \bibinfo {author}
  {\bibfnamefont {R.}~\bibnamefont {Adam}}, \bibinfo {author} {\bibfnamefont
  {C.}~\bibnamefont {Chen}}, \bibinfo {author} {\bibfnamefont {H.~C.}\
  \bibnamefont {Kapteyn}}, \bibinfo {author} {\bibfnamefont {M.~M.}\
  \bibnamefont {Murnane}}, \bibinfo {author} {\bibfnamefont {L.}~\bibnamefont
  {Plucinski}}, \bibinfo {author} {\bibfnamefont {D.}~\bibnamefont {Steil}},
  \bibinfo {author} {\bibfnamefont {B.}~\bibnamefont {Stadtm{\"u}ller}},
  \bibinfo {author} {\bibfnamefont {M.}~\bibnamefont {Cinchetti}}, \bibinfo
  {author} {\bibfnamefont {M.}~\bibnamefont {Aeschlimann}}, \bibinfo {author}
  {\bibfnamefont {C.~M.}\ \bibnamefont {Schneider}},\ and\ \bibinfo {author}
  {\bibfnamefont {S.}~\bibnamefont {Mathias}},\ }\bibfield  {title} {\bibinfo
  {title} {Band structure evolution during the ultrafast
  ferromagnetic-paramagnetic phase transition in cobalt},\ }\href
  {https://doi.org/10.1126/sciadv.1602094} {\bibfield  {journal} {\bibinfo
  {journal} {Sci. Adv.}\ }\textbf {\bibinfo {volume} {3}},\ \bibinfo {pages}
  {e1602094} (\bibinfo {year} {2017})}\BibitemShut {NoStop}%
\bibitem [{\citenamefont {Yamamoto}\ \emph {et~al.}(2019)\citenamefont
  {Yamamoto}, \citenamefont {Kubota}, \citenamefont {Suzuki}, \citenamefont
  {Hirata}, \citenamefont {Carva}, \citenamefont {Berritta}, \citenamefont
  {Takubo}, \citenamefont {Uemura}, \citenamefont {Fukaya}, \citenamefont
  {Tanaka}, \citenamefont {Nishimura}, \citenamefont {Ohkochi}, \citenamefont
  {Katayama}, \citenamefont {Togashi}, \citenamefont {Tamasaku}, \citenamefont
  {Yabashi}, \citenamefont {Tanaka}, \citenamefont {Seki}, \citenamefont
  {Takanashi}, \citenamefont {Oppeneer},\ and\ \citenamefont
  {Wadati}}]{Yamamoto2019}%
  \BibitemOpen
  \bibfield  {author} {\bibinfo {author} {\bibfnamefont {K.}~\bibnamefont
  {Yamamoto}}, \bibinfo {author} {\bibfnamefont {Y.}~\bibnamefont {Kubota}},
  \bibinfo {author} {\bibfnamefont {M.}~\bibnamefont {Suzuki}}, \bibinfo
  {author} {\bibfnamefont {Y.}~\bibnamefont {Hirata}}, \bibinfo {author}
  {\bibfnamefont {K.}~\bibnamefont {Carva}}, \bibinfo {author} {\bibfnamefont
  {M.}~\bibnamefont {Berritta}}, \bibinfo {author} {\bibfnamefont
  {K.}~\bibnamefont {Takubo}}, \bibinfo {author} {\bibfnamefont
  {Y.}~\bibnamefont {Uemura}}, \bibinfo {author} {\bibfnamefont
  {R.}~\bibnamefont {Fukaya}}, \bibinfo {author} {\bibfnamefont
  {K.}~\bibnamefont {Tanaka}}, \bibinfo {author} {\bibfnamefont
  {W.}~\bibnamefont {Nishimura}}, \bibinfo {author} {\bibfnamefont
  {T.}~\bibnamefont {Ohkochi}}, \bibinfo {author} {\bibfnamefont
  {T.}~\bibnamefont {Katayama}}, \bibinfo {author} {\bibfnamefont
  {T.}~\bibnamefont {Togashi}}, \bibinfo {author} {\bibfnamefont
  {K.}~\bibnamefont {Tamasaku}}, \bibinfo {author} {\bibfnamefont
  {M.}~\bibnamefont {Yabashi}}, \bibinfo {author} {\bibfnamefont
  {Y.}~\bibnamefont {Tanaka}}, \bibinfo {author} {\bibfnamefont
  {T.}~\bibnamefont {Seki}}, \bibinfo {author} {\bibfnamefont {K.}~\bibnamefont
  {Takanashi}}, \bibinfo {author} {\bibfnamefont {P.~M.}\ \bibnamefont
  {Oppeneer}},\ and\ \bibinfo {author} {\bibfnamefont {H.}~\bibnamefont
  {Wadati}},\ }\bibfield  {title} {\bibinfo {title} {{Ultrafast demagnetization
  of Pt magnetic moment in L10-FePt probed by magnetic circular dichroism at a
  hard x-ray free electron laser}},\ }\href
  {https://doi.org/10.1088/1367-2630/ab5ac2} {\bibfield  {journal} {\bibinfo
  {journal} {New J. Phys.}\ }\textbf {\bibinfo {volume} {21}},\ \bibinfo
  {pages} {123010} (\bibinfo {year} {2019})}\BibitemShut {NoStop}%
\bibitem [{\citenamefont {Frietsch}\ \emph {et~al.}(2020)\citenamefont
  {Frietsch}, \citenamefont {Donges}, \citenamefont {Carley}, \citenamefont
  {Teichmann}, \citenamefont {Bowlan}, \citenamefont {Döbrich}, \citenamefont
  {Carva}, \citenamefont {Legut}, \citenamefont {Oppeneer}, \citenamefont
  {Nowak},\ and\ \citenamefont {Weinelt}}]{Frietsch2020}%
  \BibitemOpen
  \bibfield  {author} {\bibinfo {author} {\bibfnamefont {B.}~\bibnamefont
  {Frietsch}}, \bibinfo {author} {\bibfnamefont {A.}~\bibnamefont {Donges}},
  \bibinfo {author} {\bibfnamefont {R.}~\bibnamefont {Carley}}, \bibinfo
  {author} {\bibfnamefont {M.}~\bibnamefont {Teichmann}}, \bibinfo {author}
  {\bibfnamefont {J.}~\bibnamefont {Bowlan}}, \bibinfo {author} {\bibfnamefont
  {K.}~\bibnamefont {Döbrich}}, \bibinfo {author} {\bibfnamefont
  {K.}~\bibnamefont {Carva}}, \bibinfo {author} {\bibfnamefont
  {D.}~\bibnamefont {Legut}}, \bibinfo {author} {\bibfnamefont {P.~M.}\
  \bibnamefont {Oppeneer}}, \bibinfo {author} {\bibfnamefont {U.}~\bibnamefont
  {Nowak}},\ and\ \bibinfo {author} {\bibfnamefont {M.}~\bibnamefont
  {Weinelt}},\ }\bibfield  {title} {\bibinfo {title} {The role of ultrafast
  magnon generation in the magnetization dynamics of rare-earth metals},\
  }\href {https://doi.org/10.1126/sciadv.abb1601} {\bibfield  {journal}
  {\bibinfo  {journal} {Sci. Adv.}\ }\textbf {\bibinfo {volume} {6}},\ \bibinfo
  {pages} {eabb1601} (\bibinfo {year} {2020})}\BibitemShut {NoStop}%
\bibitem [{\citenamefont {Rhie}\ \emph {et~al.}(2003)\citenamefont {Rhie},
  \citenamefont {D\"urr},\ and\ \citenamefont {Eberhardt}}]{Rhie2003}%
  \BibitemOpen
  \bibfield  {author} {\bibinfo {author} {\bibfnamefont {H.-S.}\ \bibnamefont
  {Rhie}}, \bibinfo {author} {\bibfnamefont {H.~A.}\ \bibnamefont {D\"urr}},\
  and\ \bibinfo {author} {\bibfnamefont {W.}~\bibnamefont {Eberhardt}},\
  }\bibfield  {title} {\bibinfo {title} {Femtosecond electron and spin dynamics
  in $\mathrm{N}\mathrm{i}/\mathrm{W}(110)$ films},\ }\href
  {https://doi.org/10.1103/PhysRevLett.90.247201} {\bibfield  {journal}
  {\bibinfo  {journal} {Phys. Rev. Lett.}\ }\textbf {\bibinfo {volume} {90}},\
  \bibinfo {pages} {247201} (\bibinfo {year} {2003})}\BibitemShut {NoStop}%
\bibitem [{\citenamefont {Cinchetti}\ \emph {et~al.}(2006)\citenamefont
  {Cinchetti}, \citenamefont {S\'anchez~Albaneda}, \citenamefont {Hoffmann},
  \citenamefont {Roth}, \citenamefont {W\"ustenberg}, \citenamefont
  {Krau\ss{}}, \citenamefont {Andreyev}, \citenamefont {Schneider},
  \citenamefont {Bauer},\ and\ \citenamefont {Aeschlimann}}]{Cinchetti2006}%
  \BibitemOpen
  \bibfield  {author} {\bibinfo {author} {\bibfnamefont {M.}~\bibnamefont
  {Cinchetti}}, \bibinfo {author} {\bibfnamefont {M.}~\bibnamefont
  {S\'anchez~Albaneda}}, \bibinfo {author} {\bibfnamefont {D.}~\bibnamefont
  {Hoffmann}}, \bibinfo {author} {\bibfnamefont {T.}~\bibnamefont {Roth}},
  \bibinfo {author} {\bibfnamefont {J.-P.}\ \bibnamefont {W\"ustenberg}},
  \bibinfo {author} {\bibfnamefont {M.}~\bibnamefont {Krau\ss{}}}, \bibinfo
  {author} {\bibfnamefont {O.}~\bibnamefont {Andreyev}}, \bibinfo {author}
  {\bibfnamefont {H.~C.}\ \bibnamefont {Schneider}}, \bibinfo {author}
  {\bibfnamefont {M.}~\bibnamefont {Bauer}},\ and\ \bibinfo {author}
  {\bibfnamefont {M.}~\bibnamefont {Aeschlimann}},\ }\bibfield  {title}
  {\bibinfo {title} {Spin-flip processes and ultrafast magnetization dynamics
  in {C}o: Unifying the microscopic and macroscopic view of femtosecond
  magnetism},\ }\href {https://doi.org/10.1103/PhysRevLett.97.177201}
  {\bibfield  {journal} {\bibinfo  {journal} {Phys. Rev. Lett.}\ }\textbf
  {\bibinfo {volume} {97}},\ \bibinfo {pages} {177201} (\bibinfo {year}
  {2006})}\BibitemShut {NoStop}%
\bibitem [{\citenamefont {Koopmans}\ \emph {et~al.}(2010)\citenamefont
  {Koopmans}, \citenamefont {Malinowski}, \citenamefont {Dalla~Longa},
  \citenamefont {Steiauf}, \citenamefont {F{\"a}hnle}, \citenamefont {Roth},
  \citenamefont {Cinchetti},\ and\ \citenamefont {Aeschlimann}}]{Koopmans2010}%
  \BibitemOpen
  \bibfield  {author} {\bibinfo {author} {\bibfnamefont {B.}~\bibnamefont
  {Koopmans}}, \bibinfo {author} {\bibfnamefont {G.}~\bibnamefont
  {Malinowski}}, \bibinfo {author} {\bibfnamefont {F.}~\bibnamefont
  {Dalla~Longa}}, \bibinfo {author} {\bibfnamefont {D.}~\bibnamefont
  {Steiauf}}, \bibinfo {author} {\bibfnamefont {M.}~\bibnamefont {F{\"a}hnle}},
  \bibinfo {author} {\bibfnamefont {T.}~\bibnamefont {Roth}}, \bibinfo {author}
  {\bibfnamefont {M.}~\bibnamefont {Cinchetti}},\ and\ \bibinfo {author}
  {\bibfnamefont {M.}~\bibnamefont {Aeschlimann}},\ }\bibfield  {title}
  {\bibinfo {title} {Explaining the paradoxical diversity of ultrafast
  laser-induced demagnetization},\ }\href {https://doi.org/10.1038/nmat2593}
  {\bibfield  {journal} {\bibinfo  {journal} {Nature Mater.}\ }\textbf
  {\bibinfo {volume} {9}},\ \bibinfo {pages} {259} (\bibinfo {year}
  {2010})}\BibitemShut {NoStop}%
\bibitem [{\citenamefont {Schellekens}\ and\ \citenamefont
  {Koopmans}(2013)}]{Schellekens2013b}%
  \BibitemOpen
  \bibfield  {author} {\bibinfo {author} {\bibfnamefont {A.~J.}\ \bibnamefont
  {Schellekens}}\ and\ \bibinfo {author} {\bibfnamefont {B.}~\bibnamefont
  {Koopmans}},\ }\bibfield  {title} {\bibinfo {title} {Comparing ultrafast
  demagnetization rates between competing models for finite temperature
  magnetism},\ }\href {https://doi.org/10.1103/PhysRevLett.110.217204}
  {\bibfield  {journal} {\bibinfo  {journal} {Phys. Rev. Lett.}\ }\textbf
  {\bibinfo {volume} {110}},\ \bibinfo {pages} {217204} (\bibinfo {year}
  {2013})}\BibitemShut {NoStop}%
\bibitem [{\citenamefont {Griepe}\ and\ \citenamefont
  {Atxitia}(2023)}]{Griepe2023}%
  \BibitemOpen
  \bibfield  {author} {\bibinfo {author} {\bibfnamefont {T.}~\bibnamefont
  {Griepe}}\ and\ \bibinfo {author} {\bibfnamefont {U.}~\bibnamefont
  {Atxitia}},\ }\bibfield  {title} {\bibinfo {title} {Evidence of
  electron-phonon mediated spin flip as driving mechanism for ultrafast
  magnetization dynamics in $3d$ ferromagnets},\ }\href
  {https://doi.org/10.1103/PhysRevB.107.L100407} {\bibfield  {journal}
  {\bibinfo  {journal} {Phys. Rev. B}\ }\textbf {\bibinfo {volume} {107}},\
  \bibinfo {pages} {L100407} (\bibinfo {year} {2023})}\BibitemShut {NoStop}%
\bibitem [{\citenamefont {Battiato}\ \emph {et~al.}(2010)\citenamefont
  {Battiato}, \citenamefont {Carva},\ and\ \citenamefont
  {Oppeneer}}]{Battiato2010}%
  \BibitemOpen
  \bibfield  {author} {\bibinfo {author} {\bibfnamefont {M.}~\bibnamefont
  {Battiato}}, \bibinfo {author} {\bibfnamefont {K.}~\bibnamefont {Carva}},\
  and\ \bibinfo {author} {\bibfnamefont {P.~M.}\ \bibnamefont {Oppeneer}},\
  }\bibfield  {title} {\bibinfo {title} {Superdiffusive spin transport as a
  mechanism of ultrafast demagnetization},\ }\href
  {https://doi.org/10.1103/PhysRevLett.105.027203} {\bibfield  {journal}
  {\bibinfo  {journal} {Phys. Rev. Lett.}\ }\textbf {\bibinfo {volume} {105}},\
  \bibinfo {pages} {027203} (\bibinfo {year} {2010})}\BibitemShut {NoStop}%
\bibitem [{\citenamefont {Battiato}\ \emph {et~al.}(2012)\citenamefont
  {Battiato}, \citenamefont {Carva},\ and\ \citenamefont
  {Oppeneer}}]{Battiato2012}%
  \BibitemOpen
  \bibfield  {author} {\bibinfo {author} {\bibfnamefont {M.}~\bibnamefont
  {Battiato}}, \bibinfo {author} {\bibfnamefont {K.}~\bibnamefont {Carva}},\
  and\ \bibinfo {author} {\bibfnamefont {P.~M.}\ \bibnamefont {Oppeneer}},\
  }\bibfield  {title} {\bibinfo {title} {Theory of laser-induced ultrafast
  superdiffusive spin transport in layered heterostructures},\ }\href
  {https://doi.org/10.1103/PhysRevB.86.024404} {\bibfield  {journal} {\bibinfo
  {journal} {Phys. Rev. B}\ }\textbf {\bibinfo {volume} {86}},\ \bibinfo
  {pages} {024404} (\bibinfo {year} {2012})}\BibitemShut {NoStop}%
\bibitem [{\citenamefont {Malinowski}\ \emph {et~al.}(2008)\citenamefont
  {Malinowski}, \citenamefont {Dalla~Longa}, \citenamefont {Rietjens},
  \citenamefont {Paluskar}, \citenamefont {Huijink}, \citenamefont {Swagten},\
  and\ \citenamefont {Koopmans}}]{Malinowski2008}%
  \BibitemOpen
  \bibfield  {author} {\bibinfo {author} {\bibfnamefont {G.}~\bibnamefont
  {Malinowski}}, \bibinfo {author} {\bibfnamefont {F.}~\bibnamefont
  {Dalla~Longa}}, \bibinfo {author} {\bibfnamefont {J.~H.~H.}\ \bibnamefont
  {Rietjens}}, \bibinfo {author} {\bibfnamefont {P.~V.}\ \bibnamefont
  {Paluskar}}, \bibinfo {author} {\bibfnamefont {R.}~\bibnamefont {Huijink}},
  \bibinfo {author} {\bibfnamefont {H.~J.~M.}\ \bibnamefont {Swagten}},\ and\
  \bibinfo {author} {\bibfnamefont {B.}~\bibnamefont {Koopmans}},\ }\bibfield
  {title} {\bibinfo {title} {Control of speed and efficiency of ultrafast
  demagnetization by direct transfer of spin angular momentum},\ }\href
  {https://doi.org/10.1038/nphys1092} {\bibfield  {journal} {\bibinfo
  {journal} {Nature Phys.}\ }\textbf {\bibinfo {volume} {4}},\ \bibinfo {pages}
  {855} (\bibinfo {year} {2008})}\BibitemShut {NoStop}%
\bibitem [{\citenamefont {Rudolf}\ \emph {et~al.}(2012)\citenamefont {Rudolf},
  \citenamefont {La-O-Vorakiat}, \citenamefont {Battiato}, \citenamefont
  {Adam}, \citenamefont {Shaw}, \citenamefont {Turgut}, \citenamefont
  {Maldonado}, \citenamefont {Mathias}, \citenamefont {Grychtol}, \citenamefont
  {Nembach}, \citenamefont {Silva}, \citenamefont {Aeschlimann}, \citenamefont
  {Kapteyn}, \citenamefont {Murnane}, \citenamefont {Schneider},\ and\
  \citenamefont {Oppeneer}}]{Rudolf2012}%
  \BibitemOpen
  \bibfield  {author} {\bibinfo {author} {\bibfnamefont {D.}~\bibnamefont
  {Rudolf}}, \bibinfo {author} {\bibfnamefont {C.}~\bibnamefont
  {La-O-Vorakiat}}, \bibinfo {author} {\bibfnamefont {M.}~\bibnamefont
  {Battiato}}, \bibinfo {author} {\bibfnamefont {R.}~\bibnamefont {Adam}},
  \bibinfo {author} {\bibfnamefont {J.~M.}\ \bibnamefont {Shaw}}, \bibinfo
  {author} {\bibfnamefont {E.}~\bibnamefont {Turgut}}, \bibinfo {author}
  {\bibfnamefont {P.}~\bibnamefont {Maldonado}}, \bibinfo {author}
  {\bibfnamefont {S.}~\bibnamefont {Mathias}}, \bibinfo {author} {\bibfnamefont
  {P.}~\bibnamefont {Grychtol}}, \bibinfo {author} {\bibfnamefont {H.~T.}\
  \bibnamefont {Nembach}}, \bibinfo {author} {\bibfnamefont {T.~J.}\
  \bibnamefont {Silva}}, \bibinfo {author} {\bibfnamefont {M.}~\bibnamefont
  {Aeschlimann}}, \bibinfo {author} {\bibfnamefont {H.~C.}\ \bibnamefont
  {Kapteyn}}, \bibinfo {author} {\bibfnamefont {M.~M.}\ \bibnamefont
  {Murnane}}, \bibinfo {author} {\bibfnamefont {C.~M.}\ \bibnamefont
  {Schneider}},\ and\ \bibinfo {author} {\bibfnamefont {P.~M.}\ \bibnamefont
  {Oppeneer}},\ }\bibfield  {title} {\bibinfo {title} {Ultrafast magnetization
  enhancement in metallic multilayers driven by superdiffusive spin current},\
  }\href {https://doi.org/10.1038/ncomms2029} {\bibfield  {journal} {\bibinfo
  {journal} {Nature Commun.}\ }\textbf {\bibinfo {volume} {3}},\ \bibinfo
  {pages} {1037} (\bibinfo {year} {2012})}\BibitemShut {NoStop}%
\bibitem [{\citenamefont {Bergeard}\ \emph {et~al.}(2016)\citenamefont
  {Bergeard}, \citenamefont {Hehn}, \citenamefont {Mangin}, \citenamefont
  {Lengaigne}, \citenamefont {Montaigne}, \citenamefont {Lalieu}, \citenamefont
  {Koopmans},\ and\ \citenamefont {Malinowski}}]{Bergeard2016}%
  \BibitemOpen
  \bibfield  {author} {\bibinfo {author} {\bibfnamefont {N.}~\bibnamefont
  {Bergeard}}, \bibinfo {author} {\bibfnamefont {M.}~\bibnamefont {Hehn}},
  \bibinfo {author} {\bibfnamefont {S.}~\bibnamefont {Mangin}}, \bibinfo
  {author} {\bibfnamefont {G.}~\bibnamefont {Lengaigne}}, \bibinfo {author}
  {\bibfnamefont {F.}~\bibnamefont {Montaigne}}, \bibinfo {author}
  {\bibfnamefont {M.~L.~M.}\ \bibnamefont {Lalieu}}, \bibinfo {author}
  {\bibfnamefont {B.}~\bibnamefont {Koopmans}},\ and\ \bibinfo {author}
  {\bibfnamefont {G.}~\bibnamefont {Malinowski}},\ }\bibfield  {title}
  {\bibinfo {title} {Hot-electron-induced ultrafast demagnetization in
  $\mathrm{Co}/\mathrm{Pt}$ multilayers},\ }\href
  {https://doi.org/10.1103/PhysRevLett.117.147203} {\bibfield  {journal}
  {\bibinfo  {journal} {Phys. Rev. Lett.}\ }\textbf {\bibinfo {volume} {117}},\
  \bibinfo {pages} {147203} (\bibinfo {year} {2016})}\BibitemShut {NoStop}%
\bibitem [{\citenamefont {Xu}\ \emph {et~al.}(2017)\citenamefont {Xu},
  \citenamefont {Deb}, \citenamefont {Malinowski}, \citenamefont {Hehn},
  \citenamefont {Zhao},\ and\ \citenamefont {Mangin}}]{Xu2017}%
  \BibitemOpen
  \bibfield  {author} {\bibinfo {author} {\bibfnamefont {Y.}~\bibnamefont
  {Xu}}, \bibinfo {author} {\bibfnamefont {M.}~\bibnamefont {Deb}}, \bibinfo
  {author} {\bibfnamefont {G.}~\bibnamefont {Malinowski}}, \bibinfo {author}
  {\bibfnamefont {M.}~\bibnamefont {Hehn}}, \bibinfo {author} {\bibfnamefont
  {W.}~\bibnamefont {Zhao}},\ and\ \bibinfo {author} {\bibfnamefont
  {S.}~\bibnamefont {Mangin}},\ }\bibfield  {title} {\bibinfo {title}
  {Ultrafast magnetization manipulation using single femtosecond light and
  hot-electron pulses},\ }\href
  {https://doi.org/https://doi.org/10.1002/adma.201703474} {\bibfield
  {journal} {\bibinfo  {journal} {Adv. Mater.}\ }\textbf {\bibinfo {volume}
  {29}},\ \bibinfo {pages} {1703474} (\bibinfo {year} {2017})}\BibitemShut
  {NoStop}%
\bibitem [{\citenamefont {Hofherr}\ \emph {et~al.}(2017)\citenamefont
  {Hofherr}, \citenamefont {Maldonado}, \citenamefont {Schmitt}, \citenamefont
  {Berritta}, \citenamefont {Bierbrauer}, \citenamefont {Sadashivaiah},
  \citenamefont {Schellekens}, \citenamefont {Koopmans}, \citenamefont {Steil},
  \citenamefont {Cinchetti}, \citenamefont {Stadtm\"uller}, \citenamefont
  {Oppeneer}, \citenamefont {Mathias},\ and\ \citenamefont
  {Aeschlimann}}]{Hofherr2017}%
  \BibitemOpen
  \bibfield  {author} {\bibinfo {author} {\bibfnamefont {M.}~\bibnamefont
  {Hofherr}}, \bibinfo {author} {\bibfnamefont {P.}~\bibnamefont {Maldonado}},
  \bibinfo {author} {\bibfnamefont {O.}~\bibnamefont {Schmitt}}, \bibinfo
  {author} {\bibfnamefont {M.}~\bibnamefont {Berritta}}, \bibinfo {author}
  {\bibfnamefont {U.}~\bibnamefont {Bierbrauer}}, \bibinfo {author}
  {\bibfnamefont {S.}~\bibnamefont {Sadashivaiah}}, \bibinfo {author}
  {\bibfnamefont {A.~J.}\ \bibnamefont {Schellekens}}, \bibinfo {author}
  {\bibfnamefont {B.}~\bibnamefont {Koopmans}}, \bibinfo {author}
  {\bibfnamefont {D.}~\bibnamefont {Steil}}, \bibinfo {author} {\bibfnamefont
  {M.}~\bibnamefont {Cinchetti}}, \bibinfo {author} {\bibfnamefont
  {B.}~\bibnamefont {Stadtm\"uller}}, \bibinfo {author} {\bibfnamefont {P.~M.}\
  \bibnamefont {Oppeneer}}, \bibinfo {author} {\bibfnamefont {S.}~\bibnamefont
  {Mathias}},\ and\ \bibinfo {author} {\bibfnamefont {M.}~\bibnamefont
  {Aeschlimann}},\ }\bibfield  {title} {\bibinfo {title} {Speed and efficiency
  of femtosecond spin current injection into a nonmagnetic material},\ }\href
  {https://doi.org/10.1103/PhysRevB.96.100403} {\bibfield  {journal} {\bibinfo
  {journal} {Phys. Rev. B}\ }\textbf {\bibinfo {volume} {96}},\ \bibinfo
  {pages} {100403} (\bibinfo {year} {2017})}\BibitemShut {NoStop}%
\bibitem [{\citenamefont {Gupta}\ \emph {et~al.}(2023)\citenamefont {Gupta},
  \citenamefont {Cosco}, \citenamefont {Malik}, \citenamefont {Chen},
  \citenamefont {Saha}, \citenamefont {Ghosh}, \citenamefont {Pohlmann},
  \citenamefont {Mardegan}, \citenamefont {Francoual}, \citenamefont
  {Stefanuik}, \citenamefont {S\"oderstr\"om}, \citenamefont {Sanyal},
  \citenamefont {Karis}, \citenamefont {Svedlindh}, \citenamefont {Oppeneer},\
  and\ \citenamefont {Knut}}]{Gupta2023}%
  \BibitemOpen
  \bibfield  {author} {\bibinfo {author} {\bibfnamefont {R.}~\bibnamefont
  {Gupta}}, \bibinfo {author} {\bibfnamefont {F.}~\bibnamefont {Cosco}},
  \bibinfo {author} {\bibfnamefont {R.~S.}\ \bibnamefont {Malik}}, \bibinfo
  {author} {\bibfnamefont {X.}~\bibnamefont {Chen}}, \bibinfo {author}
  {\bibfnamefont {S.}~\bibnamefont {Saha}}, \bibinfo {author} {\bibfnamefont
  {A.}~\bibnamefont {Ghosh}}, \bibinfo {author} {\bibfnamefont
  {T.}~\bibnamefont {Pohlmann}}, \bibinfo {author} {\bibfnamefont {J.~R.~L.}\
  \bibnamefont {Mardegan}}, \bibinfo {author} {\bibfnamefont {S.}~\bibnamefont
  {Francoual}}, \bibinfo {author} {\bibfnamefont {R.}~\bibnamefont
  {Stefanuik}}, \bibinfo {author} {\bibfnamefont {J.}~\bibnamefont
  {S\"oderstr\"om}}, \bibinfo {author} {\bibfnamefont {B.}~\bibnamefont
  {Sanyal}}, \bibinfo {author} {\bibfnamefont {O.}~\bibnamefont {Karis}},
  \bibinfo {author} {\bibfnamefont {P.}~\bibnamefont {Svedlindh}}, \bibinfo
  {author} {\bibfnamefont {P.~M.}\ \bibnamefont {Oppeneer}},\ and\ \bibinfo
  {author} {\bibfnamefont {R.}~\bibnamefont {Knut}},\ }\bibfield  {title}
  {\bibinfo {title} {Element-resolved evidence of superdiffusive spin current
  arising from ultrafast demagnetization process},\ }\href
  {https://doi.org/10.1103/PhysRevB.108.064427} {\bibfield  {journal} {\bibinfo
   {journal} {Phys. Rev. B}\ }\textbf {\bibinfo {volume} {108}},\ \bibinfo
  {pages} {064427} (\bibinfo {year} {2023})}\BibitemShut {NoStop}%
\bibitem [{\citenamefont {Dornes}\ \emph {et~al.}(2019)\citenamefont {Dornes},
  \citenamefont {Acremann}, \citenamefont {Savoini}, \citenamefont {Kubli},
  \citenamefont {Neugebauer}, \citenamefont {Abreu}, \citenamefont {Huber},
  \citenamefont {Lantz}, \citenamefont {Vaz}, \citenamefont {Lemke},
  \citenamefont {Bothschafter}, \citenamefont {Porer}, \citenamefont
  {Esposito}, \citenamefont {Rettig}, \citenamefont {Buzzi}, \citenamefont
  {Alberca}, \citenamefont {Windsor}, \citenamefont {Beaud}, \citenamefont
  {Staub}, \citenamefont {Zhu}, \citenamefont {Song}, \citenamefont {Glownia},\
  and\ \citenamefont {Johnson}}]{Dornes2019}%
  \BibitemOpen
  \bibfield  {author} {\bibinfo {author} {\bibfnamefont {C.}~\bibnamefont
  {Dornes}}, \bibinfo {author} {\bibfnamefont {Y.}~\bibnamefont {Acremann}},
  \bibinfo {author} {\bibfnamefont {M.}~\bibnamefont {Savoini}}, \bibinfo
  {author} {\bibfnamefont {M.}~\bibnamefont {Kubli}}, \bibinfo {author}
  {\bibfnamefont {M.~J.}\ \bibnamefont {Neugebauer}}, \bibinfo {author}
  {\bibfnamefont {E.}~\bibnamefont {Abreu}}, \bibinfo {author} {\bibfnamefont
  {L.}~\bibnamefont {Huber}}, \bibinfo {author} {\bibfnamefont
  {G.}~\bibnamefont {Lantz}}, \bibinfo {author} {\bibfnamefont {C.~A.~F.}\
  \bibnamefont {Vaz}}, \bibinfo {author} {\bibfnamefont {H.}~\bibnamefont
  {Lemke}}, \bibinfo {author} {\bibfnamefont {E.~M.}\ \bibnamefont
  {Bothschafter}}, \bibinfo {author} {\bibfnamefont {M.}~\bibnamefont {Porer}},
  \bibinfo {author} {\bibfnamefont {V.}~\bibnamefont {Esposito}}, \bibinfo
  {author} {\bibfnamefont {L.}~\bibnamefont {Rettig}}, \bibinfo {author}
  {\bibfnamefont {M.}~\bibnamefont {Buzzi}}, \bibinfo {author} {\bibfnamefont
  {A.}~\bibnamefont {Alberca}}, \bibinfo {author} {\bibfnamefont {Y.~W.}\
  \bibnamefont {Windsor}}, \bibinfo {author} {\bibfnamefont {P.}~\bibnamefont
  {Beaud}}, \bibinfo {author} {\bibfnamefont {U.}~\bibnamefont {Staub}},
  \bibinfo {author} {\bibfnamefont {D.}~\bibnamefont {Zhu}}, \bibinfo {author}
  {\bibfnamefont {S.}~\bibnamefont {Song}}, \bibinfo {author} {\bibfnamefont
  {J.~M.}\ \bibnamefont {Glownia}},\ and\ \bibinfo {author} {\bibfnamefont
  {S.~L.}\ \bibnamefont {Johnson}},\ }\bibfield  {title} {\bibinfo {title} {The
  ultrafast {Einstein--de Haas} effect},\ }\href
  {https://doi.org/10.1038/s41586-018-0822-7} {\bibfield  {journal} {\bibinfo
  {journal} {Nature}\ }\textbf {\bibinfo {volume} {565}},\ \bibinfo {pages}
  {209} (\bibinfo {year} {2019})}\BibitemShut {NoStop}%
\bibitem [{\citenamefont {Tauchert}\ \emph {et~al.}(2022)\citenamefont
  {Tauchert}, \citenamefont {Volkov}, \citenamefont {Ehberger}, \citenamefont
  {Kazenwadel}, \citenamefont {Evers}, \citenamefont {Lange}, \citenamefont
  {Donges}, \citenamefont {Book}, \citenamefont {Kreuzpaintner}, \citenamefont
  {Nowak},\ and\ \citenamefont {Baum}}]{Tauchert2022}%
  \BibitemOpen
  \bibfield  {author} {\bibinfo {author} {\bibfnamefont {S.~R.}\ \bibnamefont
  {Tauchert}}, \bibinfo {author} {\bibfnamefont {M.}~\bibnamefont {Volkov}},
  \bibinfo {author} {\bibfnamefont {D.}~\bibnamefont {Ehberger}}, \bibinfo
  {author} {\bibfnamefont {D.}~\bibnamefont {Kazenwadel}}, \bibinfo {author}
  {\bibfnamefont {M.}~\bibnamefont {Evers}}, \bibinfo {author} {\bibfnamefont
  {H.}~\bibnamefont {Lange}}, \bibinfo {author} {\bibfnamefont
  {A.}~\bibnamefont {Donges}}, \bibinfo {author} {\bibfnamefont
  {A.}~\bibnamefont {Book}}, \bibinfo {author} {\bibfnamefont {W.}~\bibnamefont
  {Kreuzpaintner}}, \bibinfo {author} {\bibfnamefont {U.}~\bibnamefont
  {Nowak}},\ and\ \bibinfo {author} {\bibfnamefont {P.}~\bibnamefont {Baum}},\
  }\bibfield  {title} {\bibinfo {title} {Polarized phonons carry angular
  momentum in ultrafast demagnetization},\ }\href
  {https://doi.org/10.1038/s41586-021-04306-4} {\bibfield  {journal} {\bibinfo
  {journal} {Nature}\ }\textbf {\bibinfo {volume} {602}},\ \bibinfo {pages}
  {73} (\bibinfo {year} {2022})}\BibitemShut {NoStop}%
\bibitem [{\citenamefont {Sun}\ \emph {et~al.}(1993)\citenamefont {Sun},
  \citenamefont {Vall\'ee}, \citenamefont {Acioli}, \citenamefont {Ippen},\
  and\ \citenamefont {Fujimoto}}]{Sun1993}%
  \BibitemOpen
  \bibfield  {author} {\bibinfo {author} {\bibfnamefont {C.-K.}\ \bibnamefont
  {Sun}}, \bibinfo {author} {\bibfnamefont {F.}~\bibnamefont {Vall\'ee}},
  \bibinfo {author} {\bibfnamefont {L.}~\bibnamefont {Acioli}}, \bibinfo
  {author} {\bibfnamefont {E.~P.}\ \bibnamefont {Ippen}},\ and\ \bibinfo
  {author} {\bibfnamefont {J.~G.}\ \bibnamefont {Fujimoto}},\ }\bibfield
  {title} {\bibinfo {title} {Femtosecond investigation of electron
  thermalization in gold},\ }\href {https://doi.org/10.1103/PhysRevB.48.12365}
  {\bibfield  {journal} {\bibinfo  {journal} {Phys. Rev. B}\ }\textbf {\bibinfo
  {volume} {48}},\ \bibinfo {pages} {12365} (\bibinfo {year}
  {1993})}\BibitemShut {NoStop}%
\bibitem [{\citenamefont {Sun}\ \emph {et~al.}(1994)\citenamefont {Sun},
  \citenamefont {Vall\'ee}, \citenamefont {Acioli}, \citenamefont {Ippen},\
  and\ \citenamefont {Fujimoto}}]{Sun1994}%
  \BibitemOpen
  \bibfield  {author} {\bibinfo {author} {\bibfnamefont {C.-K.}\ \bibnamefont
  {Sun}}, \bibinfo {author} {\bibfnamefont {F.}~\bibnamefont {Vall\'ee}},
  \bibinfo {author} {\bibfnamefont {L.~H.}\ \bibnamefont {Acioli}}, \bibinfo
  {author} {\bibfnamefont {E.~P.}\ \bibnamefont {Ippen}},\ and\ \bibinfo
  {author} {\bibfnamefont {J.~G.}\ \bibnamefont {Fujimoto}},\ }\bibfield
  {title} {\bibinfo {title} {Femtosecond-tunable measurement of electron
  thermalization in gold},\ }\href {https://doi.org/10.1103/PhysRevB.50.15337}
  {\bibfield  {journal} {\bibinfo  {journal} {Phys. Rev. B}\ }\textbf {\bibinfo
  {volume} {50}},\ \bibinfo {pages} {15337} (\bibinfo {year}
  {1994})}\BibitemShut {NoStop}%
\bibitem [{\citenamefont {Del~Fatti}\ \emph {et~al.}(1998)\citenamefont
  {Del~Fatti}, \citenamefont {Bouffanais}, \citenamefont {Vall\'ee},\ and\
  \citenamefont {Flytzanis}}]{DelFatti1998}%
  \BibitemOpen
  \bibfield  {author} {\bibinfo {author} {\bibfnamefont {N.}~\bibnamefont
  {Del~Fatti}}, \bibinfo {author} {\bibfnamefont {R.}~\bibnamefont
  {Bouffanais}}, \bibinfo {author} {\bibfnamefont {F.}~\bibnamefont
  {Vall\'ee}},\ and\ \bibinfo {author} {\bibfnamefont {C.}~\bibnamefont
  {Flytzanis}},\ }\bibfield  {title} {\bibinfo {title} {Nonequilibrium electron
  interactions in metal films},\ }\href
  {https://doi.org/10.1103/PhysRevLett.81.922} {\bibfield  {journal} {\bibinfo
  {journal} {Phys. Rev. Lett.}\ }\textbf {\bibinfo {volume} {81}},\ \bibinfo
  {pages} {922} (\bibinfo {year} {1998})}\BibitemShut {NoStop}%
\bibitem [{\citenamefont {Del~Fatti}\ \emph {et~al.}(2000)\citenamefont
  {Del~Fatti}, \citenamefont {Voisin}, \citenamefont {Achermann}, \citenamefont
  {Tzortzakis}, \citenamefont {Christofilos},\ and\ \citenamefont
  {Vall\'ee}}]{DelFatti2000}%
  \BibitemOpen
  \bibfield  {author} {\bibinfo {author} {\bibfnamefont {N.}~\bibnamefont
  {Del~Fatti}}, \bibinfo {author} {\bibfnamefont {C.}~\bibnamefont {Voisin}},
  \bibinfo {author} {\bibfnamefont {M.}~\bibnamefont {Achermann}}, \bibinfo
  {author} {\bibfnamefont {S.}~\bibnamefont {Tzortzakis}}, \bibinfo {author}
  {\bibfnamefont {D.}~\bibnamefont {Christofilos}},\ and\ \bibinfo {author}
  {\bibfnamefont {F.}~\bibnamefont {Vall\'ee}},\ }\bibfield  {title} {\bibinfo
  {title} {Nonequilibrium electron dynamics in noble metals},\ }\href
  {https://doi.org/10.1103/PhysRevB.61.16956} {\bibfield  {journal} {\bibinfo
  {journal} {Phys. Rev. B}\ }\textbf {\bibinfo {volume} {61}},\ \bibinfo
  {pages} {16956} (\bibinfo {year} {2000})}\BibitemShut {NoStop}%
\bibitem [{\citenamefont {Guo}\ \emph {et~al.}(2001)\citenamefont {Guo},
  \citenamefont {Rodriguez},\ and\ \citenamefont {Taylor}}]{Guo2001}%
  \BibitemOpen
  \bibfield  {author} {\bibinfo {author} {\bibfnamefont {C.}~\bibnamefont
  {Guo}}, \bibinfo {author} {\bibfnamefont {G.}~\bibnamefont {Rodriguez}},\
  and\ \bibinfo {author} {\bibfnamefont {A.~J.}\ \bibnamefont {Taylor}},\
  }\bibfield  {title} {\bibinfo {title} {Ultrafast dynamics of electron
  thermalization in gold},\ }\href
  {https://doi.org/10.1103/PhysRevLett.86.1638} {\bibfield  {journal} {\bibinfo
   {journal} {Phys. Rev. Lett.}\ }\textbf {\bibinfo {volume} {86}},\ \bibinfo
  {pages} {1638} (\bibinfo {year} {2001})}\BibitemShut {NoStop}%
\bibitem [{\citenamefont {Carpene}(2006)}]{Carpene2006}%
  \BibitemOpen
  \bibfield  {author} {\bibinfo {author} {\bibfnamefont {E.}~\bibnamefont
  {Carpene}},\ }\bibfield  {title} {\bibinfo {title} {Ultrafast laser
  irradiation of metals: {B}eyond the two-temperature model},\ }\href
  {https://doi.org/10.1103/PhysRevB.74.024301} {\bibfield  {journal} {\bibinfo
  {journal} {Phys. Rev. B}\ }\textbf {\bibinfo {volume} {74}},\ \bibinfo
  {pages} {024301} (\bibinfo {year} {2006})}\BibitemShut {NoStop}%
\bibitem [{\citenamefont {Mueller}\ \emph {et~al.}(2013)\citenamefont
  {Mueller}, \citenamefont {Baral}, \citenamefont {Vollmar}, \citenamefont
  {Cinchetti}, \citenamefont {Aeschlimann}, \citenamefont {Schneider},\ and\
  \citenamefont {Rethfeld}}]{Mueller2013}%
  \BibitemOpen
  \bibfield  {author} {\bibinfo {author} {\bibfnamefont {B.~Y.}\ \bibnamefont
  {Mueller}}, \bibinfo {author} {\bibfnamefont {A.}~\bibnamefont {Baral}},
  \bibinfo {author} {\bibfnamefont {S.}~\bibnamefont {Vollmar}}, \bibinfo
  {author} {\bibfnamefont {M.}~\bibnamefont {Cinchetti}}, \bibinfo {author}
  {\bibfnamefont {M.}~\bibnamefont {Aeschlimann}}, \bibinfo {author}
  {\bibfnamefont {H.~C.}\ \bibnamefont {Schneider}},\ and\ \bibinfo {author}
  {\bibfnamefont {B.}~\bibnamefont {Rethfeld}},\ }\bibfield  {title} {\bibinfo
  {title} {Feedback effect during ultrafast demagnetization dynamics in
  ferromagnets},\ }\href {https://doi.org/10.1103/PhysRevLett.111.167204}
  {\bibfield  {journal} {\bibinfo  {journal} {Phys. Rev. Lett.}\ }\textbf
  {\bibinfo {volume} {111}},\ \bibinfo {pages} {167204} (\bibinfo {year}
  {2013})}\BibitemShut {NoStop}%
\bibitem [{\citenamefont {Tveten}\ \emph {et~al.}(2015)\citenamefont {Tveten},
  \citenamefont {Brataas},\ and\ \citenamefont {Tserkovnyak}}]{Tveten2015}%
  \BibitemOpen
  \bibfield  {author} {\bibinfo {author} {\bibfnamefont {E.~G.}\ \bibnamefont
  {Tveten}}, \bibinfo {author} {\bibfnamefont {A.}~\bibnamefont {Brataas}},\
  and\ \bibinfo {author} {\bibfnamefont {Y.}~\bibnamefont {Tserkovnyak}},\
  }\bibfield  {title} {\bibinfo {title} {Electron-magnon scattering in magnetic
  heterostructures far out of equilibrium},\ }\href
  {https://doi.org/10.1103/PhysRevB.92.180412} {\bibfield  {journal} {\bibinfo
  {journal} {Phys. Rev. B}\ }\textbf {\bibinfo {volume} {92}},\ \bibinfo
  {pages} {180412} (\bibinfo {year} {2015})}\BibitemShut {NoStop}%
\bibitem [{\citenamefont {Maldonado}\ \emph {et~al.}(2017)\citenamefont
  {Maldonado}, \citenamefont {Carva}, \citenamefont {Flammer},\ and\
  \citenamefont {Oppeneer}}]{Maldonado2017}%
  \BibitemOpen
  \bibfield  {author} {\bibinfo {author} {\bibfnamefont {P.}~\bibnamefont
  {Maldonado}}, \bibinfo {author} {\bibfnamefont {K.}~\bibnamefont {Carva}},
  \bibinfo {author} {\bibfnamefont {M.}~\bibnamefont {Flammer}},\ and\ \bibinfo
  {author} {\bibfnamefont {P.~M.}\ \bibnamefont {Oppeneer}},\ }\bibfield
  {title} {\bibinfo {title} {Theory of out-of-equilibrium ultrafast relaxation
  dynamics in metals},\ }\href {https://doi.org/10.1103/PhysRevB.96.174439}
  {\bibfield  {journal} {\bibinfo  {journal} {Phys. Rev. B}\ }\textbf {\bibinfo
  {volume} {96}},\ \bibinfo {pages} {174439} (\bibinfo {year}
  {2017})}\BibitemShut {NoStop}%
\bibitem [{\citenamefont {Maldonado}\ \emph {et~al.}(2020)\citenamefont
  {Maldonado}, \citenamefont {Chase}, \citenamefont {Reid}, \citenamefont
  {Shen}, \citenamefont {Li}, \citenamefont {Carva}, \citenamefont {Payer},
  \citenamefont {Horn~von Hoegen}, \citenamefont {Sokolowski-Tinten},
  \citenamefont {Wang}, \citenamefont {Oppeneer},\ and\ \citenamefont
  {D\"urr}}]{Maldonado2020}%
  \BibitemOpen
  \bibfield  {author} {\bibinfo {author} {\bibfnamefont {P.}~\bibnamefont
  {Maldonado}}, \bibinfo {author} {\bibfnamefont {T.}~\bibnamefont {Chase}},
  \bibinfo {author} {\bibfnamefont {A.~H.}\ \bibnamefont {Reid}}, \bibinfo
  {author} {\bibfnamefont {X.}~\bibnamefont {Shen}}, \bibinfo {author}
  {\bibfnamefont {R.~K.}\ \bibnamefont {Li}}, \bibinfo {author} {\bibfnamefont
  {K.}~\bibnamefont {Carva}}, \bibinfo {author} {\bibfnamefont
  {T.}~\bibnamefont {Payer}}, \bibinfo {author} {\bibfnamefont
  {M.}~\bibnamefont {Horn~von Hoegen}}, \bibinfo {author} {\bibfnamefont
  {K.}~\bibnamefont {Sokolowski-Tinten}}, \bibinfo {author} {\bibfnamefont
  {X.~J.}\ \bibnamefont {Wang}}, \bibinfo {author} {\bibfnamefont {P.~M.}\
  \bibnamefont {Oppeneer}},\ and\ \bibinfo {author} {\bibfnamefont {H.~A.}\
  \bibnamefont {D\"urr}},\ }\bibfield  {title} {\bibinfo {title} {Tracking the
  ultrafast nonequilibrium energy flow between electronic and lattice degrees
  of freedom in crystalline nickel},\ }\href
  {https://doi.org/10.1103/PhysRevB.101.100302} {\bibfield  {journal} {\bibinfo
   {journal} {Phys. Rev. B}\ }\textbf {\bibinfo {volume} {101}},\ \bibinfo
  {pages} {100302} (\bibinfo {year} {2020})}\BibitemShut {NoStop}%
\bibitem [{\citenamefont {Wilson}\ and\ \citenamefont
  {Coh}(2020)}]{Wilson2020}%
  \BibitemOpen
  \bibfield  {author} {\bibinfo {author} {\bibfnamefont {R.~B.}\ \bibnamefont
  {Wilson}}\ and\ \bibinfo {author} {\bibfnamefont {S.}~\bibnamefont {Coh}},\
  }\bibfield  {title} {\bibinfo {title} {Parametric dependence of hot electron
  relaxation timescales on electron-electron and electron-phonon interaction
  strengths},\ }\href {https://doi.org/10.1038/s42005-020-00442-x} {\bibfield
  {journal} {\bibinfo  {journal} {Commun. Phys.}\ }\textbf {\bibinfo {volume}
  {3}},\ \bibinfo {pages} {179} (\bibinfo {year} {2020})}\BibitemShut {NoStop}%
\bibitem [{\citenamefont {Ritzmann}\ \emph {et~al.}(2020)\citenamefont
  {Ritzmann}, \citenamefont {Oppeneer},\ and\ \citenamefont
  {Maldonado}}]{Ritzmann2020}%
  \BibitemOpen
  \bibfield  {author} {\bibinfo {author} {\bibfnamefont {U.}~\bibnamefont
  {Ritzmann}}, \bibinfo {author} {\bibfnamefont {P.~M.}\ \bibnamefont
  {Oppeneer}},\ and\ \bibinfo {author} {\bibfnamefont {P.}~\bibnamefont
  {Maldonado}},\ }\bibfield  {title} {\bibinfo {title} {Theory of
  out-of-equilibrium electron and phonon dynamics in metals after femtosecond
  laser excitation},\ }\href {https://doi.org/10.1103/PhysRevB.102.214305}
  {\bibfield  {journal} {\bibinfo  {journal} {Phys. Rev. B}\ }\textbf {\bibinfo
  {volume} {102}},\ \bibinfo {pages} {214305} (\bibinfo {year}
  {2020})}\BibitemShut {NoStop}%
\bibitem [{\citenamefont {Beens}\ \emph {et~al.}(2020)\citenamefont {Beens},
  \citenamefont {Duine},\ and\ \citenamefont {Koopmans}}]{Beens2020}%
  \BibitemOpen
  \bibfield  {author} {\bibinfo {author} {\bibfnamefont {M.}~\bibnamefont
  {Beens}}, \bibinfo {author} {\bibfnamefont {R.~A.}\ \bibnamefont {Duine}},\
  and\ \bibinfo {author} {\bibfnamefont {B.}~\bibnamefont {Koopmans}},\
  }\bibfield  {title} {\bibinfo {title} {$s\text{\ensuremath{-}}d$ model for
  local and nonlocal spin dynamics in laser-excited magnetic
  heterostructures},\ }\href {https://doi.org/10.1103/PhysRevB.102.054442}
  {\bibfield  {journal} {\bibinfo  {journal} {Phys. Rev. B}\ }\textbf {\bibinfo
  {volume} {102}},\ \bibinfo {pages} {054442} (\bibinfo {year}
  {2020})}\BibitemShut {NoStop}%
\bibitem [{\citenamefont {Beens}\ \emph {et~al.}(2022)\citenamefont {Beens},
  \citenamefont {Duine},\ and\ \citenamefont {Koopmans}}]{Beens2022}%
  \BibitemOpen
  \bibfield  {author} {\bibinfo {author} {\bibfnamefont {M.}~\bibnamefont
  {Beens}}, \bibinfo {author} {\bibfnamefont {R.~A.}\ \bibnamefont {Duine}},\
  and\ \bibinfo {author} {\bibfnamefont {B.}~\bibnamefont {Koopmans}},\
  }\bibfield  {title} {\bibinfo {title} {Modeling ultrafast demagnetization and
  spin transport: The interplay of spin-polarized electrons and thermal
  magnons},\ }\href {https://doi.org/10.1103/PhysRevB.105.144420} {\bibfield
  {journal} {\bibinfo  {journal} {Phys. Rev. B}\ }\textbf {\bibinfo {volume}
  {105}},\ \bibinfo {pages} {144420} (\bibinfo {year} {2022})}\BibitemShut
  {NoStop}%
\bibitem [{\citenamefont {Shokeen}\ \emph {et~al.}(2024)\citenamefont
  {Shokeen}, \citenamefont {Heber}, \citenamefont {Kutnyakhov}, \citenamefont
  {Wang}, \citenamefont {Yaroslavtsev}, \citenamefont {Maldonado},
  \citenamefont {Berritta}, \citenamefont {Wind}, \citenamefont {Wenthaus},
  \citenamefont {Pressacco}, \citenamefont {Min}, \citenamefont {Nissen},
  \citenamefont {Mahatha}, \citenamefont {Dziarzhytski}, \citenamefont
  {Oppeneer}, \citenamefont {Rossnagel}, \citenamefont {Elmers}, \citenamefont
  {Schönhense},\ and\ \citenamefont {Dürr}}]{Shokeen2024}%
  \BibitemOpen
  \bibfield  {author} {\bibinfo {author} {\bibfnamefont {V.}~\bibnamefont
  {Shokeen}}, \bibinfo {author} {\bibfnamefont {M.}~\bibnamefont {Heber}},
  \bibinfo {author} {\bibfnamefont {D.}~\bibnamefont {Kutnyakhov}}, \bibinfo
  {author} {\bibfnamefont {X.}~\bibnamefont {Wang}}, \bibinfo {author}
  {\bibfnamefont {A.}~\bibnamefont {Yaroslavtsev}}, \bibinfo {author}
  {\bibfnamefont {P.}~\bibnamefont {Maldonado}}, \bibinfo {author}
  {\bibfnamefont {M.}~\bibnamefont {Berritta}}, \bibinfo {author}
  {\bibfnamefont {N.}~\bibnamefont {Wind}}, \bibinfo {author} {\bibfnamefont
  {L.}~\bibnamefont {Wenthaus}}, \bibinfo {author} {\bibfnamefont
  {F.}~\bibnamefont {Pressacco}}, \bibinfo {author} {\bibfnamefont {C.-H.}\
  \bibnamefont {Min}}, \bibinfo {author} {\bibfnamefont {M.}~\bibnamefont
  {Nissen}}, \bibinfo {author} {\bibfnamefont {S.~K.}\ \bibnamefont {Mahatha}},
  \bibinfo {author} {\bibfnamefont {S.}~\bibnamefont {Dziarzhytski}}, \bibinfo
  {author} {\bibfnamefont {P.~M.}\ \bibnamefont {Oppeneer}}, \bibinfo {author}
  {\bibfnamefont {K.}~\bibnamefont {Rossnagel}}, \bibinfo {author}
  {\bibfnamefont {H.-J.}\ \bibnamefont {Elmers}}, \bibinfo {author}
  {\bibfnamefont {G.}~\bibnamefont {Schönhense}},\ and\ \bibinfo {author}
  {\bibfnamefont {H.~A.}\ \bibnamefont {Dürr}},\ }\bibfield  {title} {\bibinfo
  {title} {Real-time observation of non-equilibrium phonon-electron energy and
  angular momentum flow in laser-heated nickel},\ }\href
  {https://doi.org/10.1126/sciadv.adj2407} {\bibfield  {journal} {\bibinfo
  {journal} {Sci. Adv.}\ }\textbf {\bibinfo {volume} {10}},\ \bibinfo {pages}
  {eadj2407} (\bibinfo {year} {2024})}\BibitemShut {NoStop}%
\bibitem [{\citenamefont {Wei{\ss}enhofer}\ and\ \citenamefont
  {Oppeneer}(2024)}]{Weissenhofer2024}%
  \BibitemOpen
  \bibfield  {author} {\bibinfo {author} {\bibfnamefont {M.}~\bibnamefont
  {Wei{\ss}enhofer}}\ and\ \bibinfo {author} {\bibfnamefont {P.~M.}\
  \bibnamefont {Oppeneer}},\ }\bibfield  {title} {\bibinfo {title} {Ultrafast
  demagnetization through femtosecond generation of non-thermal magnons},\
  }\href {https://doi.org/https://doi.org/10.1002/apxr.202300103} {\bibfield
  {journal} {\bibinfo  {journal} {Adv. Phys. Res.}\ }\textbf {\bibinfo {volume}
  {4}},\ \bibinfo {pages} {2300103} (\bibinfo {year} {2024})}\BibitemShut
  {NoStop}%
\bibitem [{\citenamefont {Baláž}\ \emph {et~al.}(2018)\citenamefont
  {Baláž}, \citenamefont {Žonda}, \citenamefont {Carva}, \citenamefont
  {Maldonado},\ and\ \citenamefont {Oppeneer}}]{Balaz2018}%
  \BibitemOpen
  \bibfield  {author} {\bibinfo {author} {\bibfnamefont {P.}~\bibnamefont
  {Baláž}}, \bibinfo {author} {\bibfnamefont {M.}~\bibnamefont {Žonda}},
  \bibinfo {author} {\bibfnamefont {K.}~\bibnamefont {Carva}}, \bibinfo
  {author} {\bibfnamefont {P.}~\bibnamefont {Maldonado}},\ and\ \bibinfo
  {author} {\bibfnamefont {P.~M.}\ \bibnamefont {Oppeneer}},\ }\bibfield
  {title} {\bibinfo {title} {Transport theory for femtosecond laser-induced
  spin-transfer torques},\ }\href {https://doi.org/10.1088/1361-648X/aaad95}
  {\bibfield  {journal} {\bibinfo  {journal} {J. Phys.: Condens. Matter}\
  }\textbf {\bibinfo {volume} {30}},\ \bibinfo {pages} {115801} (\bibinfo
  {year} {2018})}\BibitemShut {NoStop}%
\bibitem [{\citenamefont {Bal\'a\ifmmode~\check{z}\else \v{z}\fi{}}\ \emph
  {et~al.}(2023)\citenamefont {Bal\'a\ifmmode~\check{z}\else \v{z}\fi{}},
  \citenamefont {Zwierzycki}, \citenamefont {Cosco}, \citenamefont {Carva},
  \citenamefont {Maldonado},\ and\ \citenamefont {Oppeneer}}]{Balaz2023}%
  \BibitemOpen
  \bibfield  {author} {\bibinfo {author} {\bibfnamefont {P.}~\bibnamefont
  {Bal\'a\ifmmode~\check{z}\else \v{z}\fi{}}}, \bibinfo {author} {\bibfnamefont
  {M.}~\bibnamefont {Zwierzycki}}, \bibinfo {author} {\bibfnamefont
  {F.}~\bibnamefont {Cosco}}, \bibinfo {author} {\bibfnamefont
  {K.}~\bibnamefont {Carva}}, \bibinfo {author} {\bibfnamefont
  {P.}~\bibnamefont {Maldonado}},\ and\ \bibinfo {author} {\bibfnamefont
  {P.~M.}\ \bibnamefont {Oppeneer}},\ }\bibfield  {title} {\bibinfo {title}
  {Theory of superdiffusive spin transport in noncollinear magnetic
  multilayers},\ }\href {https://doi.org/10.1103/PhysRevB.107.174418}
  {\bibfield  {journal} {\bibinfo  {journal} {Phys. Rev. B}\ }\textbf {\bibinfo
  {volume} {107}},\ \bibinfo {pages} {174418} (\bibinfo {year}
  {2023})}\BibitemShut {NoStop}%
\bibitem [{\citenamefont {Xiao}\ \emph {et~al.}(2010)\citenamefont {Xiao},
  \citenamefont {Bauer}, \citenamefont {Uchida}, \citenamefont {Saitoh},\ and\
  \citenamefont {Maekawa}}]{Xiao2010}%
  \BibitemOpen
  \bibfield  {author} {\bibinfo {author} {\bibfnamefont {J.}~\bibnamefont
  {Xiao}}, \bibinfo {author} {\bibfnamefont {G.~E.~W.}\ \bibnamefont {Bauer}},
  \bibinfo {author} {\bibfnamefont {K.-c.}\ \bibnamefont {Uchida}}, \bibinfo
  {author} {\bibfnamefont {E.}~\bibnamefont {Saitoh}},\ and\ \bibinfo {author}
  {\bibfnamefont {S.}~\bibnamefont {Maekawa}},\ }\bibfield  {title} {\bibinfo
  {title} {{Theory of magnon-driven spin Seebeck effect}},\ }\href
  {https://doi.org/10.1103/PhysRevB.81.214418} {\bibfield  {journal} {\bibinfo
  {journal} {Phys. Rev. B}\ }\textbf {\bibinfo {volume} {81}},\ \bibinfo
  {pages} {214418} (\bibinfo {year} {2010})}\BibitemShut {NoStop}%
\bibitem [{\citenamefont {Cornelissen}\ \emph {et~al.}(2016)\citenamefont
  {Cornelissen}, \citenamefont {Peters}, \citenamefont {Bauer}, \citenamefont
  {Duine},\ and\ \citenamefont {van Wees}}]{Cornelissen2016}%
  \BibitemOpen
  \bibfield  {author} {\bibinfo {author} {\bibfnamefont {L.~J.}\ \bibnamefont
  {Cornelissen}}, \bibinfo {author} {\bibfnamefont {K.~J.~H.}\ \bibnamefont
  {Peters}}, \bibinfo {author} {\bibfnamefont {G.~E.~W.}\ \bibnamefont
  {Bauer}}, \bibinfo {author} {\bibfnamefont {R.~A.}\ \bibnamefont {Duine}},\
  and\ \bibinfo {author} {\bibfnamefont {B.~J.}\ \bibnamefont {van Wees}},\
  }\bibfield  {title} {\bibinfo {title} {Magnon spin transport driven by the
  magnon chemical potential in a magnetic insulator},\ }\href
  {https://doi.org/10.1103/PhysRevB.94.014412} {\bibfield  {journal} {\bibinfo
  {journal} {Phys. Rev. B}\ }\textbf {\bibinfo {volume} {94}},\ \bibinfo
  {pages} {014412} (\bibinfo {year} {2016})}\BibitemShut {NoStop}%
\bibitem [{\citenamefont {Zahn}\ \emph {et~al.}(2021)\citenamefont {Zahn},
  \citenamefont {Jakobs}, \citenamefont {Windsor}, \citenamefont {Seiler},
  \citenamefont {Vasileiadis}, \citenamefont {Butcher}, \citenamefont {Qi},
  \citenamefont {Engel}, \citenamefont {Atxitia}, \citenamefont {Vorberger},\
  and\ \citenamefont {Ernstorfer}}]{Zahn2021}%
  \BibitemOpen
  \bibfield  {author} {\bibinfo {author} {\bibfnamefont {D.}~\bibnamefont
  {Zahn}}, \bibinfo {author} {\bibfnamefont {F.}~\bibnamefont {Jakobs}},
  \bibinfo {author} {\bibfnamefont {Y.~W.}\ \bibnamefont {Windsor}}, \bibinfo
  {author} {\bibfnamefont {H.}~\bibnamefont {Seiler}}, \bibinfo {author}
  {\bibfnamefont {T.}~\bibnamefont {Vasileiadis}}, \bibinfo {author}
  {\bibfnamefont {T.~A.}\ \bibnamefont {Butcher}}, \bibinfo {author}
  {\bibfnamefont {Y.}~\bibnamefont {Qi}}, \bibinfo {author} {\bibfnamefont
  {D.}~\bibnamefont {Engel}}, \bibinfo {author} {\bibfnamefont
  {U.}~\bibnamefont {Atxitia}}, \bibinfo {author} {\bibfnamefont
  {J.}~\bibnamefont {Vorberger}},\ and\ \bibinfo {author} {\bibfnamefont
  {R.}~\bibnamefont {Ernstorfer}},\ }\bibfield  {title} {\bibinfo {title}
  {Lattice dynamics and ultrafast energy flow between electrons, spins, and
  phonons in a 3d ferromagnet},\ }\href
  {https://doi.org/10.1103/PhysRevResearch.3.023032} {\bibfield  {journal}
  {\bibinfo  {journal} {Phys. Rev. Res.}\ }\textbf {\bibinfo {volume} {3}},\
  \bibinfo {pages} {023032} (\bibinfo {year} {2021})}\BibitemShut {NoStop}%
\bibitem [{\citenamefont {Zahn}\ \emph {et~al.}(2022)\citenamefont {Zahn},
  \citenamefont {Jakobs}, \citenamefont {Seiler}, \citenamefont {Butcher},
  \citenamefont {Engel}, \citenamefont {Vorberger}, \citenamefont {Atxitia},
  \citenamefont {Windsor},\ and\ \citenamefont {Ernstorfer}}]{Zahn2022}%
  \BibitemOpen
  \bibfield  {author} {\bibinfo {author} {\bibfnamefont {D.}~\bibnamefont
  {Zahn}}, \bibinfo {author} {\bibfnamefont {F.}~\bibnamefont {Jakobs}},
  \bibinfo {author} {\bibfnamefont {H.}~\bibnamefont {Seiler}}, \bibinfo
  {author} {\bibfnamefont {T.~A.}\ \bibnamefont {Butcher}}, \bibinfo {author}
  {\bibfnamefont {D.}~\bibnamefont {Engel}}, \bibinfo {author} {\bibfnamefont
  {J.}~\bibnamefont {Vorberger}}, \bibinfo {author} {\bibfnamefont
  {U.}~\bibnamefont {Atxitia}}, \bibinfo {author} {\bibfnamefont {Y.~W.}\
  \bibnamefont {Windsor}},\ and\ \bibinfo {author} {\bibfnamefont
  {R.}~\bibnamefont {Ernstorfer}},\ }\bibfield  {title} {\bibinfo {title}
  {Intrinsic energy flow in laser-excited $3d$ ferromagnets},\ }\href
  {https://doi.org/10.1103/PhysRevResearch.4.013104} {\bibfield  {journal}
  {\bibinfo  {journal} {Phys. Rev. Res.}\ }\textbf {\bibinfo {volume} {4}},\
  \bibinfo {pages} {013104} (\bibinfo {year} {2022})}\BibitemShut {NoStop}%
\bibitem [{\citenamefont {Zener}(1951{\natexlab{a}})}]{Zener1951}%
  \BibitemOpen
  \bibfield  {author} {\bibinfo {author} {\bibfnamefont {C.}~\bibnamefont
  {Zener}},\ }\bibfield  {title} {\bibinfo {title} {Interaction between the $d$
  shells in the transition metals},\ }\href
  {https://doi.org/10.1103/PhysRev.81.440} {\bibfield  {journal} {\bibinfo
  {journal} {Phys. Rev.}\ }\textbf {\bibinfo {volume} {81}},\ \bibinfo {pages}
  {440} (\bibinfo {year} {1951}{\natexlab{a}})}\BibitemShut {NoStop}%
\bibitem [{\citenamefont {Zener}(1951{\natexlab{b}})}]{Zener1951b}%
  \BibitemOpen
  \bibfield  {author} {\bibinfo {author} {\bibfnamefont {C.}~\bibnamefont
  {Zener}},\ }\bibfield  {title} {\bibinfo {title} {{Interaction between the
  $d$-Shells in the Transition Metals. III. Calculation of the Weiss Factors in
  Fe, Co, and Ni}},\ }\href {https://doi.org/10.1103/PhysRev.83.299} {\bibfield
   {journal} {\bibinfo  {journal} {Phys. Rev.}\ }\textbf {\bibinfo {volume}
  {83}},\ \bibinfo {pages} {299} (\bibinfo {year}
  {1951}{\natexlab{b}})}\BibitemShut {NoStop}%
\bibitem [{\citenamefont {Mitchell}(1957)}]{Mitchell1957}%
  \BibitemOpen
  \bibfield  {author} {\bibinfo {author} {\bibfnamefont {A.~H.}\ \bibnamefont
  {Mitchell}},\ }\bibfield  {title} {\bibinfo {title} {Ferromagnetic relaxation
  by the exchange interaction between ferromagnetic electrons and conduction
  electrons},\ }\href {https://doi.org/10.1103/PhysRev.105.1439} {\bibfield
  {journal} {\bibinfo  {journal} {Phys. Rev.}\ }\textbf {\bibinfo {volume}
  {105}},\ \bibinfo {pages} {1439} (\bibinfo {year} {1957})}\BibitemShut
  {NoStop}%
\bibitem [{\citenamefont {Heinrich}\ \emph {et~al.}(1967)\citenamefont
  {Heinrich}, \citenamefont {Fraitová},\ and\ \citenamefont
  {Kamberský}}]{Heinrich1967}%
  \BibitemOpen
  \bibfield  {author} {\bibinfo {author} {\bibfnamefont {B.}~\bibnamefont
  {Heinrich}}, \bibinfo {author} {\bibfnamefont {D.}~\bibnamefont
  {Fraitová}},\ and\ \bibinfo {author} {\bibfnamefont {V.}~\bibnamefont
  {Kamberský}},\ }\bibfield  {title} {\bibinfo {title} {The influence of s-d
  exchange on relaxation of magnons in metals},\ }\href
  {https://doi.org/https://doi.org/10.1002/pssb.19670230209} {\bibfield
  {journal} {\bibinfo  {journal} {phys. status solidi (b)}\ }\textbf {\bibinfo
  {volume} {23}},\ \bibinfo {pages} {501} (\bibinfo {year} {1967})}\BibitemShut
  {NoStop}%
\bibitem [{\citenamefont {Tserkovnyak}\ \emph {et~al.}(2004)\citenamefont
  {Tserkovnyak}, \citenamefont {Fiete},\ and\ \citenamefont
  {Halperin}}]{Tserkovnyak2004}%
  \BibitemOpen
  \bibfield  {author} {\bibinfo {author} {\bibfnamefont {Y.}~\bibnamefont
  {Tserkovnyak}}, \bibinfo {author} {\bibfnamefont {G.~A.}\ \bibnamefont
  {Fiete}},\ and\ \bibinfo {author} {\bibfnamefont {B.~I.}\ \bibnamefont
  {Halperin}},\ }\bibfield  {title} {\bibinfo {title} {{Mean-field
  magnetization relaxation in conducting ferromagnets}},\ }\href
  {https://doi.org/10.1063/1.1762979} {\bibfield  {journal} {\bibinfo
  {journal} {Appl. Phys. Lett.}\ }\textbf {\bibinfo {volume} {84}},\ \bibinfo
  {pages} {5234} (\bibinfo {year} {2004})}\BibitemShut {NoStop}%
\bibitem [{\citenamefont {Zhang}\ and\ \citenamefont {Li}(2004)}]{Zhang2004}%
  \BibitemOpen
  \bibfield  {author} {\bibinfo {author} {\bibfnamefont {S.}~\bibnamefont
  {Zhang}}\ and\ \bibinfo {author} {\bibfnamefont {Z.}~\bibnamefont {Li}},\
  }\bibfield  {title} {\bibinfo {title} {Roles of nonequilibrium conduction
  electrons on the magnetization dynamics of ferromagnets},\ }\href
  {https://doi.org/10.1103/PhysRevLett.93.127204} {\bibfield  {journal}
  {\bibinfo  {journal} {Phys. Rev. Lett.}\ }\textbf {\bibinfo {volume} {93}},\
  \bibinfo {pages} {127204} (\bibinfo {year} {2004})}\BibitemShut {NoStop}%
\bibitem [{\citenamefont {Manchon}\ \emph {et~al.}(2012)\citenamefont
  {Manchon}, \citenamefont {Li}, \citenamefont {Xu},\ and\ \citenamefont
  {Zhang}}]{Manchon2012}%
  \BibitemOpen
  \bibfield  {author} {\bibinfo {author} {\bibfnamefont {A.}~\bibnamefont
  {Manchon}}, \bibinfo {author} {\bibfnamefont {Q.}~\bibnamefont {Li}},
  \bibinfo {author} {\bibfnamefont {L.}~\bibnamefont {Xu}},\ and\ \bibinfo
  {author} {\bibfnamefont {S.}~\bibnamefont {Zhang}},\ }\bibfield  {title}
  {\bibinfo {title} {Theory of laser-induced demagnetization at high
  temperatures},\ }\href {https://doi.org/10.1103/PhysRevB.85.064408}
  {\bibfield  {journal} {\bibinfo  {journal} {Phys. Rev. B}\ }\textbf {\bibinfo
  {volume} {85}},\ \bibinfo {pages} {064408} (\bibinfo {year}
  {2012})}\BibitemShut {NoStop}%
\bibitem [{\citenamefont {Brener}\ \emph {et~al.}(2017)\citenamefont {Brener},
  \citenamefont {Murzaliev}, \citenamefont {Titov},\ and\ \citenamefont
  {Katsnelson}}]{Brener2017}%
  \BibitemOpen
  \bibfield  {author} {\bibinfo {author} {\bibfnamefont {S.}~\bibnamefont
  {Brener}}, \bibinfo {author} {\bibfnamefont {B.}~\bibnamefont {Murzaliev}},
  \bibinfo {author} {\bibfnamefont {M.}~\bibnamefont {Titov}},\ and\ \bibinfo
  {author} {\bibfnamefont {M.~I.}\ \bibnamefont {Katsnelson}},\ }\bibfield
  {title} {\bibinfo {title} {Magnon activation by hot electrons via
  nonquasiparticle states},\ }\href
  {https://doi.org/10.1103/PhysRevB.95.220409} {\bibfield  {journal} {\bibinfo
  {journal} {Phys. Rev. B}\ }\textbf {\bibinfo {volume} {95}},\ \bibinfo
  {pages} {220409} (\bibinfo {year} {2017})}\BibitemShut {NoStop}%
\bibitem [{\citenamefont {Barbeau}\ \emph {et~al.}(2022)\citenamefont
  {Barbeau}, \citenamefont {Titov}, \citenamefont {Katsnelson},\ and\
  \citenamefont {Qaiumzadeh}}]{Barbeau2022}%
  \BibitemOpen
  \bibfield  {author} {\bibinfo {author} {\bibfnamefont {M.~M.~S.}\
  \bibnamefont {Barbeau}}, \bibinfo {author} {\bibfnamefont {M.}~\bibnamefont
  {Titov}}, \bibinfo {author} {\bibfnamefont {M.~I.}\ \bibnamefont
  {Katsnelson}},\ and\ \bibinfo {author} {\bibfnamefont {A.}~\bibnamefont
  {Qaiumzadeh}},\ }\href@noop {} {\bibinfo {title} {Nonequilibrium magnons from
  hot electrons in antiferromagnetic systems}} (\bibinfo {year} {2022}),\
  \Eprint {https://arxiv.org/abs/2209.03469} {arXiv:2209.03469
  [cond-mat.mes-hall]} \BibitemShut {NoStop}%
\bibitem [{\citenamefont {Remy}(2023)}]{Remy2023}%
  \BibitemOpen
  \bibfield  {author} {\bibinfo {author} {\bibfnamefont {Q.}~\bibnamefont
  {Remy}},\ }\bibfield  {title} {\bibinfo {title} {Ultrafast magnetization
  reversal in ferromagnetic spin valves: An $s\text{\ensuremath{-}}d$ model
  perspective},\ }\href {https://doi.org/10.1103/PhysRevB.107.174431}
  {\bibfield  {journal} {\bibinfo  {journal} {Phys. Rev. B}\ }\textbf {\bibinfo
  {volume} {107}},\ \bibinfo {pages} {174431} (\bibinfo {year}
  {2023})}\BibitemShut {NoStop}%
\bibitem [{\citenamefont {Allen}(1987)}]{Allen1987}%
  \BibitemOpen
  \bibfield  {author} {\bibinfo {author} {\bibfnamefont {P.~B.}\ \bibnamefont
  {Allen}},\ }\bibfield  {title} {\bibinfo {title} {Theory of thermal
  relaxation of electrons in metals},\ }\href
  {https://doi.org/10.1103/PhysRevLett.59.1460} {\bibfield  {journal} {\bibinfo
   {journal} {Phys. Rev. Lett.}\ }\textbf {\bibinfo {volume} {59}},\ \bibinfo
  {pages} {1460} (\bibinfo {year} {1987})}\BibitemShut {NoStop}%
\bibitem [{\citenamefont {Aeschlimann}\ \emph {et~al.}(1997)\citenamefont
  {Aeschlimann}, \citenamefont {Bauer}, \citenamefont {Pawlik}, \citenamefont
  {Weber}, \citenamefont {Burgermeister}, \citenamefont {Oberli},\ and\
  \citenamefont {Siegmann}}]{Aeschlimann1997}%
  \BibitemOpen
  \bibfield  {author} {\bibinfo {author} {\bibfnamefont {M.}~\bibnamefont
  {Aeschlimann}}, \bibinfo {author} {\bibfnamefont {M.}~\bibnamefont {Bauer}},
  \bibinfo {author} {\bibfnamefont {S.}~\bibnamefont {Pawlik}}, \bibinfo
  {author} {\bibfnamefont {W.}~\bibnamefont {Weber}}, \bibinfo {author}
  {\bibfnamefont {R.}~\bibnamefont {Burgermeister}}, \bibinfo {author}
  {\bibfnamefont {D.}~\bibnamefont {Oberli}},\ and\ \bibinfo {author}
  {\bibfnamefont {H.~C.}\ \bibnamefont {Siegmann}},\ }\bibfield  {title}
  {\bibinfo {title} {Ultrafast spin-dependent electron dynamics in fcc {C}o},\
  }\href {https://doi.org/10.1103/PhysRevLett.79.5158} {\bibfield  {journal}
  {\bibinfo  {journal} {Phys. Rev. Lett.}\ }\textbf {\bibinfo {volume} {79}},\
  \bibinfo {pages} {5158} (\bibinfo {year} {1997})}\BibitemShut {NoStop}%
\bibitem [{\citenamefont {Nowak}(2007)}]{Nowak2007}%
  \BibitemOpen
  \bibfield  {author} {\bibinfo {author} {\bibfnamefont {U.}~\bibnamefont
  {Nowak}},\ }\bibinfo {title} {Classical spin models},\ in\ \href
  {https://doi.org/10.1002/9780470022184} {\emph {\bibinfo {booktitle}
  {Handbook of Magnetism and Advanced Magnetic Materials}}},\ \bibinfo {editor}
  {edited by\ \bibinfo {editor} {\bibfnamefont {H.}~\bibnamefont
  {Kronm\"uller}}\ and\ \bibinfo {editor} {\bibfnamefont {S.}~\bibnamefont
  {Parkin}}}\ (\bibinfo  {publisher} {John Wiley \& Sons},\ \bibinfo {year}
  {2007})\ Chap.\ \bibinfo {chapter} {Micromagnetism}, pp.\ \bibinfo {pages}
  {858--876}\BibitemShut {NoStop}%
\bibitem [{\citenamefont {Kazantseva}\ \emph {et~al.}(2007)\citenamefont
  {Kazantseva}, \citenamefont {Nowak}, \citenamefont {Chantrell}, \citenamefont
  {Hohlfeld},\ and\ \citenamefont {Rebei}}]{Kazantseva_2008}%
  \BibitemOpen
  \bibfield  {author} {\bibinfo {author} {\bibfnamefont {N.}~\bibnamefont
  {Kazantseva}}, \bibinfo {author} {\bibfnamefont {U.}~\bibnamefont {Nowak}},
  \bibinfo {author} {\bibfnamefont {R.~W.}\ \bibnamefont {Chantrell}}, \bibinfo
  {author} {\bibfnamefont {J.}~\bibnamefont {Hohlfeld}},\ and\ \bibinfo
  {author} {\bibfnamefont {A.}~\bibnamefont {Rebei}},\ }\bibfield  {title}
  {\bibinfo {title} {Slow recovery of the magnetisation after a sub-picosecond
  heat pulse},\ }\href {https://doi.org/10.1209/0295-5075/81/27004} {\bibfield
  {journal} {\bibinfo  {journal} {Europhys. Lett.}\ }\textbf {\bibinfo {volume}
  {81}},\ \bibinfo {pages} {27004} (\bibinfo {year} {2007})}\BibitemShut
  {NoStop}%
\bibitem [{\citenamefont {Ostler}\ \emph {et~al.}(2012)\citenamefont {Ostler},
  \citenamefont {Barker}, \citenamefont {Evans}, \citenamefont {Chantrell},
  \citenamefont {Atxitia}, \citenamefont {Chubykalo-Fesenko}, \citenamefont
  {El~Moussaoui}, \citenamefont {Le~Guyader}, \citenamefont {Mengotti},
  \citenamefont {Heyderman}, \citenamefont {Nolting}, \citenamefont
  {Tsukamoto}, \citenamefont {Itoh}, \citenamefont {Afanasiev}, \citenamefont
  {Ivanov}, \citenamefont {Kalashnikova}, \citenamefont {Vahaplar},
  \citenamefont {Mentink}, \citenamefont {Kirilyuk}, \citenamefont {Rasing},\
  and\ \citenamefont {Kimel}}]{Ostler2012}%
  \BibitemOpen
  \bibfield  {author} {\bibinfo {author} {\bibfnamefont {T.~A.}\ \bibnamefont
  {Ostler}}, \bibinfo {author} {\bibfnamefont {J.}~\bibnamefont {Barker}},
  \bibinfo {author} {\bibfnamefont {R.~F.~L.}\ \bibnamefont {Evans}}, \bibinfo
  {author} {\bibfnamefont {R.~W.}\ \bibnamefont {Chantrell}}, \bibinfo {author}
  {\bibfnamefont {U.}~\bibnamefont {Atxitia}}, \bibinfo {author} {\bibfnamefont
  {O.}~\bibnamefont {Chubykalo-Fesenko}}, \bibinfo {author} {\bibfnamefont
  {S.}~\bibnamefont {El~Moussaoui}}, \bibinfo {author} {\bibfnamefont
  {L.}~\bibnamefont {Le~Guyader}}, \bibinfo {author} {\bibfnamefont
  {E.}~\bibnamefont {Mengotti}}, \bibinfo {author} {\bibfnamefont {L.~J.}\
  \bibnamefont {Heyderman}}, \bibinfo {author} {\bibfnamefont {F.}~\bibnamefont
  {Nolting}}, \bibinfo {author} {\bibfnamefont {A.}~\bibnamefont {Tsukamoto}},
  \bibinfo {author} {\bibfnamefont {A.}~\bibnamefont {Itoh}}, \bibinfo {author}
  {\bibfnamefont {D.}~\bibnamefont {Afanasiev}}, \bibinfo {author}
  {\bibfnamefont {B.~A.}\ \bibnamefont {Ivanov}}, \bibinfo {author}
  {\bibfnamefont {A.~M.}\ \bibnamefont {Kalashnikova}}, \bibinfo {author}
  {\bibfnamefont {K.}~\bibnamefont {Vahaplar}}, \bibinfo {author}
  {\bibfnamefont {J.}~\bibnamefont {Mentink}}, \bibinfo {author} {\bibfnamefont
  {A.}~\bibnamefont {Kirilyuk}}, \bibinfo {author} {\bibfnamefont
  {T.}~\bibnamefont {Rasing}},\ and\ \bibinfo {author} {\bibfnamefont {A.~V.}\
  \bibnamefont {Kimel}},\ }\bibfield  {title} {\bibinfo {title} {Ultrafast
  heating as a sufficient stimulus for magnetization reversal in a
  ferrimagnet},\ }\href {https://doi.org/10.1038/ncomms1666} {\bibfield
  {journal} {\bibinfo  {journal} {Nature Commun.}\ }\textbf {\bibinfo {volume}
  {3}},\ \bibinfo {pages} {666} (\bibinfo {year} {2012})}\BibitemShut {NoStop}%
\bibitem [{\citenamefont {Schellekens}\ \emph {et~al.}(2013)\citenamefont
  {Schellekens}, \citenamefont {Verhoeven}, \citenamefont {Vader},\ and\
  \citenamefont {Koopmans}}]{Schellekens2013}%
  \BibitemOpen
  \bibfield  {author} {\bibinfo {author} {\bibfnamefont {A.~J.}\ \bibnamefont
  {Schellekens}}, \bibinfo {author} {\bibfnamefont {W.}~\bibnamefont
  {Verhoeven}}, \bibinfo {author} {\bibfnamefont {T.~N.}\ \bibnamefont
  {Vader}},\ and\ \bibinfo {author} {\bibfnamefont {B.}~\bibnamefont
  {Koopmans}},\ }\bibfield  {title} {\bibinfo {title} {{Investigating the
  contribution of superdiffusive transport to ultrafast demagnetization of
  ferromagnetic thin films}},\ }\href {https://doi.org/10.1063/1.4812658}
  {\bibfield  {journal} {\bibinfo  {journal} {Appl. Phys. Lett.}\ }\textbf
  {\bibinfo {volume} {102}},\ \bibinfo {pages} {252408} (\bibinfo {year}
  {2013})}\BibitemShut {NoStop}%
\bibitem [{\citenamefont {Lee}\ \emph {et~al.}(2021)\citenamefont {Lee},
  \citenamefont {Lee}, \citenamefont {Yang}, \citenamefont {Mishra},
  \citenamefont {Kim}, \citenamefont {Liu}, \citenamefont {Xiong},
  \citenamefont {Kim}, \citenamefont {Lee},\ and\ \citenamefont
  {Yang}}]{Lee2021}%
  \BibitemOpen
  \bibfield  {author} {\bibinfo {author} {\bibfnamefont {K.}~\bibnamefont
  {Lee}}, \bibinfo {author} {\bibfnamefont {D.-K.}\ \bibnamefont {Lee}},
  \bibinfo {author} {\bibfnamefont {D.}~\bibnamefont {Yang}}, \bibinfo {author}
  {\bibfnamefont {R.}~\bibnamefont {Mishra}}, \bibinfo {author} {\bibfnamefont
  {D.-J.}\ \bibnamefont {Kim}}, \bibinfo {author} {\bibfnamefont
  {S.}~\bibnamefont {Liu}}, \bibinfo {author} {\bibfnamefont {Q.}~\bibnamefont
  {Xiong}}, \bibinfo {author} {\bibfnamefont {S.~K.}\ \bibnamefont {Kim}},
  \bibinfo {author} {\bibfnamefont {K.-J.}\ \bibnamefont {Lee}},\ and\ \bibinfo
  {author} {\bibfnamefont {H.}~\bibnamefont {Yang}},\ }\bibfield  {title}
  {\bibinfo {title} {Superluminal-like magnon propagation in antiferromagnetic
  {NiO} at nanoscale distances},\ }\href
  {https://doi.org/10.1038/s41565-021-00983-4} {\bibfield  {journal} {\bibinfo
  {journal} {Nature Nanotechn.}\ }\textbf {\bibinfo {volume} {16}},\ \bibinfo
  {pages} {1337 } (\bibinfo {year} {2021})}\BibitemShut {NoStop}%
\bibitem [{\citenamefont {Qiu}\ \emph {et~al.}(2025)\citenamefont {Qiu},
  \citenamefont {Franke}, \citenamefont {Tian}, \citenamefont
  {Ka\ifmmode\check{s}\else\v{s}\fi{}par}, \citenamefont {Rouzegar},
  \citenamefont {Gueckstock}, \citenamefont {Wu}, \citenamefont {Zhu},
  \citenamefont {Jin}, \citenamefont {Xu}, \citenamefont {Seifert},
  \citenamefont {Wu}, \citenamefont {Brouwer},\ and\ \citenamefont
  {Kampfrath}}]{Qiu2025}%
  \BibitemOpen
  \bibfield  {author} {\bibinfo {author} {\bibfnamefont {H.}~\bibnamefont
  {Qiu}}, \bibinfo {author} {\bibfnamefont {O.}~\bibnamefont {Franke}},
  \bibinfo {author} {\bibfnamefont {Y.}~\bibnamefont {Tian}}, \bibinfo {author}
  {\bibfnamefont {Z.}~\bibnamefont {Ka\ifmmode\check{s}\else\v{s}\fi{}par}},
  \bibinfo {author} {\bibfnamefont {R.}~\bibnamefont {Rouzegar}}, \bibinfo
  {author} {\bibfnamefont {O.}~\bibnamefont {Gueckstock}}, \bibinfo {author}
  {\bibfnamefont {J.}~\bibnamefont {Wu}}, \bibinfo {author} {\bibfnamefont
  {M.}~\bibnamefont {Zhu}}, \bibinfo {author} {\bibfnamefont {B.}~\bibnamefont
  {Jin}}, \bibinfo {author} {\bibfnamefont {Y.}~\bibnamefont {Xu}}, \bibinfo
  {author} {\bibfnamefont {T.~S.}\ \bibnamefont {Seifert}}, \bibinfo {author}
  {\bibfnamefont {D.}~\bibnamefont {Wu}}, \bibinfo {author} {\bibfnamefont
  {P.~W.}\ \bibnamefont {Brouwer}},\ and\ \bibinfo {author} {\bibfnamefont
  {T.}~\bibnamefont {Kampfrath}},\ }\bibfield  {title} {\bibinfo {title}
  {Hallmarks of ballistic terahertz magnon currents in an antiferromagnetic
  insulator},\ }\href {https://doi.org/10.1103/mk8h-yzkj} {\bibfield  {journal}
  {\bibinfo  {journal} {Phys. Rev. Lett.}\ }\textbf {\bibinfo {volume} {135}},\
  \bibinfo {pages} {176703} (\bibinfo {year} {2025})}\BibitemShut {NoStop}%
\bibitem [{\citenamefont {Chardonnet}\ \emph {et~al.}(2026)\citenamefont
  {Chardonnet}, \citenamefont {Hennes}, \citenamefont {Jarrier}, \citenamefont
  {Delaunay}, \citenamefont {Jaouen}, \citenamefont {Kuhlmann}, \citenamefont
  {Leveillé}, \citenamefont {von Korff~Schmising}, \citenamefont {Schick},
  \citenamefont {Yao}, \citenamefont {Liu}, \citenamefont {Chiuzbăian},
  \citenamefont {Lüning}, \citenamefont {Vodungbo},\ and\ \citenamefont
  {Jal}}]{Chardonnet2026}%
  \BibitemOpen
  \bibfield  {author} {\bibinfo {author} {\bibfnamefont {V.}~\bibnamefont
  {Chardonnet}}, \bibinfo {author} {\bibfnamefont {M.}~\bibnamefont {Hennes}},
  \bibinfo {author} {\bibfnamefont {R.}~\bibnamefont {Jarrier}}, \bibinfo
  {author} {\bibfnamefont {R.}~\bibnamefont {Delaunay}}, \bibinfo {author}
  {\bibfnamefont {N.}~\bibnamefont {Jaouen}}, \bibinfo {author} {\bibfnamefont
  {M.}~\bibnamefont {Kuhlmann}}, \bibinfo {author} {\bibfnamefont
  {C.}~\bibnamefont {Leveillé}}, \bibinfo {author} {\bibfnamefont
  {C.}~\bibnamefont {von Korff~Schmising}}, \bibinfo {author} {\bibfnamefont
  {D.}~\bibnamefont {Schick}}, \bibinfo {author} {\bibfnamefont
  {K.}~\bibnamefont {Yao}}, \bibinfo {author} {\bibfnamefont {X.}~\bibnamefont
  {Liu}}, \bibinfo {author} {\bibfnamefont {G.~S.}\ \bibnamefont
  {Chiuzbăian}}, \bibinfo {author} {\bibfnamefont {J.}~\bibnamefont
  {Lüning}}, \bibinfo {author} {\bibfnamefont {B.}~\bibnamefont {Vodungbo}},\
  and\ \bibinfo {author} {\bibfnamefont {E.}~\bibnamefont {Jal}},\ }\href
  {https://arxiv.org/abs/2603.11913} {\bibinfo {title} {Depth-resolved
  magnetization dynamics in {F}e thin films after ultrafast laser excitation}}
  (\bibinfo {year} {2026}),\ \Eprint {https://arxiv.org/abs/2603.11913}
  {arXiv:2603.11913 [cond-mat.mtrl-sci]} \BibitemShut {NoStop}%
\bibitem [{\citenamefont {Uehling}\ and\ \citenamefont
  {Uhlenbeck}(1933)}]{Uehling1933}%
  \BibitemOpen
  \bibfield  {author} {\bibinfo {author} {\bibfnamefont {E.~A.}\ \bibnamefont
  {Uehling}}\ and\ \bibinfo {author} {\bibfnamefont {G.~E.}\ \bibnamefont
  {Uhlenbeck}},\ }\bibfield  {title} {\bibinfo {title} {Transport phenomena in
  {Einstein-Bose and Fermi-Dirac} gases. {I}},\ }\href
  {https://doi.org/10.1103/PhysRev.43.552} {\bibfield  {journal} {\bibinfo
  {journal} {Phys. Rev.}\ }\textbf {\bibinfo {volume} {43}},\ \bibinfo {pages}
  {552} (\bibinfo {year} {1933})}\BibitemShut {NoStop}%
\bibitem [{\citenamefont {Zhou}\ \emph {et~al.}(2021)\citenamefont {Zhou},
  \citenamefont {Park}, \citenamefont {Lu}, \citenamefont {Maliyov},
  \citenamefont {Tong},\ and\ \citenamefont {Bernardi}}]{Zhou2021}%
  \BibitemOpen
  \bibfield  {author} {\bibinfo {author} {\bibfnamefont {J.-J.}\ \bibnamefont
  {Zhou}}, \bibinfo {author} {\bibfnamefont {J.}~\bibnamefont {Park}}, \bibinfo
  {author} {\bibfnamefont {I.-T.}\ \bibnamefont {Lu}}, \bibinfo {author}
  {\bibfnamefont {I.}~\bibnamefont {Maliyov}}, \bibinfo {author} {\bibfnamefont
  {X.}~\bibnamefont {Tong}},\ and\ \bibinfo {author} {\bibfnamefont
  {M.}~\bibnamefont {Bernardi}},\ }\bibfield  {title} {\bibinfo {title}
  {Perturbo: A software package for ab initio electron–phonon interactions,
  charge transport and ultrafast dynamics},\ }\href
  {https://doi.org/https://doi.org/10.1016/j.cpc.2021.107970} {\bibfield
  {journal} {\bibinfo  {journal} {Computer Physics Commun.}\ }\textbf {\bibinfo
  {volume} {264}},\ \bibinfo {pages} {107970} (\bibinfo {year}
  {2021})}\BibitemShut {NoStop}%
\bibitem [{\citenamefont {Caruso}\ and\ \citenamefont
  {Novko}(2022)}]{Caruso2022}%
  \BibitemOpen
  \bibfield  {author} {\bibinfo {author} {\bibfnamefont {F.}~\bibnamefont
  {Caruso}}\ and\ \bibinfo {author} {\bibfnamefont {D.}~\bibnamefont {Novko}},\
  }\bibfield  {title} {\bibinfo {title} {Ultrafast dynamics of electrons and
  phonons: from the two-temperature model to the time-dependent {B}oltzmann
  equation},\ }\href {https://doi.org/10.1080/23746149.2022.2095925} {\bibfield
   {journal} {\bibinfo  {journal} {Adv. Phys.: X}\ }\textbf {\bibinfo {volume}
  {7}},\ \bibinfo {pages} {2095925} (\bibinfo {year} {2022})}\BibitemShut
  {NoStop}%
\bibitem [{\citenamefont {Lee}\ \emph {et~al.}(2023)\citenamefont {Lee},
  \citenamefont {Ponc{\'e}}, \citenamefont {Bushick}, \citenamefont
  {Hajinazar}, \citenamefont {Lafuente-Bartolome}, \citenamefont {Leveillee},
  \citenamefont {Lian}, \citenamefont {Lihm}, \citenamefont {Macheda},
  \citenamefont {Mori}, \citenamefont {Paudyal}, \citenamefont {Sio},
  \citenamefont {Tiwari}, \citenamefont {Zacharias}, \citenamefont {Zhang},
  \citenamefont {Bonini}, \citenamefont {Kioupakis}, \citenamefont {Margine},\
  and\ \citenamefont {Giustino}}]{Lee2023}%
  \BibitemOpen
  \bibfield  {author} {\bibinfo {author} {\bibfnamefont {H.}~\bibnamefont
  {Lee}}, \bibinfo {author} {\bibfnamefont {S.}~\bibnamefont {Ponc{\'e}}},
  \bibinfo {author} {\bibfnamefont {K.}~\bibnamefont {Bushick}}, \bibinfo
  {author} {\bibfnamefont {S.}~\bibnamefont {Hajinazar}}, \bibinfo {author}
  {\bibfnamefont {J.}~\bibnamefont {Lafuente-Bartolome}}, \bibinfo {author}
  {\bibfnamefont {J.}~\bibnamefont {Leveillee}}, \bibinfo {author}
  {\bibfnamefont {C.}~\bibnamefont {Lian}}, \bibinfo {author} {\bibfnamefont
  {J.-M.}\ \bibnamefont {Lihm}}, \bibinfo {author} {\bibfnamefont
  {F.}~\bibnamefont {Macheda}}, \bibinfo {author} {\bibfnamefont
  {H.}~\bibnamefont {Mori}}, \bibinfo {author} {\bibfnamefont {H.}~\bibnamefont
  {Paudyal}}, \bibinfo {author} {\bibfnamefont {W.~H.}\ \bibnamefont {Sio}},
  \bibinfo {author} {\bibfnamefont {S.}~\bibnamefont {Tiwari}}, \bibinfo
  {author} {\bibfnamefont {M.}~\bibnamefont {Zacharias}}, \bibinfo {author}
  {\bibfnamefont {X.}~\bibnamefont {Zhang}}, \bibinfo {author} {\bibfnamefont
  {N.}~\bibnamefont {Bonini}}, \bibinfo {author} {\bibfnamefont
  {E.}~\bibnamefont {Kioupakis}}, \bibinfo {author} {\bibfnamefont {E.~R.}\
  \bibnamefont {Margine}},\ and\ \bibinfo {author} {\bibfnamefont
  {F.}~\bibnamefont {Giustino}},\ }\bibfield  {title} {\bibinfo {title}
  {Electron--phonon physics from first principles using the {EPW} code},\
  }\href {https://doi.org/10.1038/s41524-023-01107-3} {\bibfield  {journal}
  {\bibinfo  {journal} {npj Comput. Mater.}\ }\textbf {\bibinfo {volume} {9}},\
  \bibinfo {pages} {156} (\bibinfo {year} {2023})}\BibitemShut {NoStop}%
\bibitem [{\citenamefont {Haberman}(2012)}]{FouriersLaw}%
  \BibitemOpen
  \bibfield  {author} {\bibinfo {author} {\bibfnamefont {R.}~\bibnamefont
  {Haberman}},\ }\href@noop {} {\emph {\bibinfo {title} {{Applied Partial
  Differential Equations with Fourier Series and Boundary Value Problems}}}},\
  \bibinfo {edition} {5th}\ ed.\ (\bibinfo  {publisher} {Pearson},\ \bibinfo
  {year} {2012})\ p.~\bibinfo {pages} {7}\BibitemShut {NoStop}%
\bibitem [{\citenamefont {Wilson}\ and\ \citenamefont
  {Cahill}(2014)}]{Wilson2014}%
  \BibitemOpen
  \bibfield  {author} {\bibinfo {author} {\bibfnamefont {R.~B.}\ \bibnamefont
  {Wilson}}\ and\ \bibinfo {author} {\bibfnamefont {D.~G.}\ \bibnamefont
  {Cahill}},\ }\bibfield  {title} {\bibinfo {title} {{Anisotropic failure of
  Fourier theory in time-domain thermoreflectance experiments}},\ }\href
  {https://doi.org/10.1038/ncomms6075} {\bibfield  {journal} {\bibinfo
  {journal} {Nature Commun.}\ }\textbf {\bibinfo {volume} {5}},\ \bibinfo
  {pages} {5075} (\bibinfo {year} {2014})}\BibitemShut {NoStop}%
\bibitem [{\citenamefont {Awsaf}\ \emph {et~al.}(2026)\citenamefont {Awsaf},
  \citenamefont {Thakur}, \citenamefont {Wei\ss{}enhofer}, \citenamefont
  {G\"ordes}, \citenamefont {Walter}, \citenamefont {Mawass}, \citenamefont
  {Pontius}, \citenamefont {Sch\"u\ss{}ler-Langeheine}, \citenamefont
  {Oppeneer},\ and\ \citenamefont {Kuch}}]{Awsaf2025}%
  \BibitemOpen
  \bibfield  {author} {\bibinfo {author} {\bibfnamefont {C.~S.}\ \bibnamefont
  {Awsaf}}, \bibinfo {author} {\bibfnamefont {S.}~\bibnamefont {Thakur}},
  \bibinfo {author} {\bibfnamefont {M.}~\bibnamefont {Wei\ss{}enhofer}},
  \bibinfo {author} {\bibfnamefont {J.}~\bibnamefont {G\"ordes}}, \bibinfo
  {author} {\bibfnamefont {M.}~\bibnamefont {Walter}}, \bibinfo {author}
  {\bibfnamefont {M.-A.}\ \bibnamefont {Mawass}}, \bibinfo {author}
  {\bibfnamefont {N.}~\bibnamefont {Pontius}}, \bibinfo {author} {\bibfnamefont
  {C.}~\bibnamefont {Sch\"u\ss{}ler-Langeheine}}, \bibinfo {author}
  {\bibfnamefont {P.~M.}\ \bibnamefont {Oppeneer}},\ and\ \bibinfo {author}
  {\bibfnamefont {W.}~\bibnamefont {Kuch}},\ }\bibfield  {title} {\bibinfo
  {title} {Element-selective probing of ultrafast
  ferromagnetic-antiferromagnetic order dynamics in {F}e/{C}o{O} bilayers},\
  }\href {https://doi.org/10.1103/gcwk-tsj5} {\bibfield  {journal} {\bibinfo
  {journal} {Phys. Rev. Lett.}\ }\textbf {\bibinfo {volume} {136}},\ \bibinfo
  {pages} {126705} (\bibinfo {year} {2026})}\BibitemShut {NoStop}%
\bibitem [{\citenamefont {Streib}\ \emph {et~al.}(2018)\citenamefont {Streib},
  \citenamefont {Keshtgar},\ and\ \citenamefont {Bauer}}]{Streib2018}%
  \BibitemOpen
  \bibfield  {author} {\bibinfo {author} {\bibfnamefont {S.}~\bibnamefont
  {Streib}}, \bibinfo {author} {\bibfnamefont {H.}~\bibnamefont {Keshtgar}},\
  and\ \bibinfo {author} {\bibfnamefont {G.~E.~W.}\ \bibnamefont {Bauer}},\
  }\bibfield  {title} {\bibinfo {title} {Damping of magnetization dynamics by
  phonon pumping},\ }\href {https://doi.org/10.1103/PhysRevLett.121.027202}
  {\bibfield  {journal} {\bibinfo  {journal} {Phys. Rev. Lett.}\ }\textbf
  {\bibinfo {volume} {121}},\ \bibinfo {pages} {027202} (\bibinfo {year}
  {2018})}\BibitemShut {NoStop}%
\bibitem [{\citenamefont {Jeannin}(2023)}]{pyGTM}%
  \BibitemOpen
  \bibfield  {author} {\bibinfo {author} {\bibfnamefont {M.}~\bibnamefont
  {Jeannin}},\ }\href@noop {} {\emph {\bibinfo {title} {pyGTM}}} (\bibinfo
  {year} {2023}),\ \bibinfo {note} {available at
  \url{https://pygtm.readthedocs.io}. Accessed: 2025-11-10}\BibitemShut
  {NoStop}%
\bibitem [{\citenamefont {Passler}\ and\ \citenamefont
  {Paarmann}(2017)}]{Passler2017}%
  \BibitemOpen
  \bibfield  {author} {\bibinfo {author} {\bibfnamefont {N.~C.}\ \bibnamefont
  {Passler}}\ and\ \bibinfo {author} {\bibfnamefont {A.}~\bibnamefont
  {Paarmann}},\ }\bibfield  {title} {\bibinfo {title} {Generalized 4
  {\texttimes} 4 matrix formalism for light propagation in anisotropic
  stratified media: study of surface phonon polaritons in polar dielectric
  heterostructures},\ }\href {https://doi.org/10.1364/JOSAB.34.002128}
  {\bibfield  {journal} {\bibinfo  {journal} {J. Opt. Soc. Am. B}\ }\textbf
  {\bibinfo {volume} {34}},\ \bibinfo {pages} {2128} (\bibinfo {year}
  {2017})}\BibitemShut {NoStop}%
\bibitem [{\citenamefont {Passler}\ and\ \citenamefont
  {Paarmann}(2019)}]{Passler2019}%
  \BibitemOpen
  \bibfield  {author} {\bibinfo {author} {\bibfnamefont {N.~C.}\ \bibnamefont
  {Passler}}\ and\ \bibinfo {author} {\bibfnamefont {A.}~\bibnamefont
  {Paarmann}},\ }\bibfield  {title} {\bibinfo {title} {Generalized 4
  {\texttimes} 4 matrix formalism for light propagation in anisotropic
  stratified media: study of surface phonon polaritons in polar dielectric
  heterostructures: erratum},\ }\href {https://doi.org/10.1364/JOSAB.36.003246}
  {\bibfield  {journal} {\bibinfo  {journal} {J. Opt. Soc. Am. B}\ }\textbf
  {\bibinfo {volume} {36}},\ \bibinfo {pages} {3246} (\bibinfo {year}
  {2019})}\BibitemShut {NoStop}%
\bibitem [{\citenamefont {Passler}\ \emph {et~al.}(2020)\citenamefont
  {Passler}, \citenamefont {Jeannin},\ and\ \citenamefont
  {Paarmann}}]{Passler2020}%
  \BibitemOpen
  \bibfield  {author} {\bibinfo {author} {\bibfnamefont {N.~C.}\ \bibnamefont
  {Passler}}, \bibinfo {author} {\bibfnamefont {M.}~\bibnamefont {Jeannin}},\
  and\ \bibinfo {author} {\bibfnamefont {A.}~\bibnamefont {Paarmann}},\
  }\bibfield  {title} {\bibinfo {title} {Layer-resolved absorption of light in
  arbitrarily anisotropic heterostructures},\ }\href
  {https://doi.org/10.1103/PhysRevB.101.165425} {\bibfield  {journal} {\bibinfo
   {journal} {Phys. Rev. B}\ }\textbf {\bibinfo {volume} {101}},\ \bibinfo
  {pages} {165425} (\bibinfo {year} {2020})}\BibitemShut {NoStop}%
\bibitem [{\citenamefont {LeVeque}(2007)}]{Crank-Nicholson}%
  \BibitemOpen
  \bibfield  {author} {\bibinfo {author} {\bibfnamefont {R.~J.}\ \bibnamefont
  {LeVeque}},\ }\href@noop {} {\emph {\bibinfo {title} {Finite Difference
  Methods for Ordinary and Partial Differential Equations: {S}teady-State and
  Time-dependent Problems}}}\ (\bibinfo  {publisher} {Society for Industrial
  and Applied Mathematics},\ \bibinfo {year} {2007})\BibitemShut {NoStop}%
\bibitem [{\citenamefont {LeVeque}(2002)}]{TVD}%
  \BibitemOpen
  \bibfield  {author} {\bibinfo {author} {\bibfnamefont {R.~J.}\ \bibnamefont
  {LeVeque}},\ }\href@noop {} {\emph {\bibinfo {title} {Finite Volume Methods
  for Hyperbolic Problems}}}\ (\bibinfo  {publisher} {Cambridge University
  Press},\ \bibinfo {year} {2002})\BibitemShut {NoStop}%
\bibitem [{\citenamefont {Lin}\ \emph {et~al.}(2021)\citenamefont {Lin},
  \citenamefont {Chong}, \citenamefont {Ding}, \citenamefont {Zhou},
  \citenamefont {Gan}, \citenamefont {Xu}, \citenamefont {Wei},\ and\
  \citenamefont {Feng}}]{ThermalCond}%
  \BibitemOpen
  \bibfield  {author} {\bibinfo {author} {\bibfnamefont {Y.}~\bibnamefont
  {Lin}}, \bibinfo {author} {\bibfnamefont {X.}~\bibnamefont {Chong}}, \bibinfo
  {author} {\bibfnamefont {Y.}~\bibnamefont {Ding}}, \bibinfo {author}
  {\bibfnamefont {Y.}~\bibnamefont {Zhou}}, \bibinfo {author} {\bibfnamefont
  {M.}~\bibnamefont {Gan}}, \bibinfo {author} {\bibfnamefont {L.}~\bibnamefont
  {Xu}}, \bibinfo {author} {\bibfnamefont {S.}~\bibnamefont {Wei}},\ and\
  \bibinfo {author} {\bibfnamefont {J.}~\bibnamefont {Feng}},\ }\bibfield
  {title} {\bibinfo {title} {{First-Principles Calculations of Thermal and
  Electrical Transport Properties of bcc and fcc Dilute {Fe–X (X = Al, Co,
  Cr, Mn, Mo, Nb, Ni, Ti, V, and W}) Binary Alloys}},\ }\href
  {https://doi.org/10.3390/met11121988} {\bibfield  {journal} {\bibinfo
  {journal} {Metals}\ }\textbf {\bibinfo {volume} {11}},\ \bibinfo {pages}
  {1988} (\bibinfo {year} {2021})}\BibitemShut {NoStop}%
\bibitem [{\citenamefont {Carpene}\ \emph
  {et~al.}(2008{\natexlab{b}})\citenamefont {Carpene}, \citenamefont {Mancini},
  \citenamefont {Dallera}, \citenamefont {Brenna}, \citenamefont {Puppin},\
  and\ \citenamefont {De~Silvestri}}]{800nm_laser}%
  \BibitemOpen
  \bibfield  {author} {\bibinfo {author} {\bibfnamefont {E.}~\bibnamefont
  {Carpene}}, \bibinfo {author} {\bibfnamefont {E.}~\bibnamefont {Mancini}},
  \bibinfo {author} {\bibfnamefont {C.}~\bibnamefont {Dallera}}, \bibinfo
  {author} {\bibfnamefont {M.}~\bibnamefont {Brenna}}, \bibinfo {author}
  {\bibfnamefont {E.}~\bibnamefont {Puppin}},\ and\ \bibinfo {author}
  {\bibfnamefont {S.}~\bibnamefont {De~Silvestri}},\ }\bibfield  {title}
  {\bibinfo {title} {Dynamics of electron-magnon interaction and ultrafast
  demagnetization in thin iron films},\ }\href
  {https://doi.org/10.1103/PhysRevB.78.174422} {\bibfield  {journal} {\bibinfo
  {journal} {Phys. Rev. B}\ }\textbf {\bibinfo {volume} {78}},\ \bibinfo
  {pages} {174422} (\bibinfo {year} {2008}{\natexlab{b}})}\BibitemShut
  {NoStop}%
\bibitem [{\citenamefont {Oppeneer}\ \emph {et~al.}(1992)\citenamefont
  {Oppeneer}, \citenamefont {Maurer}, \citenamefont {Sticht},\ and\
  \citenamefont {K\"ubler}}]{Oppeneer1992}%
  \BibitemOpen
  \bibfield  {author} {\bibinfo {author} {\bibfnamefont {P.~M.}\ \bibnamefont
  {Oppeneer}}, \bibinfo {author} {\bibfnamefont {T.}~\bibnamefont {Maurer}},
  \bibinfo {author} {\bibfnamefont {J.}~\bibnamefont {Sticht}},\ and\ \bibinfo
  {author} {\bibfnamefont {J.}~\bibnamefont {K\"ubler}},\ }\bibfield  {title}
  {\bibinfo {title} {Ab initio calculated magneto-optical kerr effect of
  ferromagnetic metals: {F}e and {N}i},\ }\href
  {https://doi.org/10.1103/PhysRevB.45.10924} {\bibfield  {journal} {\bibinfo
  {journal} {Phys. Rev. B}\ }\textbf {\bibinfo {volume} {45}},\ \bibinfo
  {pages} {10924} (\bibinfo {year} {1992})}\BibitemShut {NoStop}%
\bibitem [{\citenamefont {Polyanskiy}(2024)}]{Refractiveindex.info}%
  \BibitemOpen
  \bibfield  {author} {\bibinfo {author} {\bibfnamefont {M.~N.}\ \bibnamefont
  {Polyanskiy}},\ }\bibfield  {title} {\bibinfo {title} {Refractiveindex.info
  database of optical constants},\ }\bibfield  {journal} {\bibinfo  {journal}
  {Scientific Data}\ }\textbf {\bibinfo {volume} {11}},\ \href
  {https://doi.org/10.1038/s41597-023-02898-2} {10.1038/s41597-023-02898-2}
  (\bibinfo {year} {2024})\BibitemShut {NoStop}%
\bibitem [{\citenamefont {Uchida}\ \emph {et~al.}(2010)\citenamefont {Uchida},
  \citenamefont {Xiao}, \citenamefont {Adachi}, \citenamefont {Ohe},
  \citenamefont {Takahashi}, \citenamefont {Ieda}, \citenamefont {Ota},
  \citenamefont {Kajiwara}, \citenamefont {Umezawa}, \citenamefont {Kawai},
  \citenamefont {Bauer}, \citenamefont {Maekawa},\ and\ \citenamefont
  {Saitoh}}]{Uchida2010}%
  \BibitemOpen
  \bibfield  {author} {\bibinfo {author} {\bibfnamefont {K.}~\bibnamefont
  {Uchida}}, \bibinfo {author} {\bibfnamefont {J.}~\bibnamefont {Xiao}},
  \bibinfo {author} {\bibfnamefont {H.}~\bibnamefont {Adachi}}, \bibinfo
  {author} {\bibfnamefont {J.}~\bibnamefont {Ohe}}, \bibinfo {author}
  {\bibfnamefont {S.}~\bibnamefont {Takahashi}}, \bibinfo {author}
  {\bibfnamefont {J.}~\bibnamefont {Ieda}}, \bibinfo {author} {\bibfnamefont
  {T.}~\bibnamefont {Ota}}, \bibinfo {author} {\bibfnamefont {Y.}~\bibnamefont
  {Kajiwara}}, \bibinfo {author} {\bibfnamefont {H.}~\bibnamefont {Umezawa}},
  \bibinfo {author} {\bibfnamefont {H.}~\bibnamefont {Kawai}}, \bibinfo
  {author} {\bibfnamefont {G.~E.~W.}\ \bibnamefont {Bauer}}, \bibinfo {author}
  {\bibfnamefont {S.}~\bibnamefont {Maekawa}},\ and\ \bibinfo {author}
  {\bibfnamefont {E.}~\bibnamefont {Saitoh}},\ }\bibfield  {title} {\bibinfo
  {title} {{Spin Seebeck insulator}},\ }\href
  {https://doi.org/10.1038/nmat2856} {\bibfield  {journal} {\bibinfo  {journal}
  {Nature Mater.}\ }\textbf {\bibinfo {volume} {9}},\ \bibinfo {pages} {894}
  (\bibinfo {year} {2010})}\BibitemShut {NoStop}%
\bibitem [{\citenamefont {Kimling}\ \emph {et~al.}(2017)\citenamefont
  {Kimling}, \citenamefont {Choi}, \citenamefont {Brangham}, \citenamefont
  {Matalla-Wagner}, \citenamefont {Huebner}, \citenamefont {Kuschel},
  \citenamefont {Yang},\ and\ \citenamefont {Cahill}}]{Kimling2017}%
  \BibitemOpen
  \bibfield  {author} {\bibinfo {author} {\bibfnamefont {J.}~\bibnamefont
  {Kimling}}, \bibinfo {author} {\bibfnamefont {G.-M.}\ \bibnamefont {Choi}},
  \bibinfo {author} {\bibfnamefont {J.~T.}\ \bibnamefont {Brangham}}, \bibinfo
  {author} {\bibfnamefont {T.}~\bibnamefont {Matalla-Wagner}}, \bibinfo
  {author} {\bibfnamefont {T.}~\bibnamefont {Huebner}}, \bibinfo {author}
  {\bibfnamefont {T.}~\bibnamefont {Kuschel}}, \bibinfo {author} {\bibfnamefont
  {F.}~\bibnamefont {Yang}},\ and\ \bibinfo {author} {\bibfnamefont {D.~G.}\
  \bibnamefont {Cahill}},\ }\bibfield  {title} {\bibinfo {title} {Picosecond
  spin {S}eebeck effect},\ }\href
  {https://doi.org/10.1103/PhysRevLett.118.057201} {\bibfield  {journal}
  {\bibinfo  {journal} {Phys. Rev. Lett.}\ }\textbf {\bibinfo {volume} {118}},\
  \bibinfo {pages} {057201} (\bibinfo {year} {2017})}\BibitemShut {NoStop}%
\bibitem [{\citenamefont {Rezende}\ \emph {et~al.}(2016)\citenamefont
  {Rezende}, \citenamefont {Rodríguez-Suárez}, \citenamefont {Cunha},
  \citenamefont {{López Ortiz}},\ and\ \citenamefont {Azevedo}}]{Rezende2016}%
  \BibitemOpen
  \bibfield  {author} {\bibinfo {author} {\bibfnamefont {S.~M.}\ \bibnamefont
  {Rezende}}, \bibinfo {author} {\bibfnamefont {R.~L.}\ \bibnamefont
  {Rodríguez-Suárez}}, \bibinfo {author} {\bibfnamefont {R.~O.}\ \bibnamefont
  {Cunha}}, \bibinfo {author} {\bibfnamefont {J.~C.}\ \bibnamefont {{López
  Ortiz}}},\ and\ \bibinfo {author} {\bibfnamefont {A.}~\bibnamefont
  {Azevedo}},\ }\bibfield  {title} {\bibinfo {title} {Bulk magnon spin current
  theory for the longitudinal spin {S}eebeck effect},\ }\href
  {https://doi.org/https://doi.org/10.1016/j.jmmm.2015.07.102} {\bibfield
  {journal} {\bibinfo  {journal} {J. Magn. Magn. Mater.}\ }\textbf {\bibinfo
  {volume} {400}},\ \bibinfo {pages} {171} (\bibinfo {year}
  {2016})}\BibitemShut {NoStop}%
\bibitem [{\citenamefont {Schmidt}\ and\ \citenamefont
  {Brouwer}(2021)}]{Schmidt2021}%
  \BibitemOpen
  \bibfield  {author} {\bibinfo {author} {\bibfnamefont {R.}~\bibnamefont
  {Schmidt}}\ and\ \bibinfo {author} {\bibfnamefont {P.~W.}\ \bibnamefont
  {Brouwer}},\ }\bibfield  {title} {\bibinfo {title} {Theory of the
  low-temperature longitudinal spin {S}eebeck effect},\ }\href
  {https://doi.org/10.1103/PhysRevB.103.014412} {\bibfield  {journal} {\bibinfo
   {journal} {Phys. Rev. B}\ }\textbf {\bibinfo {volume} {103}},\ \bibinfo
  {pages} {014412} (\bibinfo {year} {2021})}\BibitemShut {NoStop}%
\bibitem [{\citenamefont {Danan}\ \emph {et~al.}(1968)\citenamefont {Danan},
  \citenamefont {Herr},\ and\ \citenamefont {Meyer}}]{Danan1968}%
  \BibitemOpen
  \bibfield  {author} {\bibinfo {author} {\bibfnamefont {H.}~\bibnamefont
  {Danan}}, \bibinfo {author} {\bibfnamefont {A.}~\bibnamefont {Herr}},\ and\
  \bibinfo {author} {\bibfnamefont {A.~J.~P.}\ \bibnamefont {Meyer}},\
  }\bibfield  {title} {\bibinfo {title} {New determinations of the saturation
  magnetization of nickel and iron},\ }\href
  {https://doi.org/10.1063/1.2163571} {\bibfield  {journal} {\bibinfo
  {journal} {J. Appl. Phys.}\ }\textbf {\bibinfo {volume} {39}},\ \bibinfo
  {pages} {669} (\bibinfo {year} {1968})}\BibitemShut {NoStop}%
\bibitem [{\citenamefont {Mohn}(2006)}]{Magnetism}%
  \BibitemOpen
  \bibfield  {author} {\bibinfo {author} {\bibfnamefont {P.}~\bibnamefont
  {Mohn}},\ }\href {https://doi.org/10.1007/3-540-30981-0} {\emph {\bibinfo
  {title} {Magnetism in the Solid State: An Introduction}}},\ Springer Series
  in Solid-State Sciences\ (\bibinfo  {publisher} {Springer, Berlin
  Heidelberg},\ \bibinfo {year} {2006})\ p.~\bibinfo {pages} {41}\BibitemShut
  {NoStop}%
\bibitem [{\citenamefont {Cheng}\ \emph {et~al.}(2018)\citenamefont {Cheng},
  \citenamefont {Xiao},\ and\ \citenamefont {Zhu}}]{Cheng2018}%
  \BibitemOpen
  \bibfield  {author} {\bibinfo {author} {\bibfnamefont {R.}~\bibnamefont
  {Cheng}}, \bibinfo {author} {\bibfnamefont {D.}~\bibnamefont {Xiao}},\ and\
  \bibinfo {author} {\bibfnamefont {J.-G.}\ \bibnamefont {Zhu}},\ }\bibfield
  {title} {\bibinfo {title} {Antiferromagnet-based magnonic spin-transfer
  torque},\ }\href {https://doi.org/10.1103/PhysRevB.98.020408} {\bibfield
  {journal} {\bibinfo  {journal} {Phys. Rev. B}\ }\textbf {\bibinfo {volume}
  {98}},\ \bibinfo {pages} {020408} (\bibinfo {year} {2018})}\BibitemShut
  {NoStop}%
\bibitem [{\citenamefont {Metzler}\ and\ \citenamefont
  {Klafter}(2004)}]{Metzler2004}%
  \BibitemOpen
  \bibfield  {author} {\bibinfo {author} {\bibfnamefont {R.}~\bibnamefont
  {Metzler}}\ and\ \bibinfo {author} {\bibfnamefont {J.}~\bibnamefont
  {Klafter}},\ }\bibfield  {title} {\bibinfo {title} {The restaurant at the end
  of the random walk: recent developments in the description of anomalous
  transport by fractional dynamics},\ }\href
  {https://doi.org/10.1088/0305-4470/37/31/R01} {\bibfield  {journal} {\bibinfo
   {journal} {J. Phys. A: Math. and General}\ }\textbf {\bibinfo {volume}
  {37}},\ \bibinfo {pages} {R161} (\bibinfo {year} {2004})}\BibitemShut
  {NoStop}%
\bibitem [{\citenamefont {Oppeneer}(2001)}]{OppeneerMOKE}%
  \BibitemOpen
  \bibfield  {author} {\bibinfo {author} {\bibfnamefont {P.~M.}\ \bibnamefont
  {Oppeneer}},\ }\bibinfo {title} {Magneto-optical {Kerr} spectra},\ in\
  \href@noop {} {\emph {\bibinfo {booktitle} {Handbook of Magnetic
  Materials}}},\ Vol.~\bibinfo {volume} {13}\ (\bibinfo  {publisher}
  {Elsevier},\ \bibinfo {address} {Amsterdam},\ \bibinfo {year} {2001})\ pp.\
  \bibinfo {pages} {229--422}\BibitemShut {NoStop}%
\bibitem [{\citenamefont {Koopmans}\ \emph {et~al.}(2000)\citenamefont
  {Koopmans}, \citenamefont {Van~Kampen}, \citenamefont {Kohlhepp},\ and\
  \citenamefont {De~Jonge}}]{Koopmans2000}%
  \BibitemOpen
  \bibfield  {author} {\bibinfo {author} {\bibfnamefont {B.}~\bibnamefont
  {Koopmans}}, \bibinfo {author} {\bibfnamefont {M.}~\bibnamefont
  {Van~Kampen}}, \bibinfo {author} {\bibfnamefont {J.~T.}\ \bibnamefont
  {Kohlhepp}},\ and\ \bibinfo {author} {\bibfnamefont {W.~J.~M.}\ \bibnamefont
  {De~Jonge}},\ }\bibfield  {title} {\bibinfo {title} {Ultrafast magneto-optics
  in nickel: magnetism or optics?},\ }\href
  {https://doi.org/10.1103/PhysRevLett.85.844} {\bibfield  {journal} {\bibinfo
  {journal} {Phys. Rev. Lett.}\ }\textbf {\bibinfo {volume} {85}},\ \bibinfo
  {pages} {844 – 847} (\bibinfo {year} {2000})}\BibitemShut {NoStop}%
\bibitem [{\citenamefont {Oppeneer}\ and\ \citenamefont
  {Liebsch}(2004)}]{Oppeneer2004}%
  \BibitemOpen
  \bibfield  {author} {\bibinfo {author} {\bibfnamefont {P.~M.}\ \bibnamefont
  {Oppeneer}}\ and\ \bibinfo {author} {\bibfnamefont {A.}~\bibnamefont
  {Liebsch}},\ }\bibfield  {title} {\bibinfo {title} {{Ultrafast
  demagnetization in Ni: Theory of magneto-optics for non-equilibrium electron
  distributions}},\ }\href {https://doi.org/10.1088/0953-8984/16/30/013}
  {\bibfield  {journal} {\bibinfo  {journal} {J. Phys: Condens. Matter}\
  }\textbf {\bibinfo {volume} {16}},\ \bibinfo {pages} {5519} (\bibinfo {year}
  {2004})}\BibitemShut {NoStop}%
\bibitem [{\citenamefont {Richter}\ \emph {et~al.}(2024)\citenamefont
  {Richter}, \citenamefont {Jana}, \citenamefont {Hennecke}, \citenamefont
  {Schick}, \citenamefont {von Korff~Schmising},\ and\ \citenamefont
  {Eisebitt}}]{Richter2024}%
  \BibitemOpen
  \bibfield  {author} {\bibinfo {author} {\bibfnamefont {J.}~\bibnamefont
  {Richter}}, \bibinfo {author} {\bibfnamefont {S.}~\bibnamefont {Jana}},
  \bibinfo {author} {\bibfnamefont {M.}~\bibnamefont {Hennecke}}, \bibinfo
  {author} {\bibfnamefont {D.}~\bibnamefont {Schick}}, \bibinfo {author}
  {\bibfnamefont {C.}~\bibnamefont {von Korff~Schmising}},\ and\ \bibinfo
  {author} {\bibfnamefont {S.}~\bibnamefont {Eisebitt}},\ }\bibfield  {title}
  {\bibinfo {title} {Relationship between magnetic asymmetry and magnetization
  in ultrafast transverse magneto-optical {K}err effect spectroscopy in the
  extreme ultraviolet spectral range},\ }\href
  {https://doi.org/10.1103/PhysRevB.109.184440} {\bibfield  {journal} {\bibinfo
   {journal} {Phys. Rev. B}\ }\textbf {\bibinfo {volume} {109}},\ \bibinfo
  {pages} {184440} (\bibinfo {year} {2024})}\BibitemShut {NoStop}%
\bibitem [{\citenamefont {Wieczorek}\ \emph {et~al.}(2015)\citenamefont
  {Wieczorek}, \citenamefont {Eschenlohr}, \citenamefont {Weidtmann},
  \citenamefont {R\"osner}, \citenamefont {Bergeard}, \citenamefont
  {Tarasevitch}, \citenamefont {Wehling},\ and\ \citenamefont
  {Bovensiepen}}]{Wieczorek2015}%
  \BibitemOpen
  \bibfield  {author} {\bibinfo {author} {\bibfnamefont {J.}~\bibnamefont
  {Wieczorek}}, \bibinfo {author} {\bibfnamefont {A.}~\bibnamefont
  {Eschenlohr}}, \bibinfo {author} {\bibfnamefont {B.}~\bibnamefont
  {Weidtmann}}, \bibinfo {author} {\bibfnamefont {M.}~\bibnamefont {R\"osner}},
  \bibinfo {author} {\bibfnamefont {N.}~\bibnamefont {Bergeard}}, \bibinfo
  {author} {\bibfnamefont {A.}~\bibnamefont {Tarasevitch}}, \bibinfo {author}
  {\bibfnamefont {T.~O.}\ \bibnamefont {Wehling}},\ and\ \bibinfo {author}
  {\bibfnamefont {U.}~\bibnamefont {Bovensiepen}},\ }\bibfield  {title}
  {\bibinfo {title} {{Separation of ultrafast spin currents and spin-flip
  scattering in Co/Cu(001) driven by femtosecond laser excitation employing the
  complex magneto-optical Kerr effect}},\ }\href
  {https://doi.org/10.1103/PhysRevB.92.174410} {\bibfield  {journal} {\bibinfo
  {journal} {Phys. Rev. B}\ }\textbf {\bibinfo {volume} {92}},\ \bibinfo
  {pages} {174410} (\bibinfo {year} {2015})}\BibitemShut {NoStop}%
\bibitem [{\citenamefont {Razdolski}\ \emph
  {et~al.}(2017{\natexlab{a}})\citenamefont {Razdolski}, \citenamefont
  {Alekhin}, \citenamefont {Martens}, \citenamefont {B{\"u}stel}, \citenamefont
  {Diesing}, \citenamefont {M{\"u}nzenberg}, \citenamefont {Bovensiepen},\ and\
  \citenamefont {Melnikov}}]{Razdolski2017-MOKE}%
  \BibitemOpen
  \bibfield  {author} {\bibinfo {author} {\bibfnamefont {I.}~\bibnamefont
  {Razdolski}}, \bibinfo {author} {\bibfnamefont {A.}~\bibnamefont {Alekhin}},
  \bibinfo {author} {\bibfnamefont {U.}~\bibnamefont {Martens}}, \bibinfo
  {author} {\bibfnamefont {D.}~\bibnamefont {B{\"u}stel}}, \bibinfo {author}
  {\bibfnamefont {D.}~\bibnamefont {Diesing}}, \bibinfo {author} {\bibfnamefont
  {M.}~\bibnamefont {M{\"u}nzenberg}}, \bibinfo {author} {\bibfnamefont
  {U.}~\bibnamefont {Bovensiepen}},\ and\ \bibinfo {author} {\bibfnamefont
  {A.}~\bibnamefont {Melnikov}},\ }\bibfield  {title} {\bibinfo {title}
  {Analysis of the time-resolved magneto-optical {K}err effect for ultrafast
  magnetization dynamics in ferromagnetic thin films},\ }\href
  {https://doi.org/10.1088/1361-648X/aa63c6} {\bibfield  {journal} {\bibinfo
  {journal} {J. Phys.: Condensed Matter}\ }\textbf {\bibinfo {volume} {29}},\
  \bibinfo {pages} {174002} (\bibinfo {year} {2017}{\natexlab{a}})}\BibitemShut
  {NoStop}%
\bibitem [{\citenamefont {You}\ \emph {et~al.}(2018)\citenamefont {You},
  \citenamefont {Tengdin}, \citenamefont {Chen}, \citenamefont {Shi},
  \citenamefont {Zusin}, \citenamefont {Zhang}, \citenamefont {Gentry},
  \citenamefont {Blonsky}, \citenamefont {Keller}, \citenamefont {Oppeneer},
  \citenamefont {Kapteyn}, \citenamefont {Tao},\ and\ \citenamefont
  {Murnane}}]{You2018}%
  \BibitemOpen
  \bibfield  {author} {\bibinfo {author} {\bibfnamefont {W.}~\bibnamefont
  {You}}, \bibinfo {author} {\bibfnamefont {P.}~\bibnamefont {Tengdin}},
  \bibinfo {author} {\bibfnamefont {C.}~\bibnamefont {Chen}}, \bibinfo {author}
  {\bibfnamefont {X.}~\bibnamefont {Shi}}, \bibinfo {author} {\bibfnamefont
  {D.}~\bibnamefont {Zusin}}, \bibinfo {author} {\bibfnamefont
  {Y.}~\bibnamefont {Zhang}}, \bibinfo {author} {\bibfnamefont
  {C.}~\bibnamefont {Gentry}}, \bibinfo {author} {\bibfnamefont
  {A.}~\bibnamefont {Blonsky}}, \bibinfo {author} {\bibfnamefont
  {M.}~\bibnamefont {Keller}}, \bibinfo {author} {\bibfnamefont {P.~M.}\
  \bibnamefont {Oppeneer}}, \bibinfo {author} {\bibfnamefont {H.}~\bibnamefont
  {Kapteyn}}, \bibinfo {author} {\bibfnamefont {Z.}~\bibnamefont {Tao}},\ and\
  \bibinfo {author} {\bibfnamefont {M.}~\bibnamefont {Murnane}},\ }\bibfield
  {title} {\bibinfo {title} {Revealing the nature of the ultrafast magnetic
  phase transition in ni by correlating extreme ultraviolet magneto-optic and
  photoemission spectroscopies},\ }\href
  {https://doi.org/10.1103/PhysRevLett.121.077204} {\bibfield  {journal}
  {\bibinfo  {journal} {Phys. Rev. Lett.}\ }\textbf {\bibinfo {volume} {121}},\
  \bibinfo {pages} {077204} (\bibinfo {year} {2018})}\BibitemShut {NoStop}%
\bibitem [{\citenamefont {Hamrle}\ \emph {et~al.}(2002)\citenamefont {Hamrle},
  \citenamefont {Ferr\'e}, \citenamefont {N\'yvlt},\ and\ \citenamefont
  {Vi\ifmmode\check{s}\else\v{s}\fi{}\ifmmode\check{n}\else\v{n}\fi{}ovsk\'y}}]{Hamrle2002}%
  \BibitemOpen
  \bibfield  {author} {\bibinfo {author} {\bibfnamefont {J.}~\bibnamefont
  {Hamrle}}, \bibinfo {author} {\bibfnamefont {J.}~\bibnamefont {Ferr\'e}},
  \bibinfo {author} {\bibfnamefont {M.}~\bibnamefont {N\'yvlt}},\ and\ \bibinfo
  {author} {\bibfnamefont {S.}~\bibnamefont
  {Vi\ifmmode\check{s}\else\v{s}\fi{}\ifmmode\check{n}\else\v{n}\fi{}ovsk\'y}},\
  }\bibfield  {title} {\bibinfo {title} {In-depth resolution of the
  magneto-optical {K}err effect in ferromagnetic multilayers},\ }\href
  {https://doi.org/10.1103/PhysRevB.66.224423} {\bibfield  {journal} {\bibinfo
  {journal} {Phys. Rev. B}\ }\textbf {\bibinfo {volume} {66}},\ \bibinfo
  {pages} {224423} (\bibinfo {year} {2002})}\BibitemShut {NoStop}%
\bibitem [{\citenamefont {Ashok}\ \emph {et~al.}(2025)\citenamefont {Ashok},
  \citenamefont {Hoefer}, \citenamefont {Stiehl}, \citenamefont {Aeschlimann},
  \citenamefont {Schneider}, \citenamefont {Rethfeld},\ and\ \citenamefont
  {Stadtm{\"u}ller}}]{Ashok2025}%
  \BibitemOpen
  \bibfield  {author} {\bibinfo {author} {\bibfnamefont {S.}~\bibnamefont
  {Ashok}}, \bibinfo {author} {\bibfnamefont {J.}~\bibnamefont {Hoefer}},
  \bibinfo {author} {\bibfnamefont {M.}~\bibnamefont {Stiehl}}, \bibinfo
  {author} {\bibfnamefont {M.}~\bibnamefont {Aeschlimann}}, \bibinfo {author}
  {\bibfnamefont {H.~C.}\ \bibnamefont {Schneider}}, \bibinfo {author}
  {\bibfnamefont {B.}~\bibnamefont {Rethfeld}},\ and\ \bibinfo {author}
  {\bibfnamefont {B.}~\bibnamefont {Stadtm{\"u}ller}},\ }\bibfield  {title}
  {\bibinfo {title} {Signatures of ballistic and diffusive transport in the
  time-dependent {K}err-response of magnetic materials},\ }\href
  {https://doi.org/10.1088/1367-2630/adce23} {\bibfield  {journal} {\bibinfo
  {journal} {New J. Phys.}\ }\textbf {\bibinfo {volume} {27}},\ \bibinfo
  {pages} {063001} (\bibinfo {year} {2025})}\BibitemShut {NoStop}%
\bibitem [{\citenamefont {Razdolski}\ \emph
  {et~al.}(2017{\natexlab{b}})\citenamefont {Razdolski}, \citenamefont
  {Alekhin}, \citenamefont {Ilin}, \citenamefont {Meyburg}, \citenamefont
  {Roddatis}, \citenamefont {Diesing}, \citenamefont {Bovensiepen},\ and\
  \citenamefont {Melnikov}}]{Razdolski2017}%
  \BibitemOpen
  \bibfield  {author} {\bibinfo {author} {\bibfnamefont {I.}~\bibnamefont
  {Razdolski}}, \bibinfo {author} {\bibfnamefont {A.}~\bibnamefont {Alekhin}},
  \bibinfo {author} {\bibfnamefont {N.}~\bibnamefont {Ilin}}, \bibinfo {author}
  {\bibfnamefont {J.~P.}\ \bibnamefont {Meyburg}}, \bibinfo {author}
  {\bibfnamefont {V.}~\bibnamefont {Roddatis}}, \bibinfo {author}
  {\bibfnamefont {D.}~\bibnamefont {Diesing}}, \bibinfo {author} {\bibfnamefont
  {U.}~\bibnamefont {Bovensiepen}},\ and\ \bibinfo {author} {\bibfnamefont
  {A.}~\bibnamefont {Melnikov}},\ }\bibfield  {title} {\bibinfo {title}
  {Nanoscale interface confinement of ultrafast spin transfer torque driving
  non-uniform spin dynamics},\ }\href {https://doi.org/10.1038/ncomms15007}
  {\bibfield  {journal} {\bibinfo  {journal} {Nature Commun.}\ }\textbf
  {\bibinfo {volume} {8}},\ \bibinfo {pages} {15007} (\bibinfo {year}
  {2017}{\natexlab{b}})}\BibitemShut {NoStop}%
\bibitem [{\citenamefont {Hamrle}\ \emph {et~al.}(2010)\citenamefont {Hamrle},
  \citenamefont {Pištora}, \citenamefont {Hillebrands}, \citenamefont {Lenk},\
  and\ \citenamefont {M{\"u}nzenberg}}]{MOKEAnalytical}%
  \BibitemOpen
  \bibfield  {author} {\bibinfo {author} {\bibfnamefont {J.}~\bibnamefont
  {Hamrle}}, \bibinfo {author} {\bibfnamefont {J.}~\bibnamefont {Pištora}},
  \bibinfo {author} {\bibfnamefont {B.}~\bibnamefont {Hillebrands}}, \bibinfo
  {author} {\bibfnamefont {B.}~\bibnamefont {Lenk}},\ and\ \bibinfo {author}
  {\bibfnamefont {M.}~\bibnamefont {M{\"u}nzenberg}},\ }\bibfield  {title}
  {\bibinfo {title} {{Analytical expression of the magneto-optical Kerr effect
  and Brillouin light scattering intensity arising from dynamic
  magnetization}},\ }\href {https://doi.org/10.1088/0022-3727/43/32/325004}
  {\bibfield  {journal} {\bibinfo  {journal} {J. Physics D: Applied Phys.}\
  }\textbf {\bibinfo {volume} {43}},\ \bibinfo {pages} {325004} (\bibinfo
  {year} {2010})}\BibitemShut {NoStop}%
\bibitem [{\citenamefont {Wu}\ \emph {et~al.}(2018)\citenamefont {Wu},
  \citenamefont {Liu},\ and\ \citenamefont {Luo}}]{Xufei2018}%
  \BibitemOpen
  \bibfield  {author} {\bibinfo {author} {\bibfnamefont {X.}~\bibnamefont
  {Wu}}, \bibinfo {author} {\bibfnamefont {Z.}~\bibnamefont {Liu}},\ and\
  \bibinfo {author} {\bibfnamefont {T.}~\bibnamefont {Luo}},\ }\bibfield
  {title} {\bibinfo {title} {Magnon and phonon dispersion, lifetime, and
  thermal conductivity of iron from spin-lattice dynamics simulations},\ }\href
  {https://doi.org/10.1063/1.5020611} {\bibfield  {journal} {\bibinfo
  {journal} {J. Appl. Phys.}\ }\textbf {\bibinfo {volume} {123}},\ \bibinfo
  {pages} {085109} (\bibinfo {year} {2018})}\BibitemShut {NoStop}%
\bibitem [{\citenamefont {Taktikos}\ \emph {et~al.}(2013)\citenamefont
  {Taktikos}, \citenamefont {Stark},\ and\ \citenamefont
  {Zaburdaev}}]{MSDAnalytical}%
  \BibitemOpen
  \bibfield  {author} {\bibinfo {author} {\bibfnamefont {J.}~\bibnamefont
  {Taktikos}}, \bibinfo {author} {\bibfnamefont {H.}~\bibnamefont {Stark}},\
  and\ \bibinfo {author} {\bibfnamefont {V.}~\bibnamefont {Zaburdaev}},\
  }\bibfield  {title} {\bibinfo {title} {How the motility pattern of bacteria
  affects their dispersal and chemotaxis},\ }\href
  {https://doi.org/10.1371/journal.pone.0081936} {\bibfield  {journal}
  {\bibinfo  {journal} {PLOS ONE}\ }\textbf {\bibinfo {volume} {8}},\ \bibinfo
  {pages} {e81936} (\bibinfo {year} {2013})}\BibitemShut {NoStop}%
\bibitem [{\citenamefont {Grosso}\ and\ \citenamefont
  {Parravicini}(2014)}]{FourierDerivation}%
  \BibitemOpen
  \bibfield  {author} {\bibinfo {author} {\bibfnamefont {G.}~\bibnamefont
  {Grosso}}\ and\ \bibinfo {author} {\bibfnamefont {G.~P.}\ \bibnamefont
  {Parravicini}},\ }\href@noop {} {\emph {\bibinfo {title} {Solid state
  physics}}},\ \bibinfo {edition} {2nd}\ ed.\ (\bibinfo  {publisher} {Academic
  Press},\ \bibinfo {year} {2014})\ pp.\ \bibinfo {pages}
  {509--511}\BibitemShut {NoStop}%
\end{thebibliography}
%\bibliographystyle{apsrev4-1}

\end{document}